\documentclass[11pt,expanded,copyright]{fsuthesis}

\usepackage{amsfonts}
\usepackage[utf8]{inputenc}
\usepackage{graphicx}
\usepackage{subcaption}
\usepackage{mathtools}      
\usepackage{amssymb}	      
\usepackage{bm}
\usepackage{tikz}
\usepackage{units}
\usepackage{cleveref}
\usepackage[numbers]{natbib} 
\usepackage{url}
\usepackage{booktabs}
\usepackage{algorithm}
\usepackage{algorithmic}

\graphicspath{/data1/blg13/Projects/Dissertation/figs/}
\newcommand{\TP}{\mathrm{TP}}
\newcommand{\FP}{\mathrm{FP}}
\newcommand{\TN}{\mathrm{TN}}
\newcommand{\FN}{\mathrm{FN}}
\newcommand{\U}{U}
\newcommand{\F}{F}
\newcommand{\Source}{S}
\newcommand{\Q}{\phi}
\newcommand{\etal}{et al. }

\newcommand*{\I}{\imath}

\title{Toward Data-Driven Subgrid-Scale Modeling of the Zel'dovich Deflagration-To-Detonation Mechanism in Dense Stellar Plasmas}
\author{Brandon L. Gusto}
\college{College of Arts and Sciences}
\department{Department of Scientific Computing}  
\manuscripttype{Dissertation}              
\degree{Doctor of Philosophy}               
\degreeyear{2023}
\defensedate{April 3rd, 2023}

\committeeperson{Tomasz Plewa}{Professor Directing Traffic}
\committeeperson{Mark Sussman}{University Representative}
\committeeperson{Adrian Barbu}{Committee Member}
\committeeperson{Bryan Quaife}{Committee Member}
\committeeperson{Olmo Zavala}{Committee Member}

\DeclareMathOperator*{\argmin}{arg\,min}

\begin{document}

\frontmatter
\maketitle
\makecommitteepage

\begin{dedication}
  This dissertation is dedicated to my loving family: to my mother and father,
  Renee and Jeffrey, my older brother, Cody, and my little sister, Jenna. Also
  to my wonderful fianc\'e, Marie. Finally, of course to my Oma who never
  expected anything less than a doctorate degree. Without their love and
  support this manuscript would not have been possible.
\end{dedication}

\begin{acknowledgments}
  The author would like to first acknowledge the Science, Mathematics, and
  Research for Transformation (SMART) Scholarship-for-Service Program for
  providing support for this work. Additionally, he would like to thank his
  advisor, Dr. Tomasz Plewa for his unwavering support and guidance; Tomasz's
  dedication to his students and his passion for scientific truth has been a
  constant source of inspiration. The author also would like to thank
  Dr. Christoph Federrath from the Australian National University for his help
  in establishing turbulence driving parameters over many constructive virtual
  meetings.
\end{acknowledgments}

\tableofcontents
\listoftables
\listoffigures

\begin{listofsymbols}
  \begin{center}
    \begin{tabular}{r l}
      $t$ & time \\
      $x$ & spatial coordinate \\
      $S_{L}$ & the laminar flame speed \\
      $\tau$ & induction time \\
      $u_{sp}$ & reactive wave speed \\
      $z$ & reactive wave speed normalized by local soundspeed \\
      $c$ & speed of sound \\
      $D_{CJ}$ & velocity of a Chapman-Jouget detonation \\
      $\rho$ & density \\
      $\rho_{\mathrm{amb}}$ & ambient density \\
      $T_{\mathrm{amb}}$ & ambient temperature \\
      $\delta \rho$ & amplitude of density perturbation \\
      $R$ & width of density perturbation \\
      $A$ & normalized amplitude of density perturbation \\
      $u$ & velocity \\
      $p$ & pressure \\
      $\bm{X}$ & species concentrations \\
      $X_{f}$ & fuel concentration \\
      $\bm{R}$ & species reaction rates \\
      $\dot{Q}$ & nuclear energy generation term \\
      $\Delta x$ & mesh resolution \\
      $N$ & number of computational cells\\
      $\bm{U}$ & approximate solution \\
      $\hat{\bm{F}}$ & numerical fluxes \\
      $\hat{\bm{S}}$ & numerical source terms \\
      $\bm{\phi}$ & exact solution \\
      $r_{0}$ & radius of hotspot region \\
      $r_{s}$ & distance from the center of the hotspot to the point where the reactive wave speed equals the soundspeed \\
      $r_{s}^{*}$ & the critical value of $r_{s}$ below which detonation cannot occur \\
      $\sigma_{0}$ & standard deviation of induction times in hotspot region \\
      $c_{0}$ & soundspeed in hotspot region \\
      $\varepsilon$ & threshold for defining size of hotspot region \\
      $\Delta_{\mathrm{wd}}$ & ILES filter cutoff for the full-star explosion scale of the white dwarf \\
      $\Delta_{\mathrm{tb}}$ & ILES filter cutoff for the turbulence-in-a-box scale \\
      $q$ & actual label of a neural network input sample \\
      $\hat{q}$ & predicted label of a neural network input sample \\
      $\bm{\chi}$ & data of a neural network input sample \\
      $\varphi$ & activation function \\
      $b$ & layer biases \\
      $w$ & layer weights \\
      $\ell$ & network layer number \\
      $n_{s}$ & number of training samples \\
      $n_{x}$ & number of spatial points in the input data \\
      $n_{ch}$ & number of input channels in a convolutional network \\
      $\delta_{t}$ & time delay between network prediction and actual detonation \\
      $\delta_{x}$ & spatial distance between loction where network prediction made and estimated detonation origin \\
    \end{tabular}
  \end{center}
\end{listofsymbols}


\begin{abstract}
  A novel, data-driven model of deflagration-to-detonation
  transition (DDT) is constructed for application to explosions of
  thermonuclear supernovae (SN Ia). The DDT mechanism has been suggested as the
  necessary physics process to obtain qualitative agreement between SN Ia
  observations and computational explosion models. This work builds upon a
  series of studies of turbulent combustion that develops during the final
  stages of the SN explosion. These studies suggest that DDT can occur in the
  turbulerized flame of the white dwarf via the Zel'dovich reactivity gradient
  mechanism when hotspots are formed. We construct a large database of direct
  numerical simulations that explore the parameter space of the Zel'dovich
  initiated detonation. We use this database to construct a neural network
  classifier for hotspots. The classifier is integrated into our supernova
  simulation code, FLASH/Proteus, and is used as the basis for a subgrid-scale
  model for DDT. The classifier is evaluated both in the training environment
  and in reactive turbulence simulations to verify its accuracy in realistic
  conditions.
\end{abstract}

\mainmatter

\chapter{Introduction}
\label{chp:intro}
 
  \section{Motivation and outline}

    The rapid transition of a subsonic flame, also known as a  deflagration, to
    a supersonic and more energetic detonation can occur suddenly in numerous
    industrial scenarios including, but not limited to, fuel transportation
    \citep{osti_7351990}, hazardous chemical storage \citep{pasman2020}, and
    loss-of-cooling incidents in nuclear reactors with H$_{2}$-air-steam
    mixtures \citep{etde_20148318}. This phenomenon is known as the
    deflagration-to-detonation transition (DDT), and it is a topic of active
    research in the combustion community. While a DDT in the aforementioned
    instances may have devestating effects, the phenomenon can also be utilized
    for constructive purposes in engineering applications. For example, DDT can
    be exploited in pulse detonation engines as an alternative to direct
    detonation initiation \citep{roy2004}.

    Beyond terrestrial applications, DDT is suspected to have contributed to
    the makeup of the universe through the process of nucleosynthesis during
    thermonuclear stellar explosions, also known as supernovae (SNe). These
    fascinating events result in the release of an enormous amount of energy;
    on the order of $\unit[10^{51}]{erg}$.  The exact mechanism of the
    explosion is not yet entirely understood. This dissertation aims to provide
    a computational model for the possible DDT in thermonuclear SNe.

    In this chapter the reader is provided with a brief introduction to the
    basic combustion physics relevant to DDT and to the topic of thermonuclear
    explosions. \Cref{chp:background} reviews the prior work done in modeling
    the deflagration-to-detonation transition in thermonuclear SNe and provides
    an overview of the modeling approach used in the present work. In
    \Cref{chp:hotspot_analysis} the critical conditions for detonation
    initiation are investigated.  \Cref{chp:machine_learning} introduces a novel,
    data-driven approach to subgrid-scale (SGS) modeling of DDT.  Concluding
    remarks are provided in \Cref{chp:summary}, as well as a critical
    discussion of the aspects of the present work that can be improved or
    expanded upon.

  \section{Relevant combustion physics}



    The deflagration-to-detonation transition requires the presence of
    combustion waves. Combustion waves are defined by a perturbation in fuel
    consumption rate and energy release propagating through a fuel mixture.
    Several viable conditions for the idealized, one-dimensional combustion
    wave are described by the well-known Rankine-Hugoniot relation (see
    Chapter 4 of \cite{kuo2005}). In particular, the weak deflagration and the
    detonation are the two modes of combustion most frequently encountered in
    actual physical systems.

    The deflagration is characterized by a subsonic propagation velocity. The
    deflagration front results in an increase in fluid velocity, and a mild
    decrease in gas density and pressure. The velocity of a one-dimensional
    deflagration is known as the laminar flame speed, $S_{L}$, and it is
    determined by thermal conduction and mass diffusion \cite{kuo2005}.

    The detonation is a supersonic mode of combustion that has a shock-reaction
    structure. The first theoretical descriptions of the one-dimensional
    detonation wave were provided independently by Chapman \citep{chapman1899}
    and Jouget \cite{jouget1905}. While the Chapman-Jouget (CJ) solution is
    fairly accurate, it considers the reaction zone as infinitesimally thin,
    whereas in reality the reaction zone has some finite width.

    A more detailed description of the detonation wave was provided by
    Zel'dovich \cite{zeldovich1940}, von Nuemann \cite{von1943} and D\"orring
    \cite{dorring1943} (ZND). The ZND model describes the one-dimensional
    detonation as a leading shockwave with a narrow reaction zone behind it.
    The shockwave serves to compress the upstream fuel, rapidly raising the
    temperature, pressure, and density closer to the point of ignition. The
    reaction zone behind the shockwave can be divided into two regions: the
    induction zone and the reaction zone. The induction zone is a region inside
    which induction times, the estimated amount of time for the fuel in a small
    mass element to burn, and reaction rates are both low. In the reaction
    zone, reaction rates increase substantially away from the shockwave. In a
    self-sustaining detonation, the reaction zone is responsible for a
    significant amount of thermal pressure due to the energy release. This
    thermal pressure acts like a piston behind the leading shockwave,
    propelling the shockwave into the unburnt region.


    The initiation of a detonation requires the coupling of a compressive wave
    and a reactive wave. If a sufficiently strong ignition source is present a
    detonation can be directly initiated \cite{abouseif1982}. This source must
    ignite the fuel and provide a strong shockwave. Such a strong ignition
    source is rarely present in nature, yet detonations are known to arise
    nevertheless. Another means of detonation initiation was proposed by
    Zel'dovich \etal \cite{zeldovich1970} via the mechanism of
    \textit{spontaneous propagation}.



    \subsection{Spontaneous detonation}
    \label{sec:zeldovich_regimes}

      Consider a region of fuel in one-dimensional space with a nonuniform
      distribution of the induction time field, $\tau(x)$, where $x$ is the
      spatial coordinate. Such regions are referred to hereafter as `hotspots'.
      The nonuniform spatial profile in the induction time gives the potential
      for a `spontaneous' wave due to the successive release of energy as each
      mass element undergoes burning and energy release. This wave is
      independent of the soundspeed and thermal conductivity of the gas. The
      phase velocity of the spontaneous wave is given as
      \begin{equation}
        u_{sp} \coloneqq \left( \frac{\partial \tau}{\partial x} \right)^{-1}.
        \label{eqn:usp}
      \end{equation}
      This relation indicates that regions of low induction time gradient have
      a higher phase velocity and vice versa.

      For a shock-reaction complex to form, both a compressive and reactive
      wave must coexist for some period of time to allow the feedback mechanism
      to strengthen (and steepen) the compressive wave.  The velocity of the
      compressive wave is primarily governed by the soundspeed, $c$, but the
      compressive wave may also be accelerated by overpressure due to burning
      behind it. Meanwhile, the speed of the emerging reactive wave is governed
      primarily by \Cref{eqn:usp}. Thus, the speed of the reactive wave should
      be in the vicinity of the local soundspeed if a coupled burning process
      is to have any chance of formation.
      
      Zel'dovich \citep{zeldovich1980} describes several regimes of spontaneous
      propagation. These regimes have also been neatly summarized in the
      textbook of Kuo \cite{kuo2005}, and they are only briefly reviewed here.
      Given the speed of a CJ detonation, $D_{CJ}$, the regimes are given by:
      \begin{enumerate}
        \item If $u_{sp} > D_{CJ}$, the result is an under-driven detonation wave.
          In the limit $u_{sp} \rightarrow \infty$ this case corresponds to a constant-volume
          explosion.
        \item If $u_{sp} \leq D_{CJ}$ then a shock wave can form ahead of the
          reactive wave and ultimately transform into a detonation through the
          feedback process.
        \item If $S_{L} < u_{sp} \ll c < D_{CJ}$, then a weak deflagration wave
          is formed which travels faster than the laminar flame speed.
        \item If $u_{sp} < S_{L}$, then a deflagration forms which travels at
          the laminar flame speed.
      \end{enumerate}
      Essentially, if the magnitude of $u_{sp}$ is too large, then the
      successive energy release happens on a timescale smaller than the
      acoustic timescale such that no coupled feedback mechanism can occur. Too
      small, and the acoustic wave will outrun the reactive wave, leaving
      either a slightly overdriven deflagration or a deflagration travelling at
      speed $S_{L}$.

      The aforementioned regimes are adequate to describe only the most simple
      initial hydrodynamic states and do not consider either the initial
      hydrodynamic state or complex thermo-fluid dynamics, as Zel'dovich
      himself points out. In an effort to compensate for these effects the
      authors of \cite{gu2003} introduce a \textit{range} of critical
      reactivity gradients which may lead to detonation in their analysis of
      hotspots in chemical fuel mixtures. Similarly, the existence of critical
      reactivity gradient values was also suggested in a large numerical study
      by Oran \& Gamezo \cite{oran2007}. The authors suggest that the
      reactivity gradient mechanism has some `universality' across many regimes
      and mixtures.

      Numerous experimental studies have examined the process of DDT, and some
      of these have provided evidence in support of the role of the reactivity
      gradient mechanism in detonation formation. Urtiew, Oppenheim, \& Saunders
      \cite{urtiew1966} studied the transition to detonation of flames inside
      of a long enclosed tube. Lee, Knystautas, \& Yoshikawaet \cite{lee1978}
      studied detonations initiated via a photochemical energy source and found
      that a certain range of gradient values are necessary for the shock wave
      to amplify and form a detonation. For a useful review of the spontaneous
      detonation of hotspots covering the mathematical, experimental, and
      computational modeling frames of reference, see \cite{bartenev2000}.

      While laboratory experiments have managed to initiate detonations either
      by boundary-induced turbulence or direct initiation methods, the
      formation of detonations in \textit{unconfined} turbulence does not occur
      as readily.  The mechanism that is believed to lead to the formation of
      detonation-prone, nonuniformly heated regions in unconfined settings is
      discussed next. This mechanism is known as flame acceleration.

    \subsection{Flame acceleration}
    \label{sec:flame_acc}


      The process of flame acceleration is essential for the
      deflagration-to-detonation transition in an unconfined setting. There are
      numerous mechanisms which can cause a laminar flame to become turbulent.
      In confined settings, surface roughness or obstacles typically play a
      significant role (see \cite{shepherd1992}). Without solid boundaries
      however, turbulence can occur along the flame surface due to hydrodynamic
      instabilities such as Rayleigh-Taylor (RT) or Kelvin Helmholtz (KH)
      instabilities, or it can be generated by the intrinsic flame instability
      at the flame surface itself \citep{shepherd1992}. Turbulence can
      significantly modify the structure of the laminar flame front depending
      on its intensity and length scales.  Higher local reaction rates, flame
      speeds, and enthalpy can be achieved in areas with a high degree of flame
      stretching, or curvature.

      The turbulent flame can be classified into at least five distinct regions
      \cite{shepherd1992} based on the ratios of the velocity and
      characteristic length scales of turbulence to those of the laminar flame.
      As mentioned, the \textit{laminar flame regime} is rarely achieved in
      natural settings due to various instabilities. If the velocity
      fluctuations are small, and the turbulent length scales are large with
      respect to the flame thickness, then the flame is only gently distorted
      by the velocity field. This regime is known as the \textit{wrinkled flame
      regime}.
      
      As the turbulent velocity fluctuations increase, the length scales tend
      to become smaller, and the disturbances to the flame become more
      pronounced. This has the effect of increasing the strain on the flame
      surface. If the flame surface remains unbroken, then this regime is
      referred to as the \textit{corrugated flame regime}. When the smallest
      eddies become comparable to the flame thickness they can enter and
      broaden the preheat zone \cite{peters2000}. This regime is known as the
      \textit{thin reaction zone}.  These regimes result, to varying degrees,
      in increased flame surface area, but more importantly in strain rate
      effects that further increase flame velocities.

      Then there is a regime in which turbulent eddies shatter the flame
      surface. This regime, known as the \textit{distributed reaction zone}, is
      realized when the timescale of the turbulent eddies is smaller than the
      timescale of the burning. In other words, eddies are able to mix the
      flame surface with the upstream fuel more quickly than the flame can burn
      away the eddy. This has important consequences for the existence of
      hotspot regions.
      

  \section{Thermonuclear supernova explosion scenario}

    The evolution of a main sequence star such as the Sun draws to an end when
    its hydrogen and helium fuels have been exhausted, leaving behind an
    extremely dense, hot, electron degenerate plasma core known as a white
    dwarf (WD) star \citep{hayashi1962}.  In this final stage of the star's
    evolution, the WD slowly burns its remaining carbon and oxygen fuels over a
    period of several billion years as the star gently cools. The thermal
    pressure due to the mild burning of carbon, as well as the electron
    degeneracy pressure serve to resist the weight of the material in the outer
    layers, keeping the star in hydrostatic equilibrium. 

    If a WD is in a close binary system then there is the potential for an
    enormous and rapid release of energy via a thermonuclear explosion. There
    are two accepted progenitor systems: single degenerate (SD), consisting of
    a WD and a non-degenerate main sequence giant star, and double degenerate
    (DD), consisting of two degenerate, intermediate mass white dwarves.

\begin{figure}[!ht]
      \center
      \includegraphics[width=0.65\textwidth]{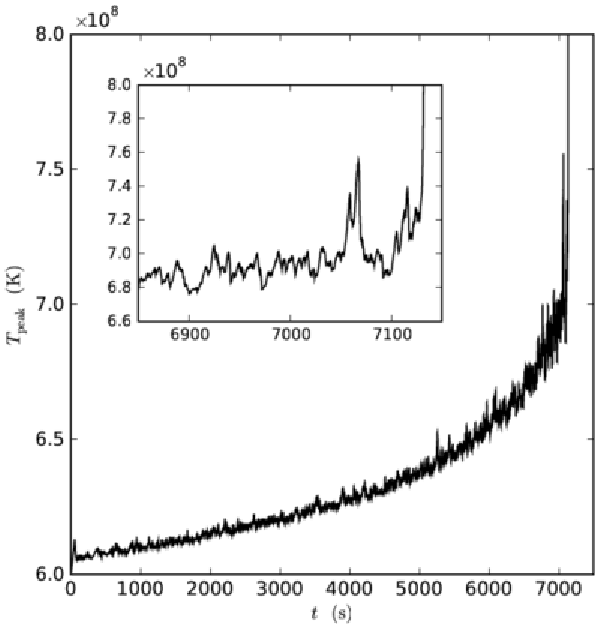}
      \caption{Simulated evolution of peak temperature within the core region
      of a white dwarf up to the moments of ignition. The inset shows the rapid
      temperature rise occuring in the last few hundred seconds prior to
      ignition. Figure borrowed from \cite{nonaka2012}.}
      \label{fig:simmering}
    \end{figure}

    In the SD scenario, the interior of the rotating WD is mostly composed of
    carbon and oxygen (C/O) fuel due to the burning of hydrogen.  This C/O
    core experiences compressive heating due to the massive outer layers, but
    also radiative cooling. While these forces are closely balanced for some time, there is a
    trend toward heating in the star that lasts for centuries due to the
    accumulation of mass from the donor star \citep{zingale2009}. This phase is
    known as the ``simmering'' phase. When the radiative cooling processes are
    no longer able to cool the star, convection in the core is tought to take
    over. Eventually, the core temperature rises to a point where ignition of
    carbon takes place, and a runaway process begins. The convection during the
    simmering stage was studied by Nonaka \etal \cite{nonaka2012}, and the
    authors observed the evolution of temperature as shown in
    \Cref{fig:simmering}.
    In this study the peak temperature in the system rapidly rises
    within a matter of minutes (note the scale of the inset of
    \Cref{fig:simmering}).

\begin{figure}[!ht]
      \center
      \includegraphics[width=0.82\textwidth]{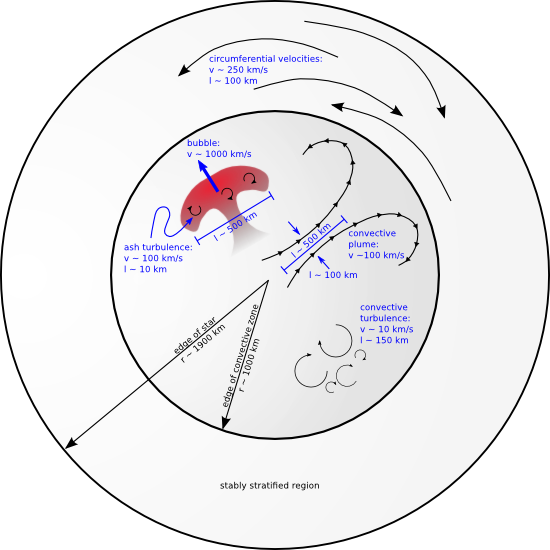}
        \caption{Illustration showing the regimes within the white dwarf. Note 
          the ash plume (the mushroom-like
          shape drawn in red) which rises very quickly due to bouyancy, and
          escapes the core region. The rising ash plume induces intense
          small-scale turbulence due to RT and KH instabilities. Figure
          borrowed from \cite{nonaka2012}.}
      \label{fig:wdcartoon}
    \end{figure}

    It is supposed that the ignition occurs at one or more regions near the
    core. The location of the first ignition within the star is important in
    determining the observational characteristics of the explosion
    \citep{niemeyer1997}.  Whether the first ignition occurs in the center of
    the core or off-center has been debated in the literature. An off-center
    scenario has been determined to be most likely in several works
    \citep{nienmeyer1996, nonaka2012}. After ignition, ash plumes (also
    referred to as flame ``bubbles"), rise toward the surface of the star due
    to bouyancy forces.  An illustrative diagram is borrowed from
    \cite{nonaka2012} and included here in \Cref{fig:wdcartoon}.
    The rapidly rising plume is unstable to Landau-Darrieus, Rayleigh-Taylor,
    and Kelvin-Helmholtz instabilities \cite{bell2004}. The result is the
    generation of turbulence along the edges of the plume.

    Turbulence along the surface of the hot ash plume mixes cold fuel from
    outside of the WD core with hot ash from inside the plume. The mixing
    between cold fuel and hot ash serves to warm the fuel, bringing it closer
    to the point of ignition. At this point it is speculated if a
    transition from deflagration to detonation can occur. The so-called
    ``delayed detonation'' model was introduced by Khokhlov
    \cite{khokhlov1991a}, arguing that spontaneous detonation (i.e.  detonation
    powered by the Zel'dovich reactivity gradient mechansim) can occur within
    the turbulent regions of a deflagrating WD.

    The delayed detonation model was proposed to address the inability of 
    pure detonation models \cite{arnett1969} or pure deflagration models
    \cite{nomoto1976} to match observations \cite{khokhlov1997}. The
    pure detonation model is known to produce an insufficient amount of
    intermediate mass elements \cite{khokhlov1991a}, while the pure
    deflagration models tend to produce an insufficient amount of energy,
    evidenced by an excess of $\prescript{54}{}{\mathrm{Fe}}$
    \cite{nomoto1984}.

    The spectral properties of the ejecta of delayed detonation models were
    investigated in \cite{hoflich1995}. Other theories have also been proposed.
    Plewa \etal \cite{plewa2004} introduced the gravitationally confined
    detonation model, which posits that after an off-center ignition, the
    resulting ash plume bursts through the surface layers, and a mix of cold
    fuel from the stably stratisfied region, and hot ash from the plume
    converge at a point opposite from the surface burst location. The
    converging flow is supersonic, and the interaction between shockwaves and
    the hot fuel has been shown to produce a detonation in computational
    models.

    Even though these models may be successful at qualitatively matching
    observations, it is important to point out that their maximum resolution
    (usually on the $\unit[1\times10^{5}]{cm}$ scale) is not resolving
    anywhere near the DDT-controlling scales. The actual SGS conditions in all
    of these scenarios are largely unknown. To date, a criteria that can
    predict the onset of DDT in Type Ia SNe is not available.  There remains a
    significant degree of uncertainty about the turbulent conditions during the
    deflagration phase.
    
    Even if the large-scale conditions are known, the accurate modeling of
    flame-turbulence interactions on smaller scales is challenging
    \cite{hicks2015}. Is a transition to detonation possible along the
    interface between cold fuel and hot ash from the rising plume? If so, what
    is the mechanism? Can this mechanism be parameterized? In this dissertation
    the available knowledge on these topics will be collected and a path toward
    a useful metric for predicting the onset of DDT will be proposed. Physical
    insights from the existing literature will be used to guide
    state-of-the-art, data-driven techniques.


\chapter{Challenges of Modeling Thermonuclear Supernovae}
\label{chp:background}

  \section{Introduction}

    The primary limitation of computational models for SN Ia is the so-called
    ``tyranny of scales''; that is to say that some physics processes are known
    to occur on very large spatial and temporal scales, while others may occur
    simultaneously on significantly smaller scales. For example, the
    characteristic size of turbulent eddies in the convectively unstable core
    is thought to range from hundreds of kilometers down to several
    centimeters. Likewise, the disparity in timescales between the large-scale
    convective motions and nuclear burning spans roughly seven orders of
    magnitude.

\begin{figure}[!ht]
      \center
      \includegraphics[scale=0.55]{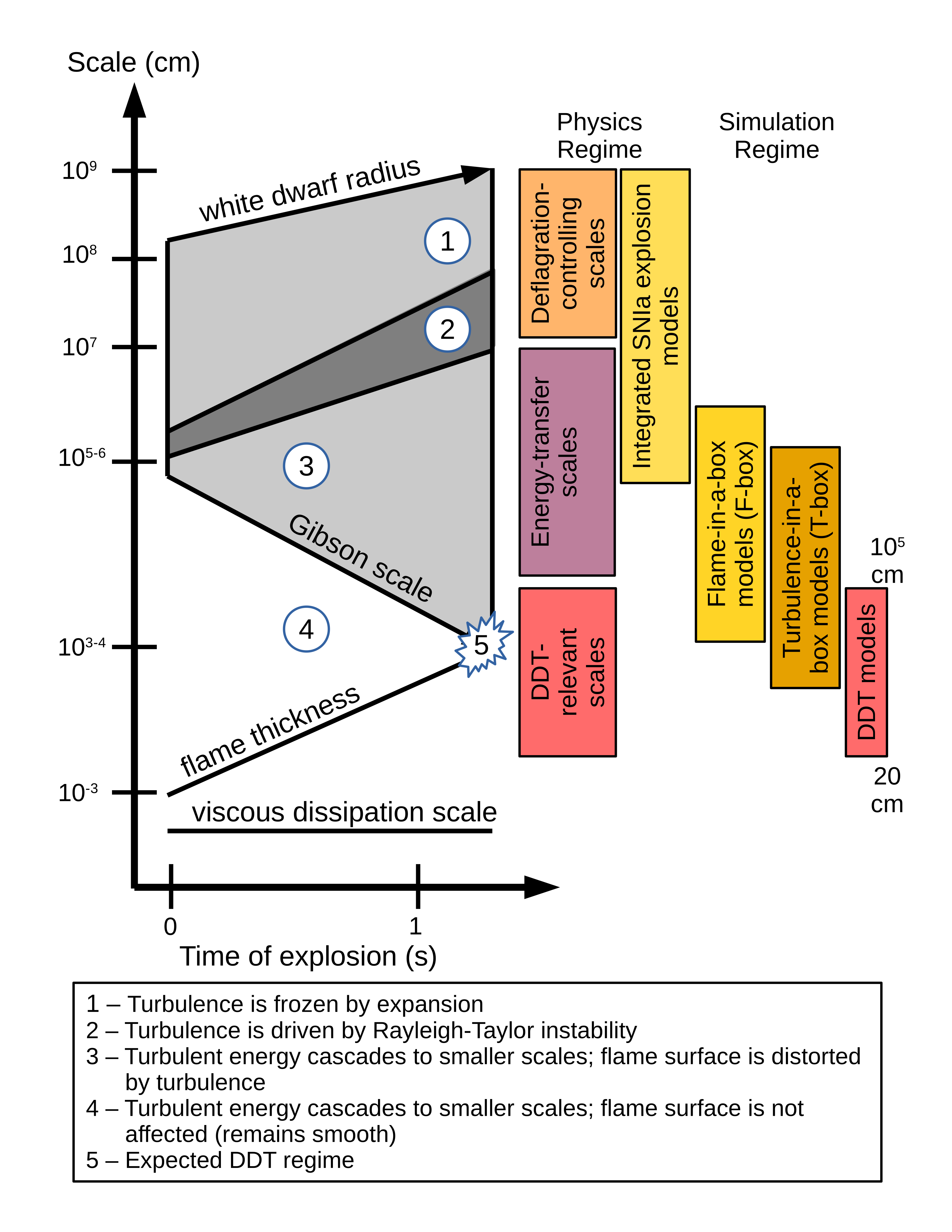}
      \caption{Spatial and temporal scales of the white dwarf during the
        explosion process. Physical regimes and their corresponding computational modeling
        regimes are highlighted. The spatial scales range from the
        deflagration-controlling scales on the order of the size of the white
        dwarf, down to the viscous dissipation scale. DDT occurs on scales
        slightly larger than the viscous dissipation scale. See text for
        details.}
      \label{fig:wdscales}
    \end{figure}

    \Cref{fig:wdscales}
    illustrates the extreme challenge of scales in this problem.  The upper
    limit of the spatial range (the vertical axis in \Cref{fig:wdscales}) is
    defined by the radius of the WD, which is typically on the order of
    hundreds of kilometers (or about $10^{8}$ to $\unit[10^{9}]{cm}$). On
    slightly smaller scales, roughly $10^{7}$ to $\unit[10^{8}]{cm}$, the
    initial modes of the RT unstable deflagration plumes are determined. Once
    the star begins exploding, deflagration plumes travel toward the outer
    layer of the star where they expand and lose momentum (region $1$). As the
    deflagration plumes expand outward they induce mixing between cold fuel and
    hot ash, producing turbulence (region $2$). The energy from these
    large-scale motions is transferred to smaller scales through the turbulent
    energy cascade. Eddies larger than the Gibson scale \citep{peters1988} are
    able to distort the flame surface with their momentum before they burn away
    (region $3$), while smaller eddies are not (region $4$). The Gibson scale
    is a function of the laminar flame speed and thus decreases during the
    evolution as density decreases (line separating regions $3$ and $4$).
    Meanwhile, the flame thickness increases with decreasing density. At the
    point where the spatial scales of the Gibson scale and the flame thickness
    become comparable (region $5$), the distributed burning regime is entered.
    In this regime, DDT is thought to be most probable.

    On the right side of the diagram in \Cref{fig:wdscales} the
    physics regime is summarized in one column, and the simulation regime in the other.
    The scales between roughly $\unit[10^7]{cm}$ and $\unit[10^9]{cm}$ are
    defined as the deflagration-controlling scales. On these scales, the star's
    radius, compositional makeup, Atwood number (a dimensionless ratio of
    densities across an interface), and other properties determine the
    characteristics of the deflagrating ash plume. Scales smaller than this
    range, from about $\unit[10^4]{cm}$ to $\unit[10^7]{cm}$, can be considered
    as the energy-transfer scales in which turbulent kinetic energy cascades
    from large to small scales as per the theory of isotropic turbulence
    \cite{kolmogorov1941}. Below these scales are the scales relevant to DDT,
    from the viscous dissipation scale of about $\unit[10^{-3}]{cm}$ to
    $\unit[10^4]{cm}$.

    Unfortunately the ability to resolve the roughly 10 orders of magnitude
    separating the spatial scales in the WD explosion is currently
    technologically out of reach. The only feasible solution at present is to
    consider smaller computational domains. In the right text column of the
    figure, four simulation regimes are introduced: the full-star explosion
    model, the flame-in-a-box models which study the RT deflagration plumes,
    the turbulence-in-a-box models, which study finer details within the
    flame brush of the RT deflagration plume, and finally the direct numerical
    simulation of detonation formation. In particular the turbulence-in-a-box
    models study how hotspot regions can form and possibly detonate.

    The approach by other researchers in the computational astrophysics
    community is not dissimilar from the present one. Typically either the full
    star is modeled and turbulence and/or flame physics are handled by scaling
    relations or SGS models such as in \cite{schoolmann2013}, or the flame
    physics on smaller scales are solved directly \cite{aspden2010} in order to
    better design SGS models or refine model parameters for full-star models.
    The hope by researchers is that the manual iteration back and forth between
    full-star models and micro-physics models may achieve convergence on some
    aspects of the problem.

  \section{Modeling of DDT}

    As discussed in \Cref{chp:intro}, the pure deflagration and pure detonation
    models of Type Ia SNe are both known to produce discrepancies with
    observations. This has led researchers to consider delayed detonation
    models, among others. In the delayed detonation scenario, again, DDT is
    suspected to occur when the ash plume breaks out of the core region of the
    star.

    According to Kohkhlov, Oran, \& Wheeler \cite{khokhlov1997},
    determining whether a transition can occur in this scenario basically
    amounts to (1) determining which characteristics allow a hotspot to form a
    self-sustaining detonation wave via the Zel'dovich reactivity gradient
    mechanism, and (2) whether the outgoing ash plume and subsequent reactive
    turbulence can produce hotspots with those characteristics.

    \subsection{Conditions for the transition to detonation}

      Several prominent works that produced some estimates of the conditions
      necessary for a hotspot to spontaneously produce a detonation wave are
      now briefly reviewed. As discussed in \Cref{chp:intro}, one of the
      earliest analytical works is by Zel'dovich \etal \citep{zeldovich1970} in
      which the authors determine critical gradients of temperature for an
      initially linear temperature profile in a chemical gas mixture. In the
      context of thermonuclear flames, Khokhlov \citep{khokhlov1991b} studies
      the evolution of hotspots using a statistical approach. In his work the
      initial conditions and evolution of temperature and induction time are
      described by a distribution function.

      One of the key results of this work is the description of the necessary
      condition for detonation initiation in a region of size $r_0$ as
      \begin{equation}
        \sigma_{0} < \frac{r_{0}}{\alpha \cdot c}
        \label{eqn:khokhlov}
      \end{equation}
      where $\sigma_{0}$ is the standard deviation of induction times in the
      region, $c$ is the soundspeed in the region, and $\alpha$ is a parameter
      that in general depends on the background conditions and turbulence.  On
      the right-hand side of \Cref{eqn:khokhlov} is the approximate
      sound-crossing time within the region $r_{0}$. The sound-crossing time is
      the amount of time it takes a soundwave to travel a specified distance.
      On the left-hand side, is the variation in induction times. This value is
      indicative of the reactive wavespeed, as a smaller amount of variation
      necessarily implies that the `mini-explosions' are better synchronized in
      time.  Better synchronization in time leads to a faster phase velocity of
      the spontaneous wave and a better opportunity for coupling to occur
      between the reactive and compressive waves.

      What Khokhlov argues via \Cref{eqn:khokhlov} is that the reactive
      timescale given by the variation in induction times $\sigma_{0}$ must be
      approximately less than the sound-crossing time of the region in order
      for a detonation to occur. If this condition is not satisfied then too
      much of the energy produced via burning can escape the region in the form
      of acoustic energy for the coupling process to become self-sustaining.

\begin{figure}[!ht]
          \center
          \includegraphics[width=0.55\textwidth]{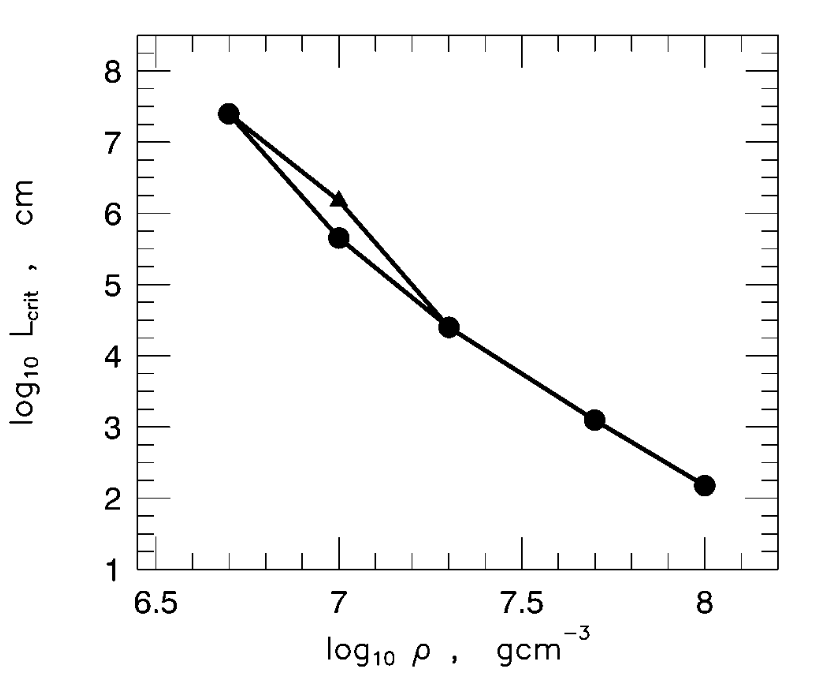}
          \caption{Critical length of hotspot region versus density for two
            nuclear energy release rates ($\unit[4 \times 10^{17}]{ergs\ g^{-1}}$,
            circles; $\unit[3 \times 10^{17}]{ergs\ g^{-1}}$, triangles). Figure
            borrowed from \citep{khokhlov1997}.}
          \label{fig:lcritvsdens}
      \end{figure}

      The previously mentioned work of Khokhlov, Oran, \& Wheeler
      \cite{khokhlov1997} analyzed hotspot configurations with linear profiles
      in fuel concentration, each beginning with zero fuel abundance at the
      center of the region, and reaching the ambient abundance at some length,
      $L$.  The assumption of linearity of the fuel concentration is a
      simplistic model of the situation in which cold fuel and hot ash mix.
      In the study, the length of the perturbed region is varied along with the
      density in order to determine the critical length and its dependence on
      density.  The results of this work are summarized in
      \Cref{fig:lcritvsdens}.
      The figure indicates that the critical length required for detonation
      decreases with increasing density. The authors attribute this to the
      specific heat of the matter being lower at higher densities, meaning that
      less heat is required to raise the temperature of the fuel toward ignition.

      A later study by Seitenzahl \etal \cite{seitenzahl2009} investigated in
      much more detail the formation of spontaneous detonations from hotspots
      in the context of SN Ia. Their study introduced several novel aspects to
      the modeling of hotspots: a variety of initial temperature profiles,
      including linear (as in \cite{khokhlov1997}), exponential, Gaussian, and
      several non-physically motivated profiles; grid convergence studies; and
      some consideration of multi-dimensionality.

      The estimates by Khokhlov \etal of the critical size for a mixed region
      to produce a detonation were used as constraints in the construction of a
      SGS model for DDT in the context of SN Ia by Ciaraldi-Schoolmann,
      Seitenzahl, \& R\"opke \cite{schoolmann2013}. The authors' model also
      incorporates estimated constraints on the turbulent velocity
      fluctuations, fuel composition, density, and fractal characteristics of
      the flame surface known to produce DDT.  These quantities are computed on
      the resolved scale.

    \subsection{Turbulent conditions in the deflagrating white dwarf}


      Whether the properties of the reactive turbulence in the deflagrating WD
      can support the formation and survival of hotspot regions long enough for
      them to form detonation waves is still somewhat of an open question. In
      fact, not only must these hotspot regions remain unperturbed for some
      time following their formation, but they must contain a significant
      amount of fuel as well. Questions related to the formation of hotspots on
      scales relevant to DDT have been investigated in a few works.

      Lisewski, Hillebrandt, \& Woosley \cite{lisewski2000} investigated the
      ability of small-scale turbulence with varying densities and velocity
      fluctuations to create detonations.  The authors found detonations to
      occur at turbulent velocity fluctuations of between $\unit[0.5
      \times10^{8}]{cm\ s^{-1}}$ and $\unit[10^{8}]{cm\ s^{-1}}$
      depending on the density and fuel composition.      

      Niemeyer \& Woosley \cite{niemeyer1997} argue that the existence of
      sufficiently large hotspot regions with preferable induction time
      gradients is unlikely. The authors conclude that if viable hotspots are
      created at all, they likely occur once the deflagration enters the
      distributed burning regime.


    \subsection{Modeling domains}
    \label{sec:modeling_domains}

        As described in \Cref{chp:intro}, it is not feasible to resolve all of
        the relevant spatial scales in a single simulation. The preferred
        strategy for modeling is to decompose the problem into smaller problem
        domains, with each domain attempting to model a particular physics
        process.

        \paragraph{Full-star explosion models}
        The largest scale range is the full-star. The full star model
        simulates from the end of the carbon 'simmering' phase prior to the
        thermal runaway in the convective core to the subsequent explosion. Due
        to the limited resolution in this type of large scale model, it is
        common to manually ignite a parcel of fluid in either the core region
        (central ignition) or in the RT flame plume (delayed ignition). This
        problem setup is given the name \texttt{wd} (short for white dwarf). In
        the context of the ILES approach, the cutoff length is defined as
        $\Delta_{\mathrm{wd}}$. This value is typically near $\unit[10^5]{cm}$. 

        \paragraph{Rayleigh-Taylor deflagration models}
        At the intermediate spatial ranges, a single deflagration bubble of the
        exploding WD is modeled to determine the turbulent kinetic energy and
        scales of the turbulent driving in the flame brush. This problem setup is
        referred to as \texttt{flame-in-a-box}. In this model the cutoff
        length is denoted by $\Delta_{\mathrm{rt}}$, and the value in the highest
        resolved cases is on the order of $\unit[10^3]{cm}$.

\begin{figure}[!ht]
          \center
          \includegraphics[width=0.9\textwidth]{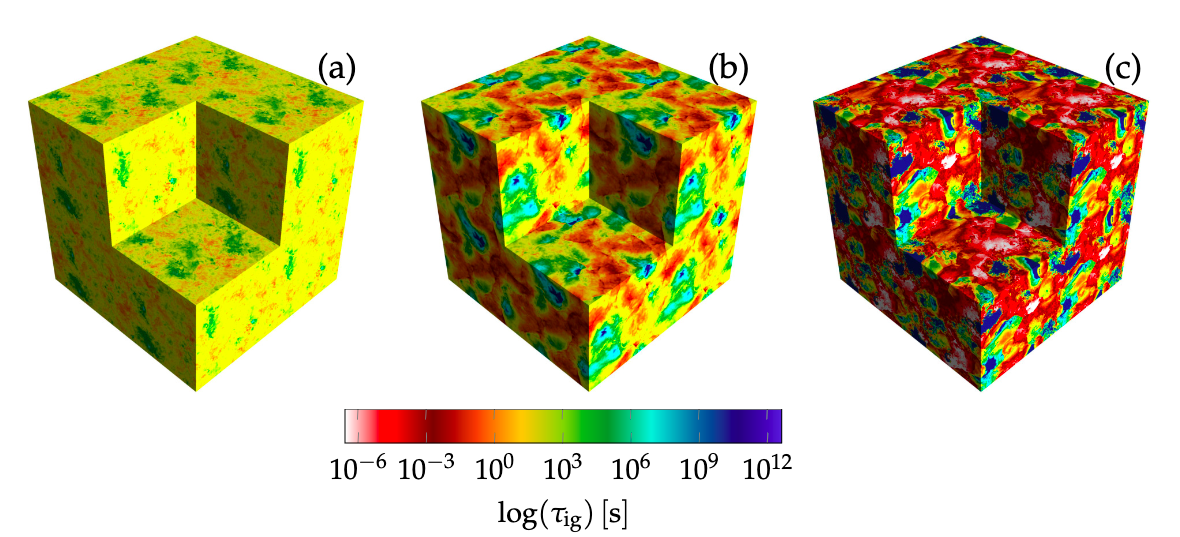}
          \caption{Pseudocolor plots of the induction time variable from three
            three-dimensional simulations of homogeneous and isotropic turbulence
            in a periodic domain (figure borrowed from \cite{fenn2017}).  The
            three plots differ in the degree of compressibility of the forcing,
            with the compressibility of the drive being (a) solenoidal, or
            divergence-free, (b) $0.75$ percent compressive, and (c) $0.875$ percent compressive.}
          \label{fig:fenn}
        \end{figure}

        \paragraph{Reactive turbulence simulations} Further refining the model
        range, the turbulent conditions in the flame brush may be examined more
        closely. In a problem setup referred to as \texttt{tburn}, the
        conditions in the flame brush are emulated in a three-dimensional
        periodic cube domain with a length of $\unit[32 \times 10^{5}]{cm}$ on
        each side. The filter cutoff scale in these models is typically set to
        $\Delta_{\mathrm{tb}} = \unit[3.125 \times 10^{3}]{cm}$. The turbulence
        is generated numerically with a spectral driving routine
        \cite{federrath2010}. The degree of compressibility of the
        drive ranges from purely solenoidal (divergence-free) or purely
        compressible (curl-free). A comparison of the nature of turbulence for
        different driving compressibility values in the \texttt{tburn} models
        \cite{fenn2017} is shown in \Cref{fig:fenn},
        These models are useful for modeling the small scale perturbations in
        induction time and pressure that may lead to deflagrations,
        hotspots, and either shock-initiated detonations, or detonation
        initiation via the Zel'dovich mechanism \citep{brooker2020}.

        \paragraph{Direct numerical simulation} Direct numberical simulation
        may be used if even finer resolution of the turbulence-flame
        interaction is desired, or if more accuracy is required in studies of
        hotspot detonation formation. Notably, the resolution of
        $\Delta_{\mathrm{tb}}$ is low for resolving detonation formation.
  \section{Computational modeling}

    In this work, the Euler equations are used to model compressible, reactive
    flows:
    \begin{subequations}\label{eqn:euler}
    \begin{align}
        \rho_{t} + \nabla \cdot (\rho \bm{u}) &= 0, \label{eqn:mass} \\
        (\rho \bm{u})_{t} + \nabla \cdot (\rho \bm{u}
            \bm{u}^{T} + p\bm{I}) &= \bm{0}, \label{eqn:momentum} \\
        (\rho \bm{X})_{t} + \nabla \cdot \left( \rho \bm{X}
        \bm{u} \right) &= \bm{R}, \label{eqn:species} \\
        (\rho E)_{t} + \nabla \cdot \left( \rho E +
            p \right) \bm{u} &= \dot{Q}, \label{eqn:energy}
    \end{align}
    \end{subequations}
    where $\rho$ is the mass density of the gas mixture, $\bm{u}$ is the
    velocity vector, $p$ is the pressure, $E$ is the specific total energy,
    $\bm{X}$ is the vector of species mass fractions, $\bm{R}$ is the vector of
    species reaction rates, and $\dot{Q}$ is the energy source term.  The mass
    fraction for the $i^{\mathrm{th}}$ species can be written as $X_{i} =
    \rho_{i} / \rho$, where $\rho_{i}$ is the corresponding mass density,
    meaning that the species must collectively satisfy the constraint
    $\sum_{i=1}^{\mathcal{N}_{\mathrm{species}}} X_{i} = 1$, where
    $\mathcal{N}_{\mathrm{species}}$ is the number of species considered. A
    special notation, $X_{f}$, is used for the abundance of carbon fuel.

    The total specific energy is calculated as the sum of the specific internal
    energy and the kinetic energy as  $E = e + \frac{1}{2} \bm{u} \bm{u}^{T}$.
    The above system of PDEs is closed using a suitable EoS which relates the
    pressure to density, internal energy, and composition.

    \subsection{Finite volume discretization}

      The system of partial differential equations \Cref{eqn:euler} are
      discretized using the standard finite volume (FV) approach. Given a
      uniform discretization of the domain $x \in \left[x_{a}, x_{b}\right]$
      with cell size $\Delta x = (x_b - x_a) / N$, where $N$ is the number of
      cells, the following one-dimensional semi-discrete scheme is considered:
      \begin{equation}
        \bm{\U}_{i}^{n+1} = \bm{\U}_{i}^{n} - \frac{1}{\Delta x}
              \int_{t_{n}}^{t_{n+1}} \left(
              \hat{\bm{\F}}_{i} -
              \hat{\bm{\F}}_{i-1} \right) dt +
              \int_{t_{n}}^{t_{n+1}}
              \hat{\bm{\Source}}_{i} d t.
          \label{eqn:fv_scheme}
      \end{equation}
      Here $\bm{\U}_{i}^{n}$ is an approximation to the average of the exact
      solution, which is denoted as $bm{\Q}(x,t)$, in the control volume
      $\left[ x_{i-1}, x_{i}\right]$ at $t_{n}$. The numerical flux
      approximates the exact flux function as
      \begin{equation}
          \hat{\bm{\F}}_{i} \coloneqq \hat{\bm{\F}} \left( \bm{\U}_{i-k}, \dots,
              \bm{\U}_{i+k+1} \right) \approx \bm{\F} \left( \bm{\Q}(x_{i},t) \right),
          \label{eqn:fluxes}
      \end{equation}
      where $2k$ is the number of cells comprising the reconstruction stencil,
      and the term
      \begin{equation}
          \hat{\bm{\Source}}_{i} \coloneqq \bm{\Source}(\bm{\U}_{i}) \approx
              \frac{1}{\Delta x} \int_{x_{i-1}}^{x_{i}} \bm{\Source}(\bm{\Q}(x,t))
              dx
          \label{eqn:source}
      \end{equation}
      is an approximate average of the source term within the control volume.
      The fluxes and source terms may be handled separately in time using
      standard operator splitting approaches (see, \cite{macnamara2016}, and
      references therein).

    \subsection{Multiresolution based adaptive mesh refinement}

      In this work the mesh is adapted using multiresolution (MR) based adaptive
      mesh refinement (AMR). The application of MR analysis to FV schemes was
      first introduced by Harten \cite{harten1994} to adaptively compute
      expensive flux calculations in simulations of compressible gas dynamics.
      Since then it has gained popularity and been used for designing fully
      adaptive schemes \cite{cohen2003}.

      A (MR) analysis is a set of nested subspaces of $L^{2}(\mathbb{R})$ that
      forms an orthonormal basis for $L^{2}(\mathbb{R})$ \cite{shima2016}. An
      MR analysis $\left\{\mathcal{V}_{j}\right\}$ satisfies the relation
      \begin{equation}
        \mathcal{V}_{0} \subset \mathcal{V}_{1} \subset \mathcal{V}_{2} \subset \cdots \subset L^{2}.
        \label{eqn:setrepfull}
      \end{equation}
      By introducing the orthogonal complement of $\mathcal{V}_{0}$ and
      $\mathcal{V}_{1}$ as $\mathcal{W}_{0}$, $\mathcal{V}_{1}$ can be written as
      \begin{equation}
        \mathcal{V}_{1} = \mathcal{V}_{0} \oplus \mathcal{W}_{0},
      \end{equation}
      where $\oplus$ denotes the direct sum. Naturally via
      \Cref{eqn:setrepfull} then $L^{2}$ may be represented as
      \begin{equation}
        L^{2} = \mathcal{V}_{0} \oplus \mathcal{W}_{0} \oplus \mathcal{W}_{1}
          \oplus \mathcal{W}_{2} \oplus \cdots.
        \label{eqn:setrep}
      \end{equation}
      The spaces $\mathcal{W}_{j}$ are called the detail spaces. The vector
      spaces $\mathcal{V}_{0}$ and $\mathcal{W}_{j}$ are spanned by a set of
      functions known as scaling functions and wavelet functions, respectively.
      Wavelets on finer scales may be defined by translations and dilations of
      the mother wavelet. Then any arbitrary function $f \in L^{2}$ may be
      expressed as linear combinations of the scaling functions and wavelets,
      with the coefficients of the wavelet functions being indicative of local
      features. These coefficients are called the detail coefficients.

      With respect to the application of MR analysis to the solution of partial
      differential equations, the purpose of the \Cref{eqn:setrep} is to
      represent the discrete solution on the finest level as a sum of the
      discrete solution values on the coarsest level plus a series of
      differences between adjacent levels. Then the detail coefficients of low
      magnitude indicate where solution features may necessitate less mesh
      resolution.
      
      In practice, the wavelet functions are never actually constructed, and
      detail coefficients are obtained using a transform. The detail
      coefficients are thresholded according to a user-defined tolerance. This
      results in a compressed solution and (usually significantly) fewer mesh
      cells. For more details on the implementation of the MR-based AMR used in
      this work, the reader is referred to the work of Gusto \& Plewa
      \cite{gusto2022}.

    \subsection{Additional remarks}

      Some details of the present approach should be mentioned. Firstly, the
      scheme described in \Cref{eqn:fv_scheme} is implemented in a Fortran code
      called Proteus, a fork of the FLASH code \citep{fryxell2000} developed at
      the University of Chicago.  Proteus uses the piecewise parabolic method
      (PPM) \citep{colella1984} to treat the hydrodynamics, a thirteen isotope
      nuclear network \cite{fryxell2000} to solve the nuclear reactions, and
      the Helmholtz equation of state (EoS) \citep{timmes2000a} to describe the
      thermodynamic state and close the governing system of equations.  The
      parallel, block-structured adaptive mesh refinement routines are provided
      by the PARAMESH library \citep{macneice2000}.

      Then the treatment of turbulence should be discussed. Due to the very low
      viscosity of the dense plasma, the Kolmogorov length scale
      \citep{kolmogorov1941} is quite small, on the order of centimeters. For
      the uninitiated reader, this value describes the scale on which the
      dissipation of turbulent kinetic energy becomes important. This is
      important, as the cutoff of grid resolution above this scale disrupts the
      energy cascade. As discussed, to include such scales in the full-star
      models is infeasible.  Then practitioners generally resort to a
      large-eddy simulation (LES) approach \cite{lesieur1996}, where the
      governing equations are filtered at some cutoff length, and a SGS model
      for describing the stresses on the unresolved scales is implemented. The
      filtered equations may also be solved \textit{without} an explicit SGS
      model, but with some dissipation provided by numerically by the hydrodynamics
      solver. In the latter approach, known as implicit LES (ILES), it is hoped
      that the amount of numerical dissipation provides an \textit{effective}
      viscosity similar to the viscosity of the actual turbulence. In
      \cite{aspden2008}, Aspden \etal actually quantify the effect that
      numerical dissipation has on the small scales using dimensional analysis.

      The present modeling approach makes no use of any type of SGS model for
      the flame or for turbulence.  The flame is simply propagated cell-to-cell
      via diffusion. These assumptions are mentioned to highlight the
      limitations of the present work.

\chapter{Direct Numerical Simulation of the Deflagration-To-Detonation Transition}
\label{chp:hotspot_analysis}

  \section{Introduction}
  \label{sec:chp3intro}

    A substantial amount of evidence in the literature suggests that the
    Zel'dovich reactivity gradient mechanism may lead to spontaneous detonation
    initiation within small regions of relatively low induction times known as
    hotspots.  The hotspots may form from the mixing of warm ash and cold fuel
    that occurs either when a flame front is broken via turbulence, or when a
    flame is quenched, as described by Khokhlov, Oran, \& Wheeler
    \citep{khokhlov1997}.  Again, these authors consider two critical
    questions: (1) what conditions allow a detonation to form within a hotspot,
    and (2) what are the large-scale turbulence driving conditions that lead to
    such situations?  Combined, the answers to these questions may help form
    the basis of a SGS model for DDT in simulations of exploding WDs. This
    chapter deals primarily with the former of these questions.

    The study of Khokhlov \etal is limited by the small number of hotspot
    configurations considered and the size of the parameter space (i.e.
    imposing idealized conditions).  How more physically-motivated profiles in
    induction time can alter the potential for detonation initiation was
    considered by Seitenzahl \etal \cite{seitenzahl2009}. However the addition
    of velocity fields and realistic initial conditions has not been addressed
    in the literature.

    In this chapter the aforementioned works are expanded upon by using DNS to
    study a broader, physically motivated set of hotspot configurations likely
    to occur in the distributed burning regime. The chapter provides insight
    into how the criteria for detonation initiation may change when more
    realistic initial hotspot configurations are considered, and when realistic
    velocity fields are incorporated.

  \section{Methods}
  \label{sec:chp3mthds}

    As discussed in \Cref{sec:zeldovich_regimes}, the theory proposed by
    Zel'dovich posits that a region with nonuniform induction time can produce
    a detonation if and only if the reactive wave can couple to the outgoing
    compressive wave. The compressive wave is produced once electron degeneracy
    is lifted and burning can then result in an increase in thermal pressure
    within the nonuniformly preconditioned region.

    The potential for the reactive wave to couple to the outgoing compressive
    wave is mainly determined by the spontaneous propagation velocity field
    $u_{sp}$, energy generation rates, and the soundspeed of the matter.  Given
    that $u_{sp}$ is the inverse of the spatial gradient of induction time, the
    initial spatial profile of the hotspot is critically important, although
    the profile can be expected to change considerably during the evolution due
    to feedback between reactive and hydrodynamic processes. Despite this,
    however, it is safe to say that for realistic profiles, the larger the radius
    is, the more mild the gradient in induction time will be, and the more
    likely the reactive wave will be able to catch up to the outgoing acoustic
    wave.

    In the first series of investigations, the dependence of the
    successful detonation on the spontaneous propagation velocity is explored
    by modulating the spatial characteristics of a Gaussian initial profile.
    The dependence on the ambient density and temperature is also investigated.
    Besides partially determining the induction time, the density also controls
    to some extent the energy generation rates once the fuel is ignited. Given
    that the successful detonation depends on the feedback from the energy
    generation behind the compressive wave, the density can be expected to
    dramatically influence the process.

    In the second series of investigations, a similar analysis is performed but
    with the initial conditions being drawn from realistic, three-dimensional
    turbulence simulations using one-dimensional extractions (lineouts).

    \subsection{Idealized initial conditions}

      Here the solution of \Cref{eqn:mass}-\Cref{eqn:energy} is obtained using
      DNS, with initial conditions being idealized hotspot profiles. The
      initial profiles are Gaussian, with the amplitude and variation
      of the profiles being free parameters.  The ambient density and
      temperature are also free parameters in the present study.  These greatly
      affect the magnitude of the induction time and the rate of energy release
      due to burning.  The choice of a Gaussian function is made based on the
      findings in \Cref{chp:turbanalysis}, which indicate that the spatial
      profiles of fluctuations in the turbulent medium are fit by a Gaussian
      function to a high degree of accuracy.

      In these models, referred to as \textit{synthetic} models, an effort is
      made to try to isolate the spontaneous combustion mechanism to the
      greatest extent possible. To this end, a pressure gradient of zero is
      prescribed in the whole domain, as well as zero velocity.  This ensures
      that a successful detonation can only be triggered by coupling to a
      spontaneous, self-generated wave.

      For every hotspot configuration the initial density profile is defined as
      \begin{equation}
        \rho(t=0,x) \coloneqq \rho_{\mathrm{amb}} - \delta \rho \exp{\left(\frac{-x^2}{2 R^2}\right)},
        \label{eqn:dens_perturb}
      \end{equation}
      where $\rho_{\mathrm{amb}}$ is the ambient density value, $\delta \rho$ is the
      perturbation in density and $R$ determines the radius of the
      perturbation.  For notational convenience the normalized amplitude is
      introduced as $A = \delta \rho / \rho_{\mathrm{amb}}$. The ambient
      pressure is obtained using the equation of state (EoS) with the desired
      ambient density and temperature as inputs.  Then the density is perturbed
      according to \Cref{eqn:dens_perturb} and passed to the EoS routine again
      with the previously computed pressure to get the temperature profile.
      This process produces the desired Gaussian form in density, temperature,
      and induction time while also maintaining baroclinic (constant pressure)
      conditions.

      A range of each of the aforementioned free parameters is chosen to build
      a database of models for analysis.  The range of each of the free
      parameters chosen for inclusion in the database is informed by the
      statistics of 3D reactive turbulence studies introduced in
      \Cref{chp:turbanalysis}. In particular the largest normalized amplitude
      of the density perturbations in the reactive turbulent flow reaches
      roughly $A = 0.3$, while the radii of the perturbations reach about
      $\unit[1 \times 10^{5}]{cm}$. In the present database, the ambient
      density varies from $\unit[7.5 \times 10^{6}]{g\ cm^{-3}}$ to $\unit[3
      \times 10^{7}]{g\ cm^{-3}}$, and the ambient temperature varies from
      $\unit[1.55 \times 10^{9}]{K}$ to $\unit[2.2 \times 10^{9}]{K}$. The fuel
      concentration is kept constant at a mix of 50/50 C/O. Naturally, the
      fuel concentration may also be varied. In the present study however, it
      is fixed in order to keep the parameter space manageable. Sampling within
      the chosen parameter ranges is done adaptively, with a base grid of
      equispaced samples computed initially, and additional samples generated
      using a bisection approach to locate the region of discontinuity.

      The computational setup considers the solution of
      \Cref{eqn:mass}-\Cref{eqn:energy}, with the hotspot centered in the domain
      with $x \in [-1.2 \times 10^{5}, 1.2 \times 10^{5}]\ \unit[]{cm}$, and
      with outflow conditions prescribed at each boundary.  The simulation is
      run until either $t = \unit[4.0 \times 10^{-4}]{s}$ or a detonation is
      formed. The timestep is determined dynamically via the CFL condition
      \citep{courant1967} or the nuclear burning timescale depending on which
      is smaller. The maximum resolution in these models is $\Delta x \approx
      \unit[19.53]{cm}$, ensuring that spatial structures such as shocks are
      very well resolved.  Multiresolution mesh adaptation is used to ensure
      that sharp structures are resolved without wasting resources in inactive
      areas of the domain\footnote{In these studies the refinement tolerance is
      very strict.  Virtually every region with nonzero velocity is refined to
      the maximumum level. Still, there are substantial computational savings
      given that the hotspot takes up a small portion of the domain
      initially.}.


      In order to determine the outcome of each model, the maximum Mach
      number is used as an indicator. In models with detonation-forming
      hotspots the maximum Mach number reaches or exceeds one.  This value is
      reported by the simulation code at every timestep.  This offers a simple
      way to classify the results.

    \subsection{Realistic initial conditions}
    \label{sec:realistic}

      Hotspots that are not synthetically formulated but rather sampled from
      realistic turbulent conditions are considered. In these models, referred
      to as \textit{turbulence extracted} models, data is extracted directly
      from three-dimensional reactive turbulence simulations (as described in
      \Cref{sec:modeling_domains}) to create the initial conditions.

      The dynamics are much the same from the point of view of the Zel'dovich
      mechanism. In other words the system can still be described as a
      competition between energy release on one hand and acoustic energy
      propagation on the other. However the extent to which the process is
      influenced by turbulent motions remains to be seen. Aside from
      arbitrarily complex velocity fields, the effect of nontrivial initial
      profiles (i.e. not of a simple Gaussian or polynomial form) is also of
      interest.

      The extraction of hotspot profiles from the 3D turbulence data follows
      several steps, outlined below:
      \begin{enumerate}
        \item one-dimensional sweeps through the principal axes ($x$, $y$, $z$)
          of the 3D data are made;
        \item hotspots are identified within the one-dimensional sweep by analyzing fluctuations in temperature or density;
        \item a window of data is extracted with the hotspot centered within it;
        \item interpolation of the density, temperature, velocity, and carbon abundance is performed within the window;
        \item the interpolated profiles are resampled at the new resolution and may be used as initial conditions for DNS calculations.
      \end{enumerate}
      Once this procedure is complete the interpolated profiles may be used
      as-is, or they may be modified. In the present study these profiles are
      modified by increasing slightly the temperature everywhere equally,
      bringing each model closer to ignition. To perform interpolation, the
      piecewise cubic hermite interpolating polynomial (PCHIP) implementation
      in Matlab is used.


      The computational setup for these models is nearly identical to the
      synthetic models. Here though the domain is increased, with $x \in [-2
      \times 10^{5}, 2 \times 10^{5}]\ \unit[]{cm}$, and the grid resolution
      consequently being increased slightly to $\Delta x \approx
      \unit[24.41]{cm}$.

      Although the conditions for these models are extracted from forced
      turbulence simulations described in \Cref{chp:background},
      forcing is not applied in these models once they are begun. The relevant
      timescale for a detonation-developing hotspot is less than the
      characteristic timescales of the turbulence considered in this work.
      These timescales are on the order of $\unit[10^{-2}]{cm}$; this is over
      one order of magnitude larger than the timescale of interest. Then the
      injection of momentum on those timescales cannot be expected to play a
      significant role during the evolution and is consequently neglected.

      The pre-existing velocity field in these models introduces a few
      complications when it comes to labeling of the large dataset. Because
      there are some model configurations with strong non-zero mean velocities,
      the Mach number cannot be used as a reliable indicator of detonation.
      Instead a filtering function is introduced that analyzes jumps in
      pressure, temperature, density, and velocity across the wave. This
      filtering function is a core part of the automatic labeler used to
      correctly determine the outcomes of such a large database of simulations.

      In the automatic labeler, a sample is declared to produce a detonation if
      the jumps found in the vicinity of the original hotspot satisfy certain
      criteria for detonations. Given state $1$ ahead of the wave and state $2$
      behind it, a detonation is defined as having conditions in pressure of
      $p_{2}/p_{1} > 2$, in temperature of $T_{2}/T_{1} > 2$, in density of
      $\rho_{2}/\rho_{1} > 2$, and in relative velocity of $\left| u_{2} -
      u_{1} \right| / u_{1} > 0.6$. It is also required that the carbon
      abundance is reduced across the wave such that $X_{f,2} / X_{f,1} < 0.1$.
      These conditions are determined empirically.
      
      To account for advection of the hotspot in a strong mean velocity field,
      a spatially adaptive monitoring window is used that evolves with the
      hotspot in time based on the magnitude of the mean velocity.  A time
      limit for the hotspot to undergo detonation is imposed; it is defined as
      three quarters of the minimum induction time value plus one
      sound-crossing time. If a detonation is not detected within this time
      then the sample is declared as non-detonation-forming.

      Some additional logic must be applied to the automatic labeler for
      certain problematic cases. One such case occurs when divergence of the
      velocity field causes fluid to build up in another part of the domain.
      This fluid can be preconditioned such that a new rapidly burning hotspot
      is formed outside of the monitoring window. If the hotspot then
      successfuly detonates, the detonation wave may enter the monitoring
      region centered around the original hotspot and trigger a false positive
      for that sample. To avoid this when analyze the characteristics of the
      detected detonation wave - if it is to the right of the hotspot center
      and the wave is moving to the left (or vice versa), then it could not
      have been generated by the hotspot of interest and the samples is
      removed.

    \subsection{Analysis approach}

      In the present analysis, the potential for detonation initiation within
      each hotspot configuration is measured according to \Cref{eqn:khokhlov}.
      In particular the sound-crossing timescale and the reactive
      timescale are computed for each sample. It is expected that for a fixed
      reactive timescale, a critical value in sound-crossing time can be
      identified such that no detonation can occur for values smaller than the
      critical value.
      
      To measure the reactive timescale at the initial state, the
      reactive time scale is computed as $\sigma_{0}$, the standard deviation
      of induction times within some region $r_{0}$. The
      sound-crossing time is computed as $r_{0} / \langle c_{0} \rangle$, where
      $\langle c_{0} \rangle$ is the mean soundspeed in the region.

      There is some freedom in choosing the size of the sampling region
      $r_{0}$. The goal in choosing this region is to include all of the
      material reacting on timescales on the order of the smallest timescale.
      For a hotspot whose induction time minimum is located at $x = 0$, the
      radius for the side with $x>0$ is defined as
      \begin{equation}
        r_{0} = \argmin_{x>0} \left| \frac{\tau(x) - \min{\tau}}{\min{\tau}} - \varepsilon \right|,
        \label{eqn:regionsize}
      \end{equation}
      where $\varepsilon$ is some threshold\footnote{Naturally, for the side
      with $x<0$, $\argmin_{x<0}$ is used in \Cref{eqn:regionsize}.}. In other
      words, the boundary of the region is defined by a relative drop in
      induction time from the minimum point.
      In testing it has been found that the choice of $\varepsilon$ does not
      have much effect on the establishment of a critical value in the
      evaluation of \Cref{eqn:khokhlov}. Based on numerical experiments, a
      value of $\varepsilon = 1.4$ is found to be a sensible choice.

      For the symmetric synthetic models, the analysis is applied to the
      half-profiles. However for the non-symmetric \texttt{TE} models, the
      analysis is applied to both sides of the hotspot independently, with the
      center of the hotspot being defined as the point of lowest induction
      time.




  \section{Results}
  \label{sec:results}

    In this section the results of a total of $27,791$ hotspot configurations
    are presented, with $2,522$ being synthetic configurations and $25,269$
    being extracted from turbulence.  In \Cref{tbl:databases}
    \begin{table}
      \centering
      \begin{tabular}{@{}lccccl@{}}\toprule
        Name & $\rho_{\mathrm{amb}}$ & $T_{\mathrm{amb}}$ & $A$ & $R$ & Runs \\ \midrule
        \texttt{GLtHt} & $0.75 \times 10^7$ & $1.78 \times 10^9$ & $0.04 - 0.32$ & $2 \times 10^{3} - 3 \times 10^{4}$ & $138$ \\
        \texttt{GMdCd} &  $1.0 \times 10^7$ & $1.66 \times 10^9$ & $0.02 - 0.3$ & $0.02 - 0.3$ & $400$ \\
        \texttt{GMdWm} &  $1.0 \times 10^7$ & $1.72 \times 10^9$ & $0.02 - 0.3$ & $0.02 - 0.3$ & $274$ \\
        \texttt{GMdHt} &  $1.0 \times 10^7$ & $1.78 \times 10^9$ & $0.005 - 0.2$ & $0.02 - 0.2$ & $1480$ \\
        \texttt{GHvHt} &  $1.33 \times 10^7$ & $1.78 \times 10^9$ & $0.04 - 0.2$ & $2 \times 10^{3} - 2 \times 10^{4}$ & $95$ \\
        \texttt{GRND} &  $1 \times 10^7 - 3 \times 10^7$ & $1.55 \times 10^9 - 2.2 \times 10^9$ & $0.02 - 0.3$ & $0.2 \times 10^{4} - 4 \times 10^{4}$ & $135$ \\
        \texttt{TE} & $0.91 \times 10^7 - 1.04 \times 10^7$ & $1.18 \times 10^{9} - 2.04 \times 10^{9}$ & - & - & $25269$ \\ \bottomrule
      \end{tabular}
      \caption{Description of databases showing the range of ambient density
        and temperature for each database as well as the normalized amplitude and
        width (for Gaussian models). The total number of configurations is shown
        for each database.}
      \label{tbl:databases}
    \end{table}
    a summary of the executions is presented. The synthetic models are divided
    into several smaller databases based on the ambient density or temperature.
    Overall, three temperatures are tested for the synthetic models:
    $T_{\mathrm{amb}} = \unit[1.66 \times 10^{9}]{K}$, $T_{\mathrm{amb}} =
    \unit[1.72 \times 10^{9}]{K}$, and $T_{\mathrm{amb}} = \unit[1.78 \times
    10^{9}]{K}$. These are labeled as cold (\texttt{Cd}), warm (\texttt{Wm}),
    and hot (\texttt{Ht}). The density at the hottest temperature is also
    varied, using values of $\rho_{\mathrm{amb}} = \unit[0.75 \times 10^{7}]{g\
    cm^{-3}}$, $\rho_{\mathrm{amb}} = \unit[1 \times 10^{7}]{g\ cm^{-3}}$,
    $\rho_{\mathrm{amb}} = \unit[1.33 \times 10^{7}]{g\ cm^{-3}}$, while for
    the other two temperatures the density is held at $\rho_{\mathrm{amb}} =
    \unit[1 \times 10^{7}]{g\ cm^{-3}}$. These densities are labeled as light
    (\texttt{Lt}), medium (\texttt{Md}), and heavy (\texttt{Hv}). Synthetic
    models and turbulence-extracted models are differentiated by appending the
    codes of synthetic databases with a \texttt{G} for Gauss. Finally the
    database codes are combined in order of the model type, density,
    temperature.

    An additional dataset, \texttt{GRND}, is executed with the four free
    parameters being sampled randomly and nonuniformly over slightly larger
    intervals as summarized in the table. 
    
    Finally, note that when referring to a specific model the additional two
    free parameters, amplitude and radius, are included. The normalized
    amplitude is denoted by \texttt{A} followed by two digits, and the radius
    is denoted by \texttt{R} with three digits following. For example, model
    \texttt{GMdHtA05R110} is a synthetic model with an ambient density of
    $\rho_{\mathrm{amb}} = \unit[1 \times 10^{7}]{g\ cm^{-3}}$, ambient
    temperature of $T_{\mathrm{amb}} = \unit[1.78 \times 10^{9}]{K}$, an
    normalized amplitude of $A = 0.08$, and a radius determined by $R =
    \unit[1.1 \times 10^5]{cm}$.

    For the \texttt{TE} models a similar code is used for naming. The code
    begins with \texttt{TE} followed by the approximate ambient temperature,
    abbreviated \texttt{ta}, and its value, then the maximum temperature,
    \texttt{ta}, and its value, then the approximate radius, \texttt{rd}, and
    its value. For instance a \texttt{TE} model with an approximate ambient
    temperature of $T_{\mathrm{amb}} = \unit[1.72\times10^{9}]{K}$, a maximum
    temperature of $T = \unit[2.12\times10^{9}]{K}$, and an approximate radius
    of $R = \unit[4.5\times10^{4}]{cm}$ would be denoted as
    \texttt{TEta17200tm21200rd00450}.



    \subsection{Idealized initial conditions}

\begin{figure}[!ht]
        \center
        \includegraphics[width=\textwidth]{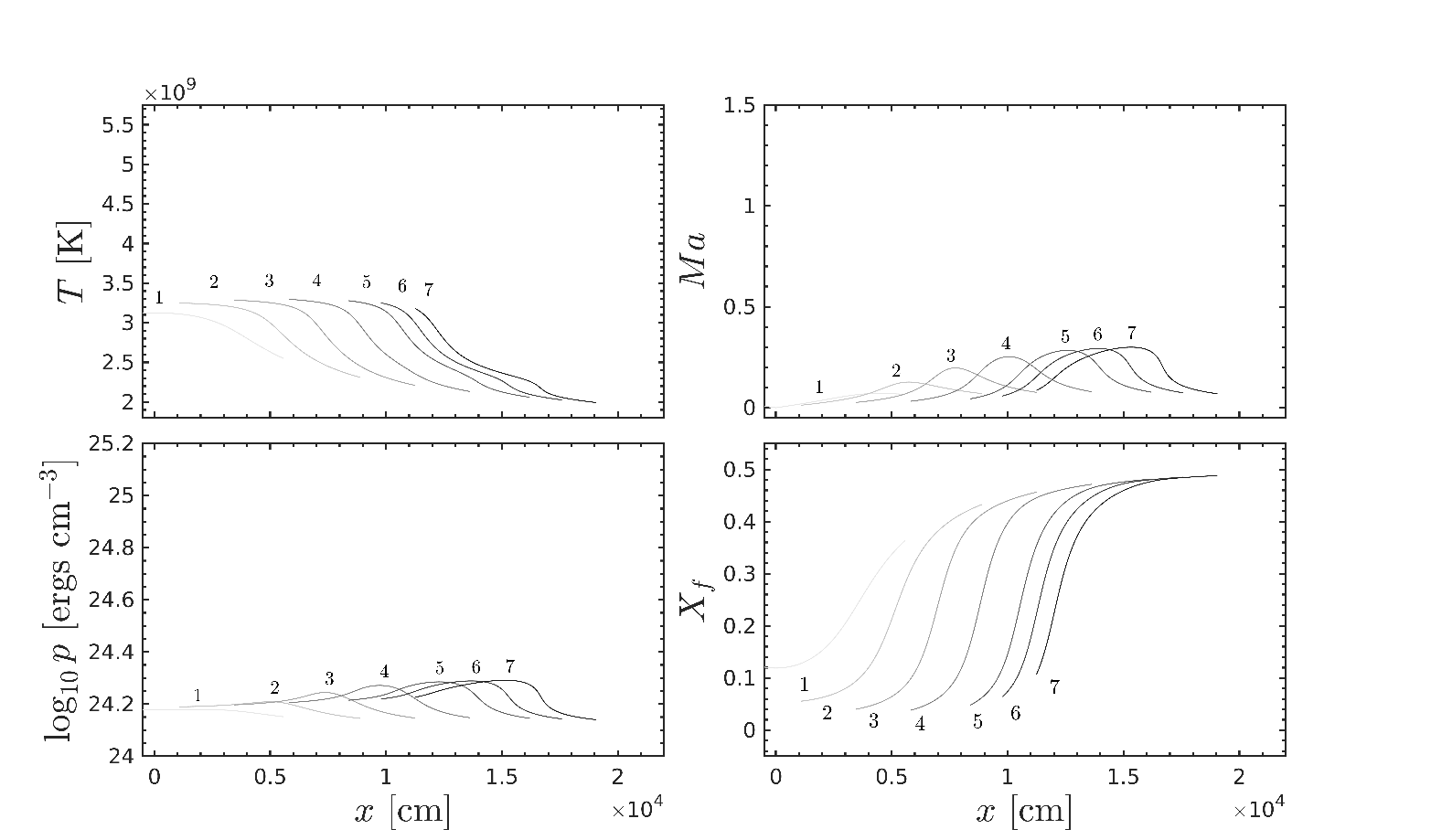}
        \caption{Evolution of a synthetic hotspot, model \texttt{GMdHtA08R130}, with $A =
          0.08$ and $R = \unit[1.3 \times 10^{4}]{cm}$ that results in a
          deflagration. Numerals $1$ through $7$ in the annotations refer to
          times $t_{1} = \unit[1.6]{ms}$, $t_{2} = \unit[1.65]{ms}$, $t_{3} =
          \unit[1.7]{ms}$, $t_{4} = \unit[1.75]{ms}$, $t_{5} = \unit[1.8]{ms}$,
          $t_{6} = \unit[1.825]{ms}$, and $t_{7} = \unit[1.85]{ms}$,
          respectively.  Highlighted portions of the temperature, Mach number,
          pressure, and carbon abundance are shown at each time.}
        \label{fig:gaussevo1}
      \end{figure}

      Idealized initial conditions are used to explore the basic conditions
      necessary for a detonation to develop from a hotspot via the Zel'dovich
      mechanism. First, two models that are adjacent in the parameter space are
      investigated.  In the first model, \texttt{GMdHtA08R130}, the normalized
      amplitude of the perturbation is $A = 0.08$ and the variation is $R =
      \unit[1.3 \times 10^{4}]{cm}$. In \Cref{fig:gaussevo1}
      the evolution of temperature, Mach number, pressure, and carbon
      abundance is shown at seven instances in time, namely at $t_{1} =
      \unit[1.6]{ms}$, $t_{2} = \unit[1.65]{ms}$, $t_{3} = \unit[1.7]{ms}$,
      $t_{4} = \unit[1.75]{ms}$, $t_{5} = \unit[1.8]{ms}$, $t_{6} =
      \unit[1.825]{ms}$, and $t_{7} = \unit[1.85]{ms}$.
      
      In profile $1$ temperature is observed reaching its maximum of around
      $\unit[3.15 \times 10^{9}]{K}$, the Mach number reaching a maximum of
      about $0.1$, pressure still roughly constant near its original isobaric
      state, and the carbon abundance in the central region at about $0.12$.
      From this point in time the system progresses rapidly. By $t_{2}$ the
      pressure has developed a definitive peak as the divergence in velocity
      due to the expanding gas has caused material to pile up. At this stage
      the carbon abundance profile (lower right panel) indicates that the width
      of the reaction zone has decreased substantially. In profiles $3$ and $4$
      the trend of pressure buildup and reaction zone steepening
      continues at a similar pace. From $t_{4}$ to $t_{5}$ however it is
      observed that the increase in pressure begins to stall. Between $t_{5}$
      and $t_{6}$, again, only a mild acceleration of the gas occurs. By
      $t_{7}$ it is evident that the process has failed to accelerate any
      further.

      This model exemplifies the situation where only a deflagration, not a
      detonation, is produced. While a clear coupling between the reactive and
      compressive waves is initially observed, resulting in an increase of
      pressure and steepness, this coupling becomes less efficient as the wave
      moves down the temperature gradient into colder, denser fuel. Then
      pressure wave can no longer provide the amount of compression needed to
      precondition the reaction zone behind it. The consequence is that the
      energy release behind the wave is no longer able to drive the pressure
      wave, resulting in the acceleration process stalling completely.

\begin{figure}[!ht]
        \center
        \includegraphics[width=\textwidth]{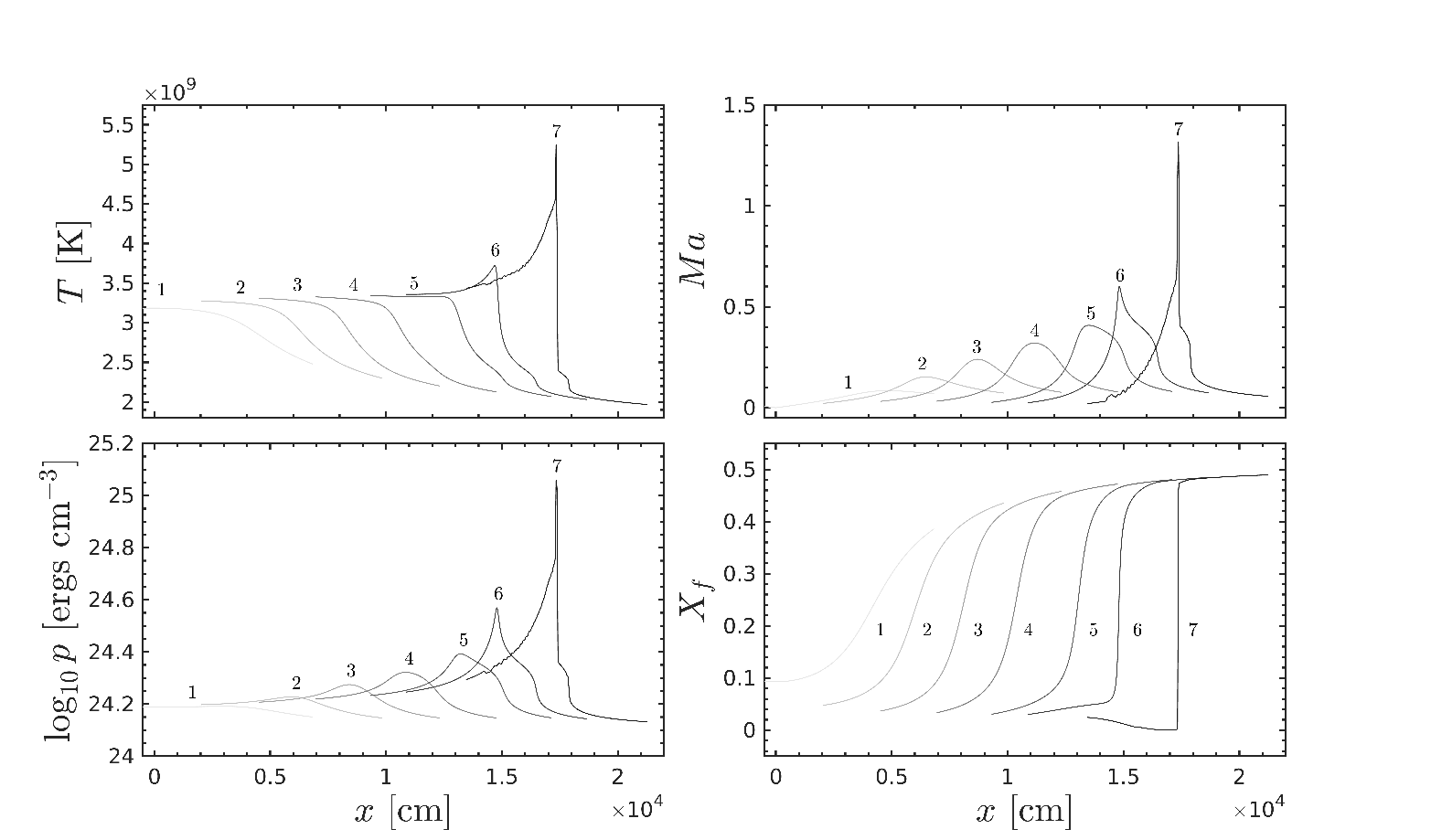}
        \caption{Evolution of a synthetic hotspot, model \texttt{GMdHtA08R140}, with $A = 0.08$ and $R =
          \unit[1.4 \times 10^{4}]{cm}$ that results in a detonation. Numerals
          $1$ through $7$ in the annotations refer to times $t_{1} =
          \unit[1.6]{ms}$, $t_{2} = \unit[1.65]{ms}$, $t_{3} = \unit[1.7]{ms}$,
          $t_{4} = \unit[1.75]{ms}$, $t_{5} = \unit[1.8]{ms}$, $t_{6} =
          \unit[1.825]{ms}$, and $t_{7} = \unit[1.85]{ms}$, respectively.
          Highlighted portions of the temperature, Mach number, pressure, and
          carbon abundance are shown at each time.}
        \label{fig:gaussevo2}
      \end{figure}

      The previous failed detonation scenario is contrasted with that of model
      \texttt{GMdHtA08R140}. This model starts with nearly identical initial
      conditions, except that the radius of the initial perturbation is
      slightly greater. In \Cref{fig:gaussevo2}
      the evolution of the same four quantities is shown on the same scales.
      Profiles are drawn at the same times as in the previous case. Profile $1$
      shows the initially burned material with a maximum temperature of around
      $\unit[3.2 \times 10^{9}]{K}$, and a carbon fraction of about $0.095$. As
      in the case of model \texttt{GMdHtA08R130}, it is observed that by $t_{2}$ a
      definitive peak in pressure has developed. Again, a steady rise in
      pressure is observed between $t_{2}$ and $t_{4}$. The difference in
      outcomes between the two models becomes evident by $t_{5}$, where one can
      clearly observe a steepening of the pressure wave. From this point in
      time the feedback process appears to occur rapidly. A sharp rise in
      temperature behind the compressive wave is visible in profile $6$, as is
      the steepening of the carbon abundance. Profile $7$ shows a Mach number
      above 1 (supersonic fluid velocity) as well as the nearly complete
      exhaustion of fuel behind the detonation wave.

      The intuition that a wider initial profile with more mild gradients of
      the induction time should lead to more favorable conditions for a
      detonation is proven correct in this case. It can be seen that the
      conditions upstream of the compressive wave vary more gradually than in
      the prior case, reducing the rate at which material must be compressed in
      order to feed the reaction zone with preconditioned fuel.

\begin{figure}[!ht]
        \center
        \includegraphics[width=0.75\textwidth]{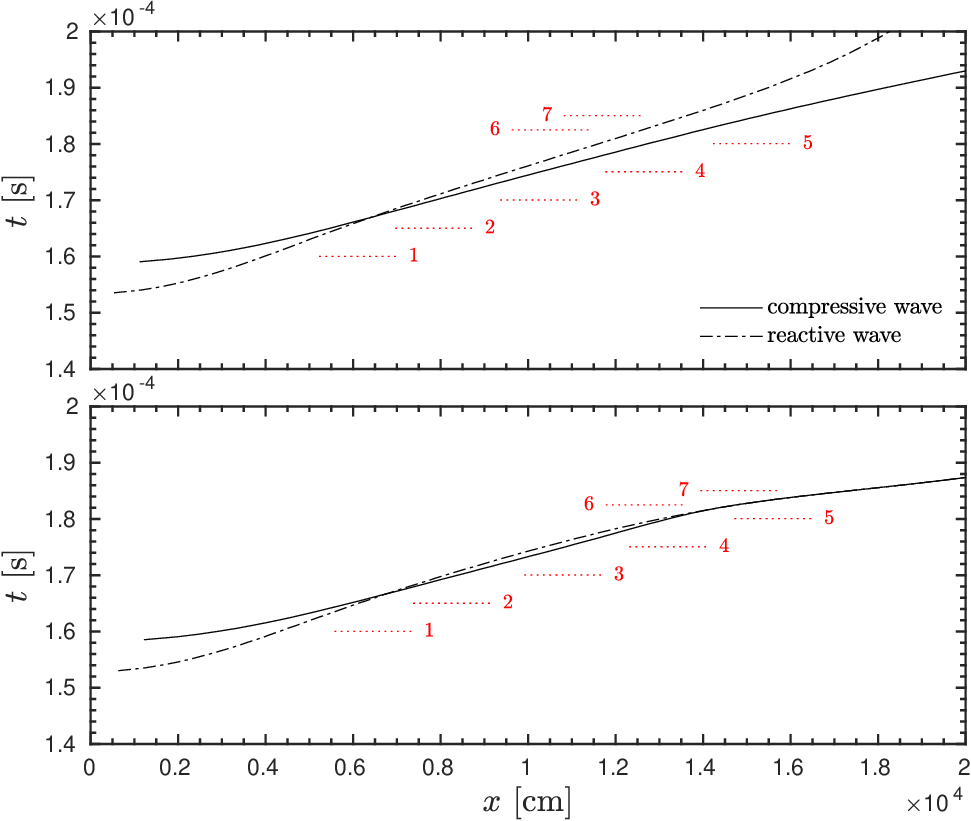}
        \caption{Space-time diagrams of the compressive and reactive waves in
          both a non-detonation-forming case (top; model \texttt{GMdHtA08R130})
          and a detonation-forming case (bottom; model \texttt{GMdHtA08R140}).
          The times $t_{1}$ through $t_{7}$ are labeled in red for added
          context.}
        \label{fig:spacetime}
      \end{figure}

      In \Cref{fig:spacetime}
      the space-time curves of the outgoing compressive and reactive waves are
      plotted for models \texttt{GMdHtA08R130} (top panel) and
      \texttt{GMdHtA08R140} (bottom panel). The position of the compressive
      wave is determined by the location of the peak in pressure of the
      outgoing wave and the position of the reactive wave is computed as the
      first point behind the wave reaching less than one half of the ambient
      fuel abundance, $X_{f}$.

      In both cases the speed of the compressive and reactive waves is
      initially high (roughly before $t_{1}$). The compressive wave slows down
      when it begins to encounter colder and more dense upstream fuel near
      $t_{1}$.  Between $t_{1}$ and $t_{2}$ the compressive wave builds
      strength, driven by the initial overpressure within the central region of
      the hotspot.  Eventually, the peak of the compressive wave overtakes the
      reactive wave (near $t_{2}$). From this point there is a chance for the
      feedback mechanism to take hold. In the case of \texttt{GMdHtA08R130},
      the reactive wave falls behind due to inefficient preconditioning of the
      upstream fuel by the compressive wave. In the case of
      \texttt{GMdHtA08R140} it is observed that the feedback process is more
      efficient, and although the waves diverge from $t_{2}$ to $t_{3}$,
      somewhere between $t_{3}$ and $t_{4}$ this trend is reversed and the
      reactive wave accelerates. By $t_{6}$ the two waves become completely
      coupled and the velocity of the combined shock-reaction complex
      accelerates significantly.

\begin{figure}[!ht]
        \center
        \includegraphics[width=0.625\textwidth]{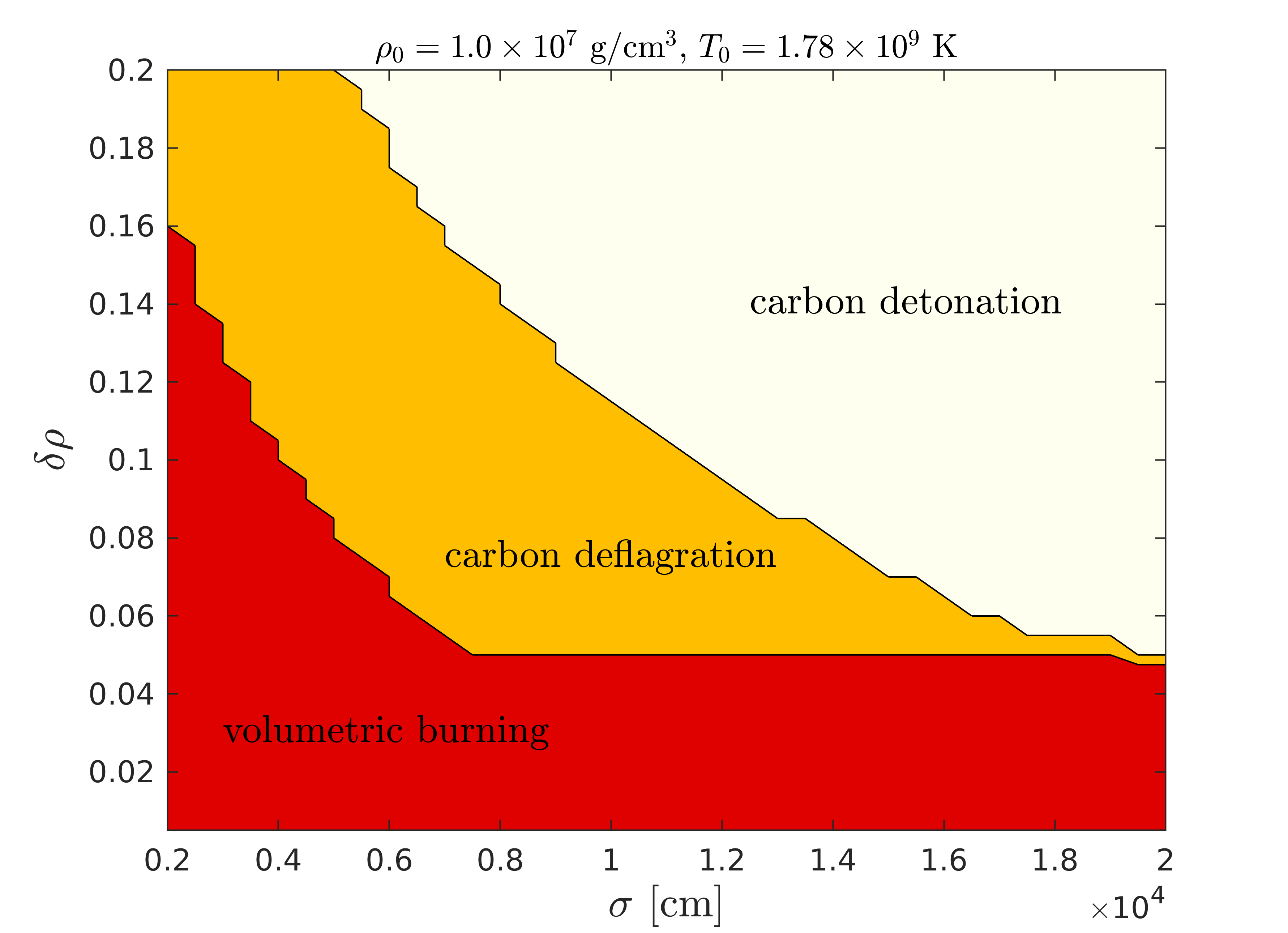}
        \caption{Phase diagram illustrating the dependence of the ability of a
          hotspot to form a detonation as a function of amplitude and width, at a
          fixed ambient density and temperature of $\rho_{\mathrm{amb}} = \unit[1 \times
          10^{7}]{g\ cm^{-3}}$ and $T_{\mathrm{amb}} = \unit[1.78 \times 10^{9}]{K}$
          (dataset \texttt{GMdHt}) respectively. Three regions are identified:
          volumetric burning, carbon deflagration, and carbon detonation.}
        \label{fig:amplvsstdvtemp178e09}
      \end{figure}

      These two cases effectively illustrate the role of the spatial
      characteristics of the hotspot in determining its viability for
      detonation initiation. Then the dependence of the ability of the hotspot
      to form a detonation as a function of amplitude, $A$, and width, $R$ is
      examined. In \Cref{fig:amplvsstdvtemp178e09}
      a phase diagram is shown that separates the simulation outcomes from
      dataset \texttt{GMdHt} into three regions: volumetric burning, carbon
      deflagration, and carbon detonation. In the volumetric burning regime, no
      significant acoustic wave is generated because the amplitude is so small
      that no significant amount of energy is released through nuclear burning.
      The threshold for volumetric burning is defined as $Ma < 0.01$. In the
      deflagration regime, a significant amount of energy is released through
      burning, however the coupling process does not occur. Still, the process
      may produce a strong outgoing shockwave and flame behind it. Finally, in
      the third regime, a detonation wave is created due to successful coupling
      between the reactive and compressive waves.
      
      Next the entire dataset is analyzed and the characteristics leading to
      detonation are quantified more precisely. The timescale analysis of
      Khokhlov is applied using \Cref{eqn:khokhlov} with the criteria
      introduced in \Cref{eqn:regionsize} to define the `width' of the region
      (as some additional specificity is required).
      
\begin{figure}[!ht]
        \center
        \includegraphics[width=0.55\textwidth]{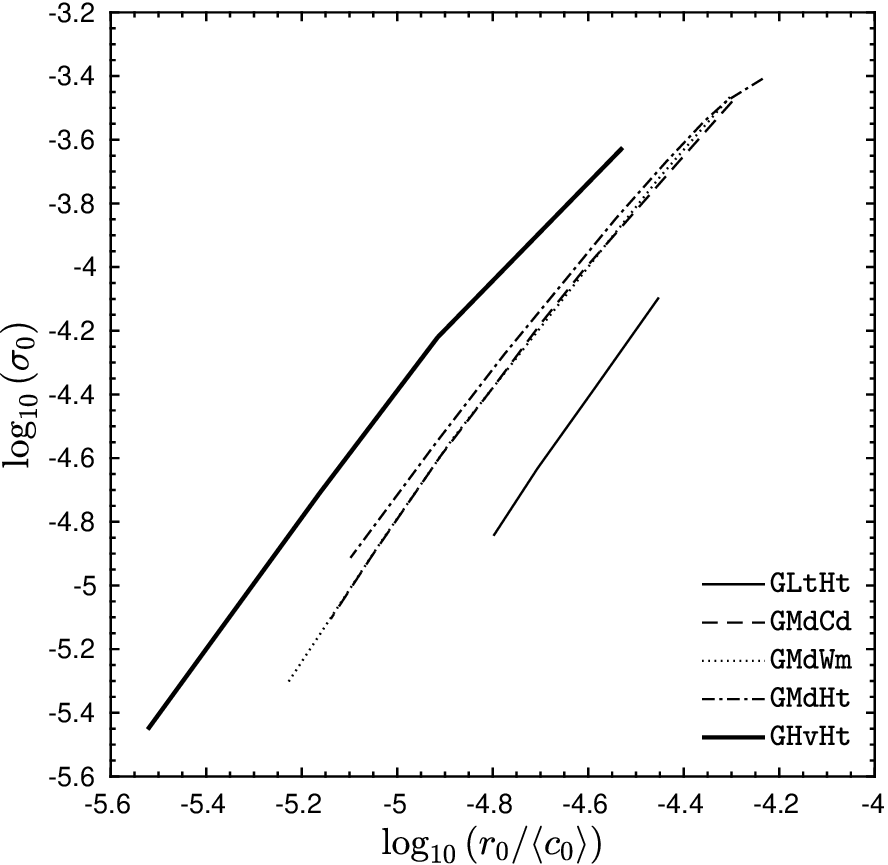}
        \caption{Critical conditions for DDT obtained by DNS of synthetically
          generated hotspot configurations. Shown are the contour lines
          separating the non-detonation-forming and detonation-forming regions
          of the reactive versus sound-crossing timescale plane for five
          databases of varying ambient density and temperature.}
        \label{fig:khokhlovplanegauss}
      \end{figure}

      \Cref{fig:khokhlovplanegauss}
      is a novel result that shows the critical values on the reactive versus
      sound-crossing timescale plot (in log scale) for various combinations of
      ambient density and temperature. Each curve divides the regions of
      non-detonation-forming and detonation-forming conditions for
      the corresponding dataset.
      
      According to \Cref{eqn:khokhlov}, a detonation is only possible when the
      reactive timescale, given by the variation in induction times
      $\sigma_{0}$, is approximately smaller than the sound-crossing timescale,
      given by $r_{0} / \langle c_{0} \rangle$. It is observed that for the three
      datasets at the intermediate ambient density of $\rho_{0} = \unit[1
      \times 10^{7}]{g\ cm^{-3}}$, \texttt{GMdCd} (dashed line), \texttt{GMdWm}
      (dotted line), and \texttt{GMdHt} (dash-dotted line), the critical values
      are closely aligned. For the datasets with varying ambient density,
      \texttt{GLtHt} (thin solid line) and \texttt{GHvHt} (thick solid line),
      a shift in the plane is observed. In particular for the dataset with
      higher density a shift of the critical values toward smaller
      sound-crossing times is observed, and for the dataset with lower density
      the opposite is observed. In other words, for hotspots at higher
      densities these results indicate that the the sound-crossing time does
      not need to be as large in order for a detonation to occur, while for
      hotspots at lower densities the opposite is true. Or put another way, for
      a given sound-crossing timescale, hotspots with higher densities require
      a smaller degree of synchronization of burning induction times (larger
      $\sigma_{0}$) to achieve coupling between the reactive and compressive
      waves.

      Given that the soundspeed is not very sensitive to density (being
      proportional to the inverse of the square root of density), these results
      do indeed seem to indicate that smaller regions ($r_{0}$) are sufficient
      for the hotspot to form a detonation in the case of dataset
      \texttt{GHvHt}. This can be attributed to the heat capacity of the
      material being lowered with increasing density, meaning that higher
      temperatures and possibly higher rates of energy generation can be
      reached during the spontaneous combustion process when compared to the
      models at the intermediate density.
      
      Again, the inverse appears to be true in the lower density case.
      Compared to the high and intermediate densities a much larger
      sound-crossing time (i.e. a larger region) is required for a detonation
      to occur. For a given $\log_{10}{(\sigma_{0})}$ value (for example
      $-4.85$), it is observed that the minimum sound-crossing time required
      for detonation increases in intervals of about $0.2$ in log-scale across
      the three models. These results corroborate the results in
      \Cref{fig:lcritvsdens}.  They have implications for the possibility of
      DDT in a delayed detonation scenario.

      Here the possible limitations of this analysis is disclosed.  First, it
      is clear that the sound-crossing time cannot extend indefinitely while
      the hotspot remains intact. In a turbulent flow, the hotspot will be
      destroyed by turbulence within the amount of time it takes eddies on a
      similar spatial scale to turn over. This natural condition is emulated in
      the present studies by the time limit imposed in the simulation. Its
      effect can be observed toward the upper right region of
      \Cref{fig:khokhlovplanegauss}, in which the curve corresponding to
      database \texttt{GMdHt} curves sharply to the right.

      Secondly, there is also a theoretical limit on the extent of the reactive
      timescale. As $\sigma_{0}$ goes toward zero, the initial explosion takes
      place faster and faster, and in the limit, the case of the
      constant-volume explosion is reached. The critical size of the region
      cannot scale accordingly because of the physical constraint of thermal
      conductivity; as the size of the region becomes extremely small the
      reactions become thermally coupled due to thermal conduction
      \citep{khokhlov1991b}. In this scenario no spontaneous propagation is
      possible.

      Finally the same analysis is applied to the dataset \texttt{GRND}.  This
      dataset samples randomly in the space of ambient density, ambient
      temperature, normalized amplitude, and width corresponding to the ranges
      presented in \Cref{tbl:databases}. Due to the large degree of variation
      in the ambient density, a clear distinction between detonation-forming
      and non-detonation-forming models in the reactive versus sound-crossing
      timescale plane is not observed.

    \subsection{Realistic initial conditions}

\begin{figure}[!ht]
        \center
        \includegraphics[width=\textwidth]{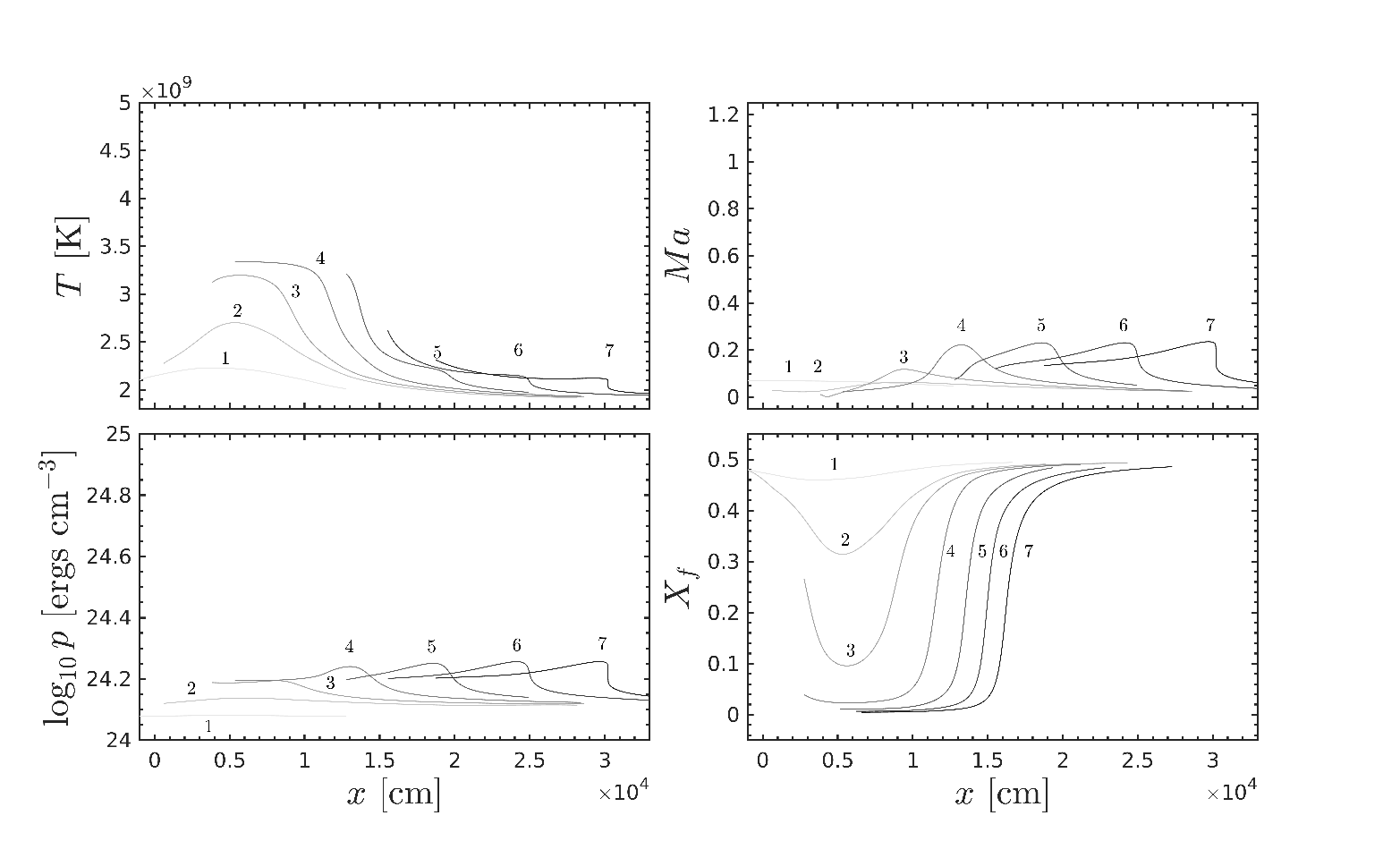}
        \caption{Evolution of a turbulence extracted hotspot, model
          \texttt{TEta17159tm21591rd00344}, that results in a detonation.
          Numerals $1$ through $7$ in the annotations refer to times $t_{1} =
          \unit[60.1]{ms}$, $t_{2} = \unit[110.1]{ms}$, $t_{3} =
          \unit[120.1]{ms}$, $t_{4} = \unit[130.1]{ms}$, $t_{5} =
          \unit[140.1]{ms}$, $t_{6} = \unit[150.1]{ms}$, and $t_{7} =
          \unit[160]{ms}$, respectively.  Highlighted portions of the
          temperature, Mach number, pressure, and carbon abundance are shown at
          each time.}
        \label{fig:ta17159tm21591rd00344}
      \end{figure}

      The same analysis approach is applied to the DNS studies of hotspot
      configurations with realistic initial conditions in the \texttt{TE}
      database. In \Cref{fig:ta17159tm21591rd00344}
      the evolution of model \texttt{TEta17159tm21591rd00344} is shown for the
      temperature, Mach number, pressure, and carbon abundance. Relevant
      portions of the profiles are shown at times $t_{1} = \unit[60.1]{ms}$,
      $t_{2} = \unit[110.1]{ms}$, $t_{3} = \unit[120.1]{ms}$, $t_{4} =
      \unit[130.1]{ms}$, $t_{5} = \unit[140.1]{ms}$, $t_{6} =
      \unit[150.1]{ms}$, and $t_{7} = \unit[160]{ms}$.

      The evolution procedes similarly to the cases in the Gaussian datasets.
      Here however it is first observed that even at $t_{1}$ there is already a
      strong velocity field in the positive $x$ direction, with a Mach number
      of around $Ma = 0.075$. From this point in time, rapid burning
      in the center of the hotspot is observed (already the center has been
      advected away from $x \approx 0$ to $x \approx 0.4$) between $t_{1}$ and
      $t_{3}$ followed by an increase in pressure within that region. Between
      $t_{3}$ and $t_{4}$ the pressure wave begins to steepen, forming a
      compressive wave that accelerates burning. The increase in burning may be
      observed in the near complete exhaustion of carbon behind the wave at
      $t_{4}$ (profile $4$ in the bottom right panel near $x \approx 1 \times
      10^{4}$).  Between $t_{4}$ and $t_{5}$ however it becomes evident by
      comparing the position of the reaction zone and the compressive wave that
      the two have failed to couple. While the compressive wave continues to
      steepen after this point, the possibility of a coupled shock-reaction
      structure is impossible.

\begin{figure}[!ht]
        \center
        \includegraphics[width=\textwidth]{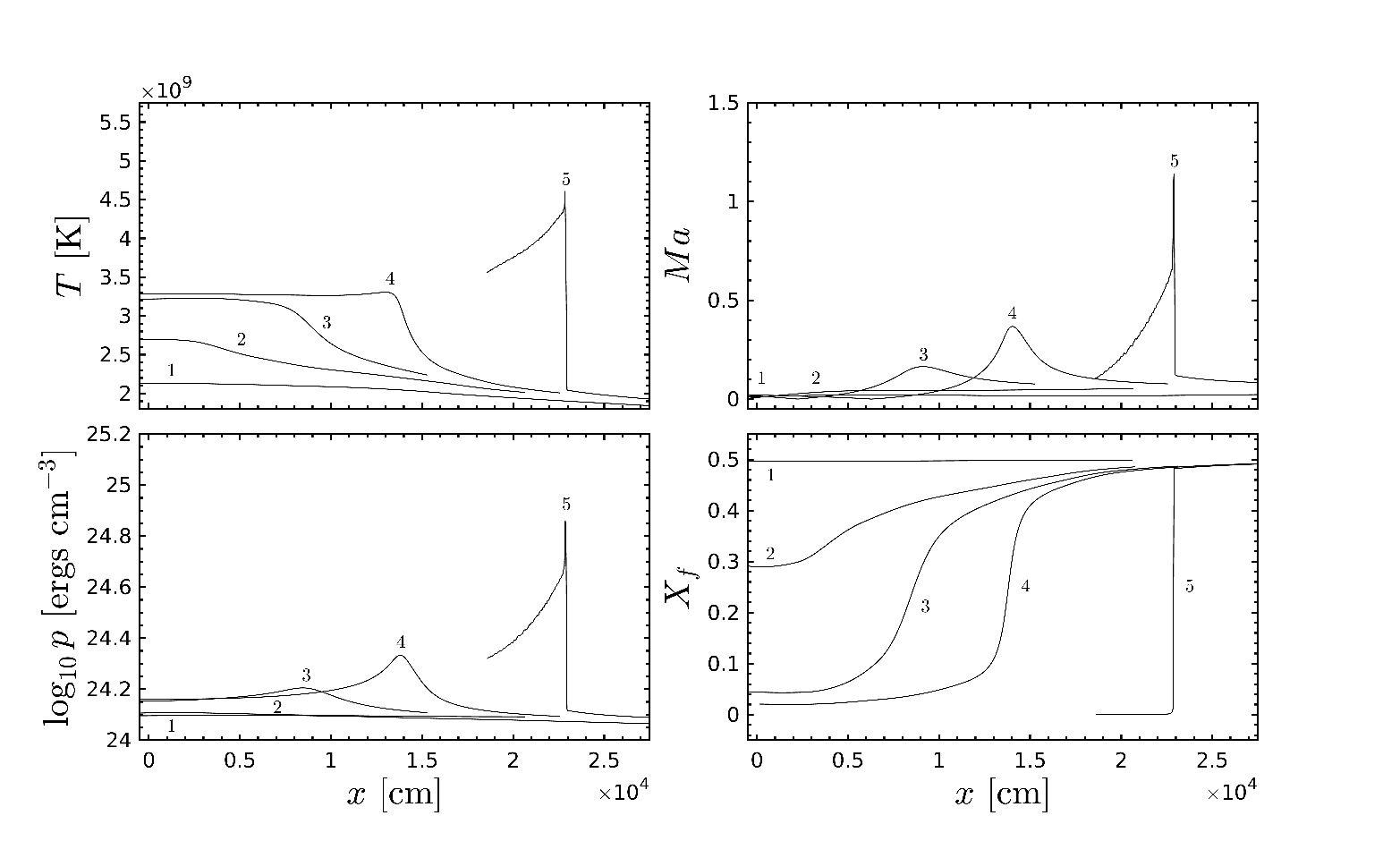}
        \caption{Evolution of a turbulence extracted hotspot, model
          \texttt{TEta16760tm21298rd00594}, that results in a detonation.
          Numerals $1$ through $7$ in the annotations refer to times $t_{1} =
          \unit[0]{ms}$, $t_{2} = \unit[120]{ms}$, $t_{3} =
          \unit[140]{ms}$, $t_{4} = \unit[150]{ms}$, and $t_{5} =
          \unit[160]{ms}$, respectively.  Highlighted portions of the
          temperature, Mach number, pressure, and carbon abundance are shown at
          each time.}
        \label{fig:ta16760tm21298rd00594}
      \end{figure}

      Next a case that successfully detonates is considered, model
      \texttt{TEta16760tm21298rd00594}. The evolution of this model is shown in
      \Cref{fig:ta16760tm21298rd00594}
      at times $t_{1} = \unit[0]{ms}$, $t_{2} = \unit[120]{ms}$, $t_{3} =
      \unit[140]{ms}$, $t_{4} = \unit[150]{ms}$, and $t_{5} = \unit[160]{ms}$
      Note that in this case the region of low induction times is wider than in
      the previous case. Additionally, note the absence of a significant
      velocity field, as evident in the first Mach number profile at $t_{1}$.

      It can be observed that between $t_{1}$ and $t_{3}$ a large region of
      material roughly $\unit[8 \times 10^{4}]{cm}$ in radius, undergoes
      complete burning, resulting in a pressure wave which can be seen in
      profile $3$ in the bottom left panel. At this time, $t_{3}$, the
      corresponding Mach number is roughly $0.2$.  The speed of the compressive
      wave dramatically increases between $t_{3}$ and $t_{4}$, reaching a value
      of about $0.35$.  This feedpack process continues to steepen the
      compressive wave and accelerate burning. By $t_{5}$ a fully fledged
      detonation wave emerges.

\begin{figure}[!ht]
        \center
        \includegraphics[width=\textwidth]{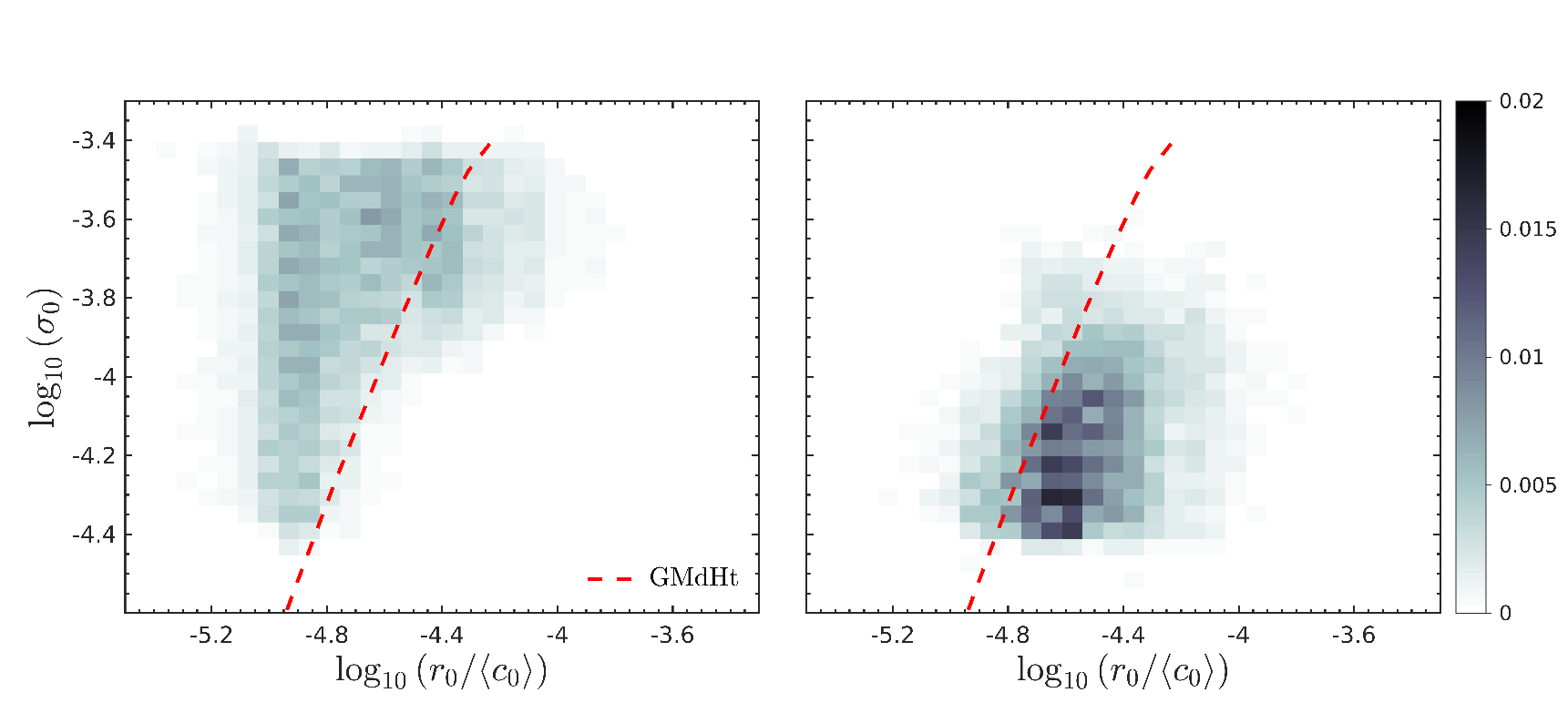}
        \caption{The \texttt{TE} dataset on the reactive versus
        sound-crossing timescale plane, separated into two panes with the left
        pane showing the bivariate histogram of non-detonation-forming points
        and the right pane showing the same but for the detonation-forming
        points. The colorscale indicates the probability of a point
        corresponding to that bin. Shown with a dotted red line is the contour
        from the \texttt{GMdHt} dataset for reference.}
        \label{fig:khokhlovplanetburn}
      \end{figure}

      Considering now the analysis of the whole set of \texttt{TE} simulations,
      the previous analysis is applied, and the reactive and sound-crossing
      timescales are computed and plotted. The data is divided into
      non-detonation-forming and detonation-forming samples, and a bivariate
      histogram is computed for each. These results are shown in
      \Cref{fig:khokhlovplanetburn}
      where it is observed that for the non-detonation-forming set (left panel),
      the vast majority of points, indicated by intensity of color, lie above
      and to the left of the curve of critical values found from the
      \texttt{GMdHt} dataset. This result is somewhat expected given that the
      density of models in the \texttt{TE} dataset fluctuates around $\unit[1
      \times 10^{7}]{g\ cm^{-3}}$ by up to about fifteen percent (again, the
      ambient density in dataset \texttt{GMdHt} is fixed at $\unit[1 \times
      10^{7}]{g\ cm^{-3}}$). Many non-detonation-forming samples with higher overall
      timescales are seemingly located in the detonation-forming region of the
      plane. These models are the result of the conservative, early termination
      approach to labeling described in \Cref{sec:realistic}.
      
      For the set of detonation-forming points (right panel) it is observed that
      the majority lie below and to the right of the reference curve,
      consistent with expectations. There is a set of detonation-forming points
      that lie across reference curve, however. It is possible that these cases
      may be explained by higher densities or a region of negative divergence
      of velocity in the region helping to compress the material in the
      potential detonation formation region.

    \section{Discussion}


      The new series of computational studies introduced in this chapter have
      been analyzed using the approach of Khokhlov, in particular by comparing
      the reactive and sound-crossing timescales. The results indicated that in
      trivially simple cases this analysis can distinguish between
      non-detonation-forming hotspots and detonation-forming hotspots. Of
      course, as observed in \Cref{fig:khokhlovplanegauss}, the critical
      conditions depend strongly on density (more so than on the ambient
      temperature).  Additionally, when non-trivial initial conditions are
      used, particularly when significant velocity fluctuations are present,
      additional analysis seems necessary to explain the exact cause of the
      deviations.

      The obvious question raised by the results in
      \Cref{fig:khokhlovplanetburn} is why the critical values of the reactive
      and sound-crossing timescales corresponding to samples from the
      \texttt{TE} dataset do not more closely align with the reference values.
      The first consideration is, as already stated, that the initial
      density in these models naturally fluctuates given that the conditions
      were extracted from turbulence with relative fluctuations of around
      $15\%$.  This explains some deviation from the reference curve. For
      instance the detonation-forming points to the left of the reference curve
      may be toward the higher end of the density range, shifting the
      critical radius toward smaller values.

      \subsection{A critical radius for detonation}


        The determination of a critical radius for detonation, whose existence
        has been alluded to in several works \cite{khokhlov1997,oran2007}, is
        not trivial in realistic conditions.  If such a thing as a critical
        radius can be meaningfully defined then it certainly is not independent
        of at least several other factors that influence the outcome of a
        reacting hotspot region, for example large-scale turbulent motions (in
        particular the divergence of velocity). Obviously the outcome is
        sensitive to the density of the matter as already observed, and
        probably also to the rates of nuclear energy generation, however this
        has not been explicitly tested (and indeed the findings of Khokhlov
        \cite{khokhlov1991b} suggest only a small dependence; see
        \Cref{fig:lcritvsdens}).

        It has been suggested in \cite{oran2007} that perhaps the most
        important criteria for detonation formation of hotspots is whether or
        not the reactive wave speed reaches the Chapman-Jouget velocity
        somewhere along the profile. In \cite{khokhlov1991b} the region where
        the reactive wave speed equals the sonic velocity is denoted by
        $r_{s}$, with some critical value $r_{s}^{*}$. Again however, the
        existence of such a value seems precarious because of the reasons just
        discussed. The reactive wave speed field is simply computed by the
        inverse of the gradient of induction time and contains basically no
        information about density, potentially adverse velocity or pressure
        gradients, or the nuclear energy generation rates.

\begin{figure}[!ht]
          \center
          \includegraphics[width=0.8\textwidth]{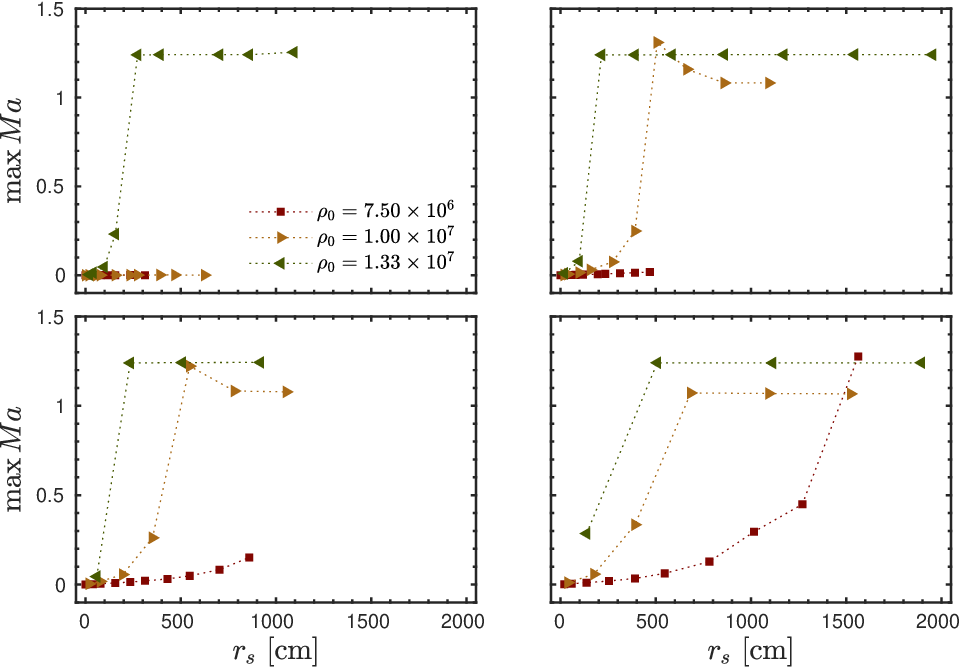}
          \caption{Comparison of maximum Mach number reached versus $r_{s}$ for
            ambient densities of $\rho_{\mathrm{amb}} = \unit[7.5 \times 10^{6}]{g\ cm^{-3}}$ (red squares),
            $\rho_{\mathrm{amb}} = \unit[1.0 \times 10^{7}]{g\ cm^{-3}}$ (gold right-arrows),
            and $\rho_{\mathrm{amb}} = \unit[1.33 \times 10^{7}]{g\ cm^{-3}}$ (green
            left-arrows), and for four normalized amplitudes: $A = 0.04$ (top
            left), $A = 0.08$ (top right), $A = 0.12$ (bottom left), and $A =
            0.16$ (bottom right).}
            \label{fig:rcrit}
        \end{figure}

        In \Cref{fig:rcrit}
        the dependence of the maximum Mach number achieved on the value $r_{s}$
        and the density across several values of relative amplitude, $A$, is
        presented.  It is found that the cases with higher ambient density
        achieve detonation with a smaller value of $r_{s}$ and vice versa. This
        finding is similar to the relationship observed between density and the
        critical sound-crossing time in \Cref{fig:khokhlovplanegauss}.

        Overall the smallest value of $r_{s}$ found to produce a detonation in
        the present synthetic studies suggests a critial value of about
        $r_{s}^{*} \approx \unit[200]{cm}$. If $r_{s}$ is any smaller than
        this value then the formation of a detonation is impossible. However it
        is noteable that these results are limited by the choice of parameters
        used to define the hotspot in the present study. Smaller values may be
        possible.
       


      \subsection{Concluding remarks}

        Overall, it seems that the semi-analytic characterization of hotspots
        using the approach of Khokhlov is useful, but it is not sufficient for
        determing the outcome of hotspots with a high degree of accuracy,
        especially when the other factors mentioned earlier in the chapter are
        considered. The present work is the first to investigate the evolution
        of hotspots in degenerate matter with \textit{realistic} turbulent
        conditions, so it is hoped that this analysis proves to be a step
        forward in this area. The present computational study and subsequent
        analysis confirms the intuition provided by Khokhlov but indeed raises
        even more questions concerning the ability of hotspots to form
        detonations in realistic conditions.  These uncertainties motivate the
        next subject of the present work: the application of machine learning
        to this challenging problem.


\chapter{A Data-Driven Model for DDT}
\label{chp:machine_learning}

  \section{Introduction}


    \Cref{chp:background} briefly reviewed past efforts at SGS modeling of DDT
    in thermonuclear SNe. These have relied on heuristic arguments for defining
    the DDT criteria.

    In this chapter a novel type of SGS model for DDT is proposed that uses DNS
    of the detonation formation process to define the conditions for detonation
    formation on the LES scales. As identified in \Cref{chp:intro}, one of the
    most likely modes of detonation formation in the unconfined turbulent flame
    is via the Zel'dovich reactivity mechanism. Thus the insight gained in
    \Cref{chp:hotspot_analysis} with respect to the Zel'dovich mechanism is
    leveraged to lay the foundation for the new model.

    One key piece of the overarching goal of producing a SGS model for the
    full-star explosion models is the development of a model for characterizing
    reactive turbulence realizations in terms of the potential for creating
    detonation-prone hotspots. One possible way to achieve this goal is to
    develop a \textit{classifier} capable of accurately classifying individual
    hotspots as either detonation-forming or non-detonation-forming.
    
    To train the model on this task, the large database of
    simulations of realistic hotspot configurations from
    \Cref{chp:hotspot_analysis} is used.  The physical insights gained in that
    chapter are used to determine which features to select for the model. At
    the end of the chapter, verification of the model's accuracy is performed
    using one-dimensional, reactive turbulence simulations on the scale
    $\Delta_{\mathrm{tb}}$ of the \texttt{tburn} models.

  \section{Methods}
    
    Developing a SGS model to predict the onset of DDT from data resolved on
    the explosion filter scales, $\Delta_{\mathrm{wd}}$, requires the
    connection of physics between this scale and the lower end of the turbulent
    flame scale, $\Delta_{\mathrm{tb}}$. In \Cref{chp:hotspot_analysis} it was
    observed that on these scales hotspots can form and lead to detonations.
    Then the input data to a model for classifying the potential for hotspots
    must be on the scale of $\Delta_{\mathrm{tb}}$.

    The collection of simulations of one-dimensional, planar hotspot
    configurations are used to construct a predictive model capable of
    identifying detonation-forming hotspots given an initial configuration.
    The classifer is given data samples of the form $\left\{ \left(
    \bm{\chi}_{i}, f(\bm{\chi}_{i}) \right) \right\}_{i=1}^{n_{s}}$, where
    $n_{s}$ is the number of samples. Here $\bm{\chi}_{i} \in \mathbb{R}^{n_{x}
    \times n_{c}}$ is the $i^{\mathrm{th}}$ input to the classifier, $n_{x}$ is
    the number of spatial cells and $n_{c}$ is the number of input channels
    (e.g. flow variables).  The label for sample $i$ is given by
    $f(\bm{\chi}_{i}) \in \left\{0,1\right\}$. A value of $0$ indicates
    non-detonation-forming, and a value of $1$ indicates detonation-forming.

    Then the task of supervised machine learning is as follows, 
    \begin{equation}
      \text{Given the measurements}\ \left\{ \left( \bm{\chi}_{i},
        f(\bm{\chi}_{i}) \right) \right\}_{i=1}^{n_{s}},\ \text{approximate}\ f.
      \label{eqn:mlequation}
    \end{equation}
    In the present work the function $f$ in \Cref{eqn:mlequation} is
    essentially mapping an initial coarsened hydrodynamic state to a physical
    outcome and can thus be thought of as a parameterization of the nonlinear
    dynamics of the governing system of partial differential equations over the
    short time window.

    The nonuniform initial conditions and highly nonlinear interaction between
    combustion and gas dynamics precludes a model based on simple
    parameterizations of the initial hotspot profile, such as its amplitude and
    width. In order to be generalizable to arbitrary initial configurations, it
    is expected that the model must be able to learn more complex features such
    as the critical gradients of induction time, the dependence on density,
    etc.  (although the amplitude and width of the hostpot may still prove to
    be valuable features for learning).

    To accomplish these objectives, the model makes use of deep neural networks
    (DNNs) which have been successful in solving certain classes of problems
    (for more information on using DNNs for function approximation, see
    \cite{adcock2021} and references therein). In particular this work
    considers both artificial neural networks (ANNs) and convolutional neural
    networks (CNNs) \cite{goodfellow2016} for the architecture of the present
    classification model.  These have their own unique tradeoffs in terms of
    performance. With a CNN the focus is on learning the salient spatial
    features of the hotspot that lead to detonation. On the other hand, the ANN
    may be sufficient if pertinent features of the problem are extracted.

    The learning process involves the training of network parameters on a set
    of training data that is representative of the problem at hand. Broadly
    speaking, the goal of the learning process is to produce a model with good
    generalizability, meaning that it is able to predict the outcome of new
    data samples not included in the training dataset. To measure how well the
    network is able to generalize, a subset of samples is selected from the
    whole dataset, remove it from the training process, and pass it through the
    network once the network has been trained. This subset of data samples is
    referred to as the validation dataset (also sometimes called the test
    data).

    The first step in building the network is to identify the appropriate
    features to use as inputs to the network. This determines the input
    dimensions and may also influence what type of architecture is to be used.
    The features chosen should effectively characterize the input data with as
    few dimensions as possible. This helps to keep the size of the network, in
    terms of the number of free parameters, as small as possible and may
    improve generalization.

    \subsection{Feature selection}

      For the feature selection step two main strategies are proposed. In the
      first strategy the spatial profiles of a select few flow quantities at
      the initial time are used. The network is tasked with learning the
      spatial features (i.e. gradients) necessary for detonation formation from
      this data. The second strategy utilizes the Khokhlov-inspired timescale
      analysis introduced in \Cref{chp:hotspot_analysis} as the input feature
      space. These two strategies are referred to as the \textit{naive
      strategy} and the \textit{Khokhlov strategy}, respectively.

      \subsubsection{Naive strategy}

        In the naive strategy, either the induction time profile or the
        reactive wave speed normalized by the soundspeed, denoted by $z =
        u_{sp} / c$, is used as the primary feature for training.  This choice
        is based on prior experience in analyzing detonation formation in the
        context of the reactivity gradient mechanism.  This strategy is
        referred to as naive because it considers only the raw flow quantity as
        a function of space and does not take into account any additional
        information that may be gathered from physical insight into the
        problem.

        Two slightly different approaches are taken with respect to the
        different datasets.  When training with the synthetic dataset, a radial
        section of the initial profile(s) is used as input, given that the
        profiles are symmetric. The dimension $n_{x}$ of the input array
        $\bm{\chi}$ is determined by the desired sampling rate. In particular,
        it will be determined by the ILES filter cutoff,
        $\Delta_{\mathrm{tn}}$, of the \texttt{tburn} models in which the
        network is to be employed.

        Five models consistent with the naive strategy are proposed:
        \begin{align}
          \mathbb{N}1 : \left\{\bm{\tau} \right\} \in \mathbb{R}^{n_{x}, 1} \mapsto q \in \left\{0,1\right\}, \label{eqn:n1} \\
          \mathbb{N}2 : \left\{\bm{\tau}, \bm{\rho}, \bm{c}, \bm{u} \right\} \in \mathbb{R}^{n_{x}, 4} \mapsto q \in \left\{0,1\right\}, \label{eqn:n2} \\
          \mathbb{N}3 : \left\{\bm{\tau}, \bm{\rho}, \bm{u} \right\} \in \mathbb{R}^{n_{x}, 3} \mapsto q \in \left\{0,1\right\}, \label{eqn:n3} \\
          \mathbb{N}4 : \left\{\bm{\tau}, \bm{c}, \bm{u} \right\} \in \mathbb{R}^{n_{x}, 3} \mapsto q \in \left\{0,1\right\}, \label{eqn:n4} \\
          \mathbb{N}5 : \left\{\bm{z}, \bm{\rho}, \bm{u} \right\} \in \mathbb{R}^{n_{x}, 3} \mapsto q \in \left\{0,1\right\}. \label{eqn:n5} 
        \end{align}
        The symbol $\mathbb{N}$ indicates that model belongs to the naive
        family of models. Model $\mathbb{N}1$ is the most simple, using as
        input only the induction time and mapping to the (scalar) output label,
        $q$. Models $\mathbb{N}2$ through $\mathbb{N}5$ use either the log of
        the induction time, $\bm{\tau}$, or the normalized reactive wavespeed,
        $\bm{z}$ as the sensor intended to capture the reactivity gradient
        mechanism.  Then they also include density, $\bm{\rho}$, and/or
        soundspeed, $\bm{c}$. The inclusion of density is intended to account
        for the density dependence observed in \Cref{chp:hotspot_analysis}. All
        models include the velocity, $\bm{u}$, to account for the effects of
        turbulent motions during the evolution.

\begin{figure}[!ht]
          \center
          \includegraphics[width=0.8\textwidth]{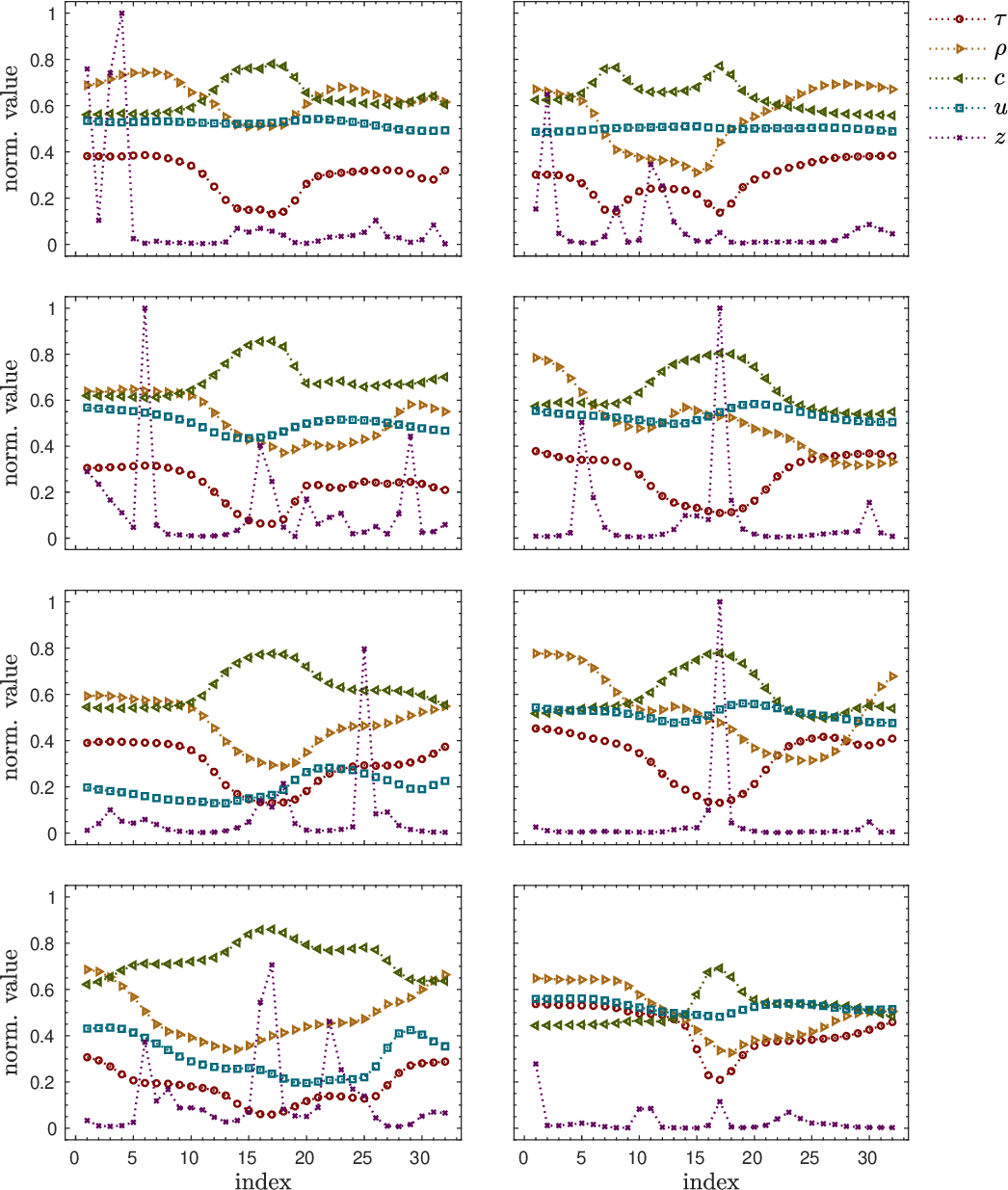}
          \caption{Normalized inputs from eight randomly chosen training samples.
            Shown in each panel is the normalized induction time (red circles),
            density (gold right-arrows), soundspeed (green left-arrows), velocity
            (blue squares), and $z$ value (purple crosses).}
          \label{fig:inputsamples}
        \end{figure}

        In \Cref{fig:inputsamples}
        the normalized inputs of eight randomly chosen training samples
        from the \texttt{TE} dataset are shown. Here the training features (flow variables)
        are normalized by the smallest and largest occuring values in the
        dataset such that the range of the inputs is between $0$ and $1$. These
        five variables are the superset of the variables available to models
        $\mathbb{N}1$ through $\mathbb{N}5$.

      \subsubsection{Khokhlov strategy}

        Another feature selection strategy developed to solve
        \Cref{eqn:mlequation} is advised by the Khokhlov-inspired analysis in
        \Cref{chp:hotspot_analysis}. The proper application of
        \Cref{eqn:khokhlov} using the new criteria \Cref{eqn:regionsize} is
        able to effectively separate non-detonation-forming and
        detonation-forming samples in the reactive timescale versus
        sound-crossing timescale phase space, as seen in
        \Cref{fig:khokhlovplanetburn}.

        In this strategy the \textbf{\textit{reactive timescale}},
        \textbf{\textit{sound-crossing timescale}}, \textbf{\textit{density}},
        and \textbf{\textit{divergence of velocity}} computed from the initial
        time are used as inputs. The latter two are taken to be the mean of the
        quantity over the region $r_{0}$. The first three choices are obvious
        from the results of \Cref{fig:khokhlovplanegauss}. The choice of the
        divergence of velocity as an additional input parameter is made to take
        into account existing expansion or compression of the gas in the region
        of interest.

        The following two models are proposed, each consistent with the
        aforementioned Khokhlov feature selection strategy:
        \begin{align}
          \mathbb{K}1 : \left\{\log_{10}\left(\sigma_{0}\right), \log_{10}\left(r_{0}/\langle c_{0} \rangle\right), \langle \rho_{0} \rangle \right\} \in \mathbb{R}^{3} \mapsto q \in \left\{0,1\right\}, \label{eqn:k1} \\
          \mathbb{K}2 : \left\{\log_{10}\left(\sigma_{0}\right), \log_{10}\left(r_{0}/\langle c_{0} \rangle\right), \langle \rho_{0} \rangle, \langle \mathrm{div}\ u_{0} \rangle \right\} \in \mathbb{R}^{4} \mapsto q \in \left\{0,1\right\}.
          \label{eqn:k2}
        \end{align}
        Here, again, $q$ is the binary output classification. The notation
        $\mathbb{K}1$ stands for `Khokhlov model $1$'. This model omits the
        divergence of velocity as input and is intended for use with the
        synthetic data only (again, all synthetic data have a velocity of zero
        everyhwere in the domain).  The model $\mathbb{K}2$ includes the
        divergence of velocity measurement and is intended for use with the
        \texttt{TE} data.

        Given that a detonation may occur on either side of the non-symmetric
        hotspot in the \texttt{TE} models, it must be decided which side to
        extract timescale data from. In the present work, the ratio of the
        reactive timescale, $\sigma_{0}$, to the sound-crossing timescale,
        $r_{0}/\langle c_{0} \rangle$, is computed for both sides, and the
        ratio with the \textit{lowest} value is selected. This value places the
        sample further toward the detonating portion of the timescale plane in
        \Cref{fig:khokhlovplanetburn}.
        




    

    \subsection{Neural network architecture}

      In terms of network architecture both ANNs and CNNs are considered. The ANN
      consists of several layers that sometimes referred to as a fully
      connected layers, as every unit in a layer applies a transformation to
      the output from each unit in the layer preceding it. The transformation
      is applied using weight values that later become the target of
      optimization. Each evaluation of a training sample produces a measure of
      error using a loss function. The updating of weights then involves the
      process of backpropagation, which calculates the gradient of the loss
      with respect to the weights and allows information about the performance
      of the network to improve the weights and biases.

      The unit output of the $\ell^{\mathrm{th}}$ fully connected layer in a
      network may be expressed mathematically as
      \begin{equation}
        q_{j}^{(\ell)} = \varphi \left( \sum_{i=1}^{n^{(\ell-1)}} w_{j,i}^{(\ell)} q_{i}^{(\ell-1)} + b_{j}^{(\ell)} \right),
        \label{eqn:unittransform}
      \end{equation}
      where $\varphi$ is a nonlinear activation function, $n^{(\ell)}$ is the
      number of units in layer $\ell$, $w_{j,i}$ is the weight for the connection
      between unit $j$ of layer $\ell$ and unit $i$ of layer $\ell-1$, and
      $b_{j}^{\ell}$ is the bias for unit $j$ of layer $\ell$ \cite{mattioli2021}.

      The unit transformation equation, \Cref{eqn:unittransform}, may be
      expressed in matrix form as $\bm{q}^{(\ell)} = \bm{w}^{(\ell)} \bm{q}^{(\ell-1)} +
      \bm{b}^{(\ell)}$. Then the feedforward ANN may be conveniently expressed as a
      series of layers acting in series such that the output of the final layer
      $L$ may be expressed as
      \begin{equation}
        \bm{q}^{(L)} = \varphi \left( \bm{w}^{(L)} \varphi \left(
          \bm{w}^{(L-1)} \varphi \left( \dots \left( \bm{w}^{(1)} \bm{\chi} +
          \bm{b}^{(1)} \right) \dots \right) + \bm{b}^{(L-1)} \right) + \bm{b}^{(L)}
          \right).
      \end{equation}

      Convolutional layers may also be included in this feedforward system.
      A convolutional layer applies a sliding filter across the input data to
      generate a feature map. The weights of this filter are learned during the
      training process. In the naive strategy the input data is strictly
      one-dimensional, thus a one-dimensional convolutional layer must be
      applied. The 1D convolutional output unit is given by
      \begin{equation}
        q_{j} = \varphi \left( \sum_{i=1}^{n_{x}^{(\ell)}} \sum_{k=1}^{n_{ch}^{(\ell)}} w_{i,k} \chi_{j+i,j+k} + b \right),
      \end{equation}
      where $n_{x}^{(\ell)}$ is the number of spatial points in the filter and
      $n_{ch}^{(\ell)}$ is the number of channels\footnote{Note that for the input
      layer the notation $n_{x}$ and $n_{ch}$ is used without superscripts for
      brevity}. Typically each channel of a filter is referred to as a kernel.

\begin{figure}[!ht]
        \center
        \includegraphics[scale=0.85]{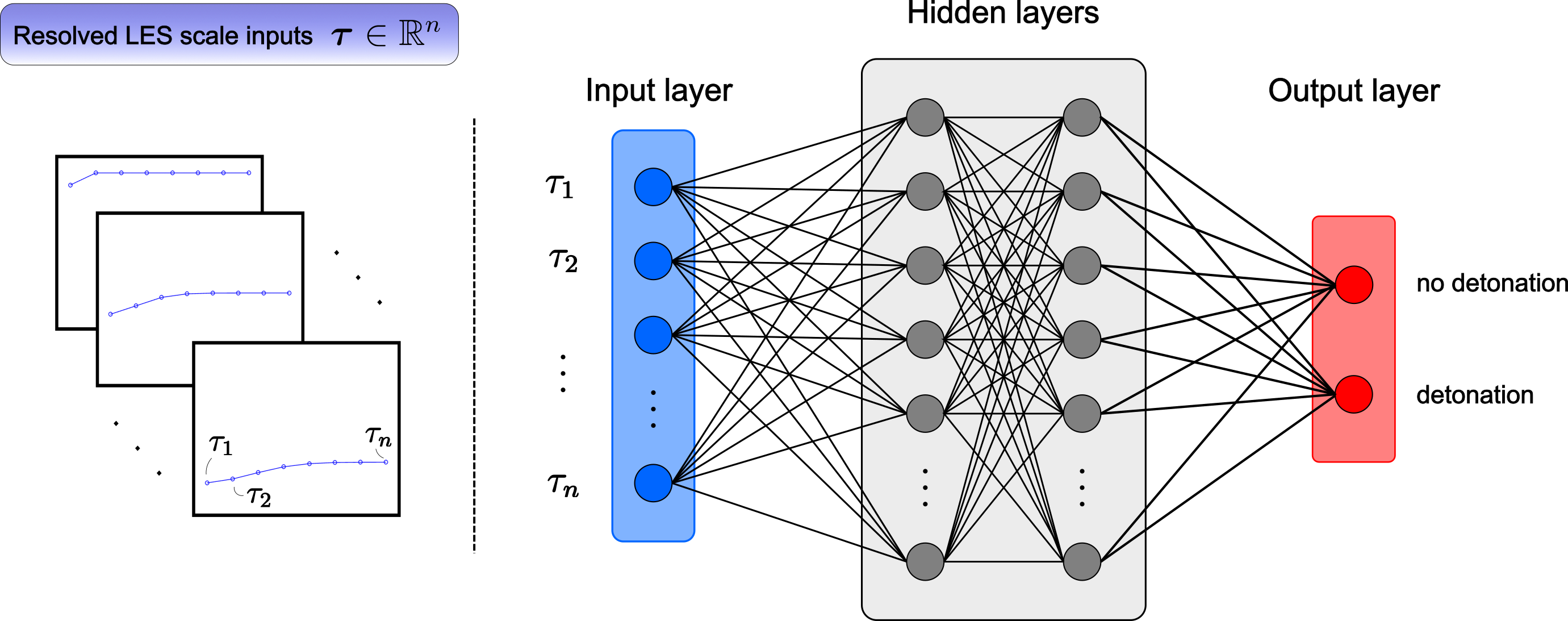}
        \caption{Overview of the neural network approach for the
          naive strategy, $\mathbb{N}1$. The network input is the induction
          time profile computed via DNS, but coarsened to the
          $\Delta_{\mathrm{tb}}$ cutoff resolution. The binary classification
          labels of `no detonation' and `detonation' are determined by the
          result of the corresponding DNS.}
        \label{fig:architecture}
      \end{figure}

      The basic architecture of the feedforward network for the naive strategy
      is illustrated in \Cref{fig:architecture}.
      The inputs to the network are are the spatial profiles of induction time,
      coarsened to the $\Delta_{\mathrm{tb}}$ filter scale, and plotted in
      blue. The hidden layers, shown in the gray region, may consist of a
      combination of convolutional and fully connected layers or only of fully
      connected layers.

      With the basic elements of network architecture established, details
      about the proposed network implementation are described next. These
      include the choice of activation function, loss function, solver, and
      more.

      \paragraph{Activation functions} Activation functions are used to
      introduce nonlinearity into the machine learning model. There are many
      choices of activation functions available. Certain qualities can make
      some activation functions superior for backpropagation such as the
      problem of vanishing gradients. The rectified linear unit (ReLU), which
      does not suffer from the issue of vanishing gradients, is widely used and
      accepted as a de facto standard for most applications.  The ReLU
      activation function is given by
      \begin{equation}
        \varphi(x) = 
        \begin{cases}
          x,\ \mathrm{if}\ x>0, \\
          0,\ \mathrm{otherwise}.
        \end{cases}
      \end{equation}
      For all of the dense layers and convolutional layers used in the present
      model, ReLU activation function is used. Finally, following the output
      layer, a softmax function is applied. This scales the activations and
      ensures that the probabilities sum to unity.

      \paragraph{Loss function} The binary cross-entropy loss function
      is used in all of the network implementations. The binary cross-entropy
      loss function is
      \begin{equation}
        \mathcal{L}_{\mathrm{bce}} = -\frac{1}{n_{s}}
          \sum_{i=1}^{n_{s}} q_{i} \log{\left(\hat{q_{i}}\right)} + \left( 1 - q_{i} \right) \log{\left( 1 - \hat{q_{i}} \right)},
      \end{equation}
      where, again, $n_{s}$ is the number of samples, $q_{i}$ are the labels and
      $\hat{q}_{i}$ are the predicted label values.

      \paragraph{Solver} Once the architecture is set, the weights
      and biases are computed using the \texttt{adam} \cite{kingma2017}
      implementation in Keras. This algorithm is a type of stochastic gradient
      descent optimization scheme.

      A scheduled learning rate is used that adjusts the learning rate,
      $\gamma$, during the learning process to reduce possibility of
      `overshoot' once a local minima in the solution space is found. Once a
      minimum number of epochs has been reached, the learning rate is adjusted
      by multiplying the previous rate by $\exp{(-0.01)}$.



    \subsection{Performance metrics}

      The performance of the neural network models is assessed using several
      commonly used metrics. Given the true-positives, $\TP$, false-positives,
      $\FP$, true-negatives, $\TN$, and false-negatives, $\FN$, the accuracy is
      given as
      \begin{equation}
        \mathrm{Accuracy} = \frac{\TP + \TN}{\TP + \TN + \FP + \FN}.
      \end{equation}
      In binary classification the accuracy metric gives the ratio of the
      number of correct predictions to the number of total predictions made.
      The next metric considers the number of positive predictions 
      (detonation-forming hotspots) that are actually true. This
      is the precision metric, given as
      \begin{equation}
        \mathrm{Precision} = \frac{\TP}{\TP + \FP}.
      \end{equation}
      Similarly, the recall (also known as the sensitivity or true positive
      rate), which is the number of actual positive samples that are correctly
      identified by the classifier as such, is also used. The recall is given
      by
      \begin{equation}
        \mathrm{Recall} = \frac{\TP}{\TP + \FN}.
      \end{equation}

      Finally the classification performance using a threshold-invariant and
      scale-invariant method known as the receiver operating characteristic
      (ROC) curve is considered. The ROC curve is comprised of many points on
      the Recall (also known as the sensitivity or true positive rate) versus
      false positive rate plane, with each point corresponding to a different
      threshold value. The integral of this function is known as the `area
      under the curve' (AUC), and is used as a common metric in testing
      accuracy. For a perfect classifier the value of the AUC is $1$.

    \subsection{Using the classifier as a subgrid-scale model}

      The primary use of the new classifier is as a component of a SGS model
      for DDT in full-star simulations of SN Ia. Referring back to
      \Cref{fig:wdscales}, the connection of scales relevant to DDT and the
      roughly $\unit[1\times10^{5}]{cm}$ resolution of the full-star models is
      required. A two-component SGS model is proposed. The first component
      requires the new classifier to assign a probability of detonation
      formation for each hotspot produced by a given reactive turbulence
      realization on the $\Delta_{\mathrm{tb}}$ scale. The predictions by the
      classifier are made during runtime of the model. The second component
      then involves another training process to assign to each reactive
      turbulence realization considered a probability of producing a DDT. Then,
      the idea is to apply this learned model during runtime of the full-star
      models to determine when and where a DDT is likely.

      To this end, a kernel tracking algorithm is developed for Proteus and
      stored in a new module named \texttt{SubGridScaleDDT}. The program
      assigns unique ID numbers to prospective DDT kernels and tracks their
      movement during the simulation.  If the properties of a prospective
      kernel meet user-defined criteria, then data is extracted from the
      vicinity of that kernel and passed to the trained neural network model of
      choice. The Proteus simulation code interfaces with the Python/Keras
      network model implementation using a package Python package known as CFFI
      (C Foreign-Function Interface). For more details on the embedding of the
      network model implementations, see \Cref{chp:sgsminproteus}.

      The \texttt{SubGridScaleDDT} module has some features that make it
      convenient to use. As stated, the module tracks the individual DDT
      kernels and their location between timesteps. Kernels are ranked in order
      of increasing minimum induction time, meaning that the kernels with the
      lowest induction time are prioritized. While the number of kernels that
      can be tracked is very large, the number of kernels allowed to be
      evaluated can be set by the user, saving computational resources. The
      frequency of evaluations with respect to the timestep number may be
      user-adjusted, as well as the starting simulation time to begin
      evaluations.

      When the naive strategy is used, the profiles of the appropriate
      quantities are extracted from the vicinity and passed to the network. In
      1D this is trivial. In 2D or 3D the procedure stays conceptually the
      same, however now multiple `lineouts', or 1D data strips are extracted at
      multiple angles for each kernel. The number of lineouts is
      user-determined. The rational for this 1D lineout procedure is given by
      the inherently one-dimensional nature of the Zel'dovich reactivity
      gradient mechanism.

  \section{Results}
  \label{sec:nnresults}

    With the neural network models defined as well as the metrics used to
    characterize their performance, the results of the training
    process are presented.  At the end of the section results are presented of
    the numerical experiments in which the learned networks are used as a SGS
    model in ILES of reactive turbulence (in the \texttt{tburn} models) to
    predict the formation of detonations.

    It is necessary to introduce a few specific training/validation datasets
    that can be used interchangeably with the neural network models. First the
    dataset \texttt{td.naive.synthetic} is defined which processes the results
    of the synthetic databases \texttt{GLtHt}, \texttt{GMdHt}, \texttt{GHvHt},
    \texttt{GMdCd}, \texttt{GMdWm}, and \texttt{GRND} into the format readable
    for the naive network models $\mathbb{N}1$ through $\mathbb{N}5$. The same
    databases are also combined into a dataset \texttt{td.khokhlov.synthetic}
    which processes the data into the format compatible with the Khokhlov
    strategy. Then the \texttt{TE} database is similarly processed into two
    training/validation datasets, \texttt{td.naive.turb} and
    \texttt{td.khokhlov.turb}.

    \subsection{Naive strategy}
      
\begin{figure}[!ht]
        \centering
        \includegraphics[scale=0.75]{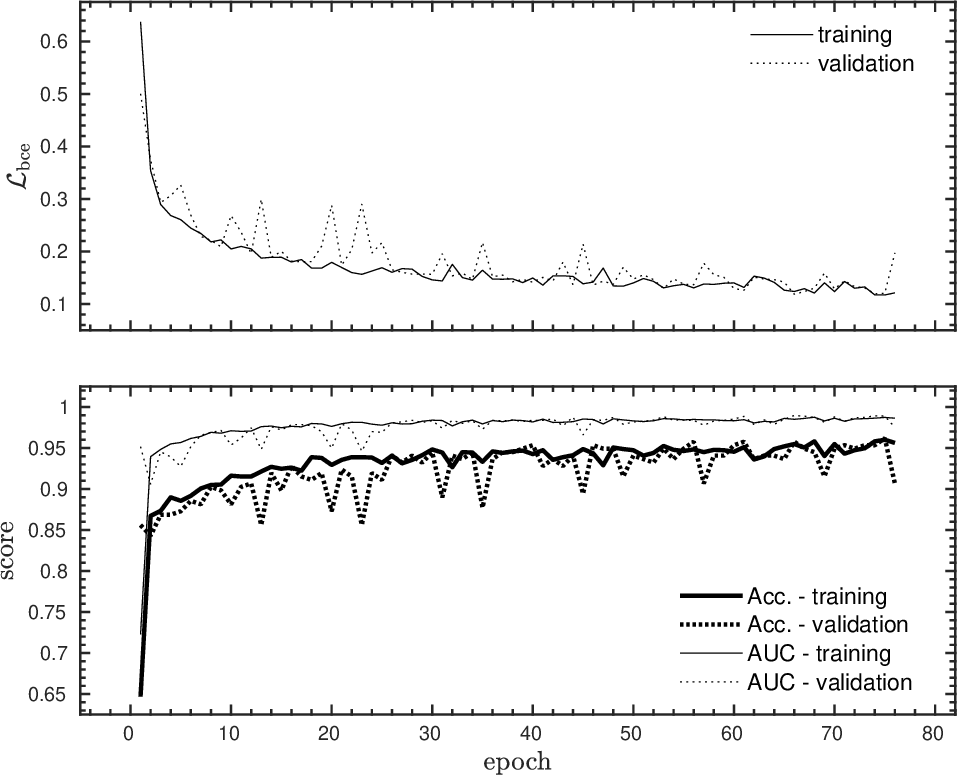}
        \caption{Traing and validation loss (top panel) and accuracy metrics (bottom panel) for the $\mathbb{N}1$ model trained on the synthetic dataset, dataset \texttt{td.naive.synthetic}.}
        \label{fig:trainlossnaivegauss}
      \end{figure}

      The communication of results begins with the training and validation
      performance of model $\mathbb{N}1$ on the synthetic dataset
      \texttt{td.naive.synthetic}. In this model the only input feature is the
      profile of induction time extracted from one half of the symmetric
      synthetic hotspot. The performance of  the model over $200$ training
      epochs is shown in \Cref{fig:trainlossnaivegauss}.
      The model achieves high scores in both Accuracy and AUC for the training
      and validation sets.

      Next the model performance when additional features are included is
      presented. The models $\mathbb{N}2$ through $\mathbb{N}5$ are trained on
      the \texttt{td.naive.turb} dataset. A subset of this training dataset is
      used to tune the hyperparameters of the model, and the
      process is documented in \Cref{chp:nnappendix}. A summary of the results
      of the hyperparameter tuning process are included in
      \Cref{tbl:bestarchitectures}.
      \begin{table}[!htb]
        \centering
        \begin{tabular}{@{}lcccccl@{}}\toprule
          Model & \texttt{nfltrs} & \texttt{fltrsz} & \texttt{dopool} & \texttt{nnodes} & \texttt{drpout} & \texttt{nparams} \\\midrule
          $\mathbb{N}2$ & - & - & - & - & - & $205,114$ \\
          \ \ $\texttt{clyr\_1}$ & $8$ & $5$ & $0$ & - & - & - \\
          \ \ $\texttt{clyr\_2}$ & $128$ & $7$ & $0$ & - & - & - \\
          \ \ $\texttt{dlyr\_1}$ & - & - & - & $256$ & $0$ & - \\
          $\mathbb{N}3$ & - & - & - & - & - & $267,362$ \\
          \ \ $\texttt{clyr\_1}$ & $8$ & $3$ & $0$ & - & - & - \\
          \ \ $\texttt{clyr\_2}$ & $128$ & $3$ & $0$ & - & - & - \\
          \ \ $\texttt{dlyr\_1}$ & - & - & - & $128$ & $0.25$ & - \\
          \ \ $\texttt{dlyr\_2}$ & - & - & - & $512$ & $0.25$ & - \\
          $\mathbb{N}4$ & & - & - & - & - & $90,082$ \\
          \ \ $\texttt{clyr\_1}$ & $8$ & $3$ & $0$ & - & - & - \\
          \ \ $\texttt{clyr\_2}$ & $128$ & $7$ & $1$ & - & - & - \\
          \ \ $\texttt{dlyr\_1}$ & - & - & - & $128$ & $0.25$ & - \\
          \ \ $\texttt{dlyr\_2}$ & - & - & - & $128$ & $0.25$ & - \\
          $\mathbb{N}5$ & - & - & - & -  &  - & $660,962$ \\
          \ \ $\texttt{clyr\_1}$ & $8$ & $3$ & $0$ & - & - & - \\
          \ \ $\texttt{clyr\_2}$ & $128$ & $3$ & $1$ & - & - & - \\
          \ \ $\texttt{dlyr\_1}$ & - & - & - & $512$ & $0$ & - \\
          \ \ $\texttt{dlyr\_2}$ & - & - & - & $512$ & $0$ & - \\\bottomrule
        \end{tabular}
        \caption{Summary of the network architectures resulting from the
        hyperparameter tuning process. For each of the four models,
        $\mathbb{N}1$, $\mathbb{N}2$, $\mathbb{N}3$, and $\mathbb{N}4$, the
        table shows the number of convolutional filters (\texttt{nfltrs}),
        filter size (\texttt{fltrsz}), use of pooling (\texttt{dopool}), number
        of dense layer nodes (\texttt{nnodes}), and dropout percentage
        (\texttt{drpout}). The $n$th convolutional layer is labeled as
        \texttt{clyr\_n}, and the $n$th dense layer is labeled as
        \texttt{dlyr\_n}.}
        \label{tbl:bestarchitectures}
      \end{table}
      From these results it seems that the preferred number of filters for the
      two convolutional layers is $8$ and $128$, respectively. The filter width
      in each of the layers varies. The number of dense layers seems also to
      fluctuate, as does the number of nodes per layer.

\begin{figure}[!ht]
        \centering
        \includegraphics[scale=0.725]{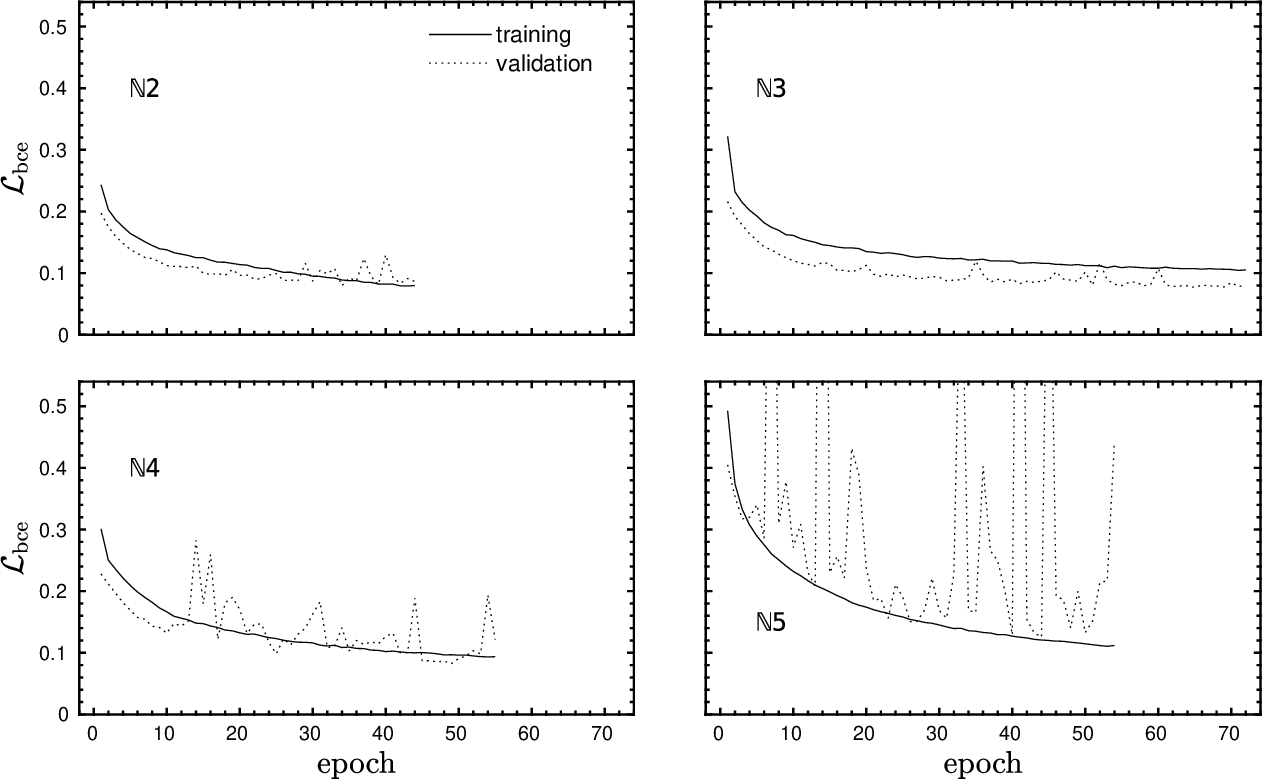}
        \caption{Comparison of the training and validation loss for the
          $\mathbb{N}2$ (top left panel), $\mathbb{N}3$ (top right panel),
          $\mathbb{N}4$ (bottom left panel), and $\mathbb{N}5$ (bottom right
          panel) models on the \texttt{td.naive.turb} dataset.}
        \label{fig:trainlossnaivetburn}
      \end{figure}

      These tuned models are trained and tested on the full
      \texttt{td.naive.turb} dataset over about $50$ to $60$ epochs, before
      early stopping is used. Early stopping is used in each of the models when
      improvement is no longer being made in the validation AUC score. The
      training and validation binary cross-entropy loss history is presented in
      \Cref{fig:trainlossnaivetburn}.
      The model $\mathbb{N}4$ achieves the lowest loss value on the validation
      set at the end of the training process, with a value of around
      $\mathcal{L}_{\mathrm{bce}} = 0.16$. In fact the histories for both
      $\mathbb{N}2$ and $\mathbb{N}3$ achieve scores around this value earlier
      in their training as well. Notably, the performance of the model
      $\mathbb{N}5$ which uses the $z$ value in place of the induction time,
      $\tau$, is quite poor.

\begin{figure}[!ht]
        \centering
        \includegraphics[scale=0.725]{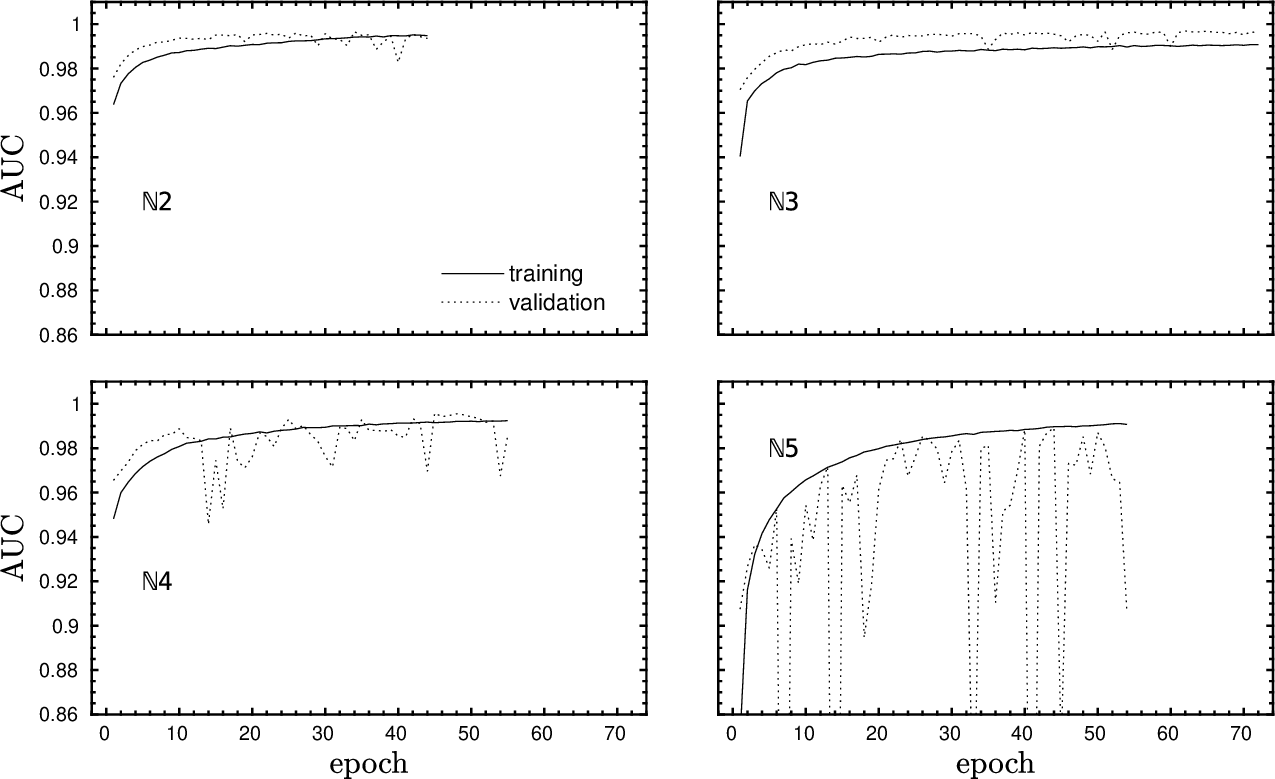}
        \caption{Comparison of the AUC metric on the training and validation
          data for the $\mathbb{N}2$ (top left panel), $\mathbb{N}3$ (top
          right panel), $\mathbb{N}4$ (bottom left panel), and $\mathbb{N}5$
          (bottom right panel) models on the \texttt{td.naive.turb} dataset.}
        \label{fig:trainaucnaivetburn}
      \end{figure}

      The training and validation AUC score histories for the four models are
      shown in \Cref{fig:trainaucnaivetburn}.
      It is observed that for the validation AUC score, again models
      $\mathbb{N}2$, $\mathbb{N}3$, and $\mathbb{N}4$ perform very similarly.
      Note that the final weights for these models are taken from the epoch
      with the best validation AUC score.

\begin{figure}[!htb]
        \centering
        \includegraphics[width=0.65\textwidth]{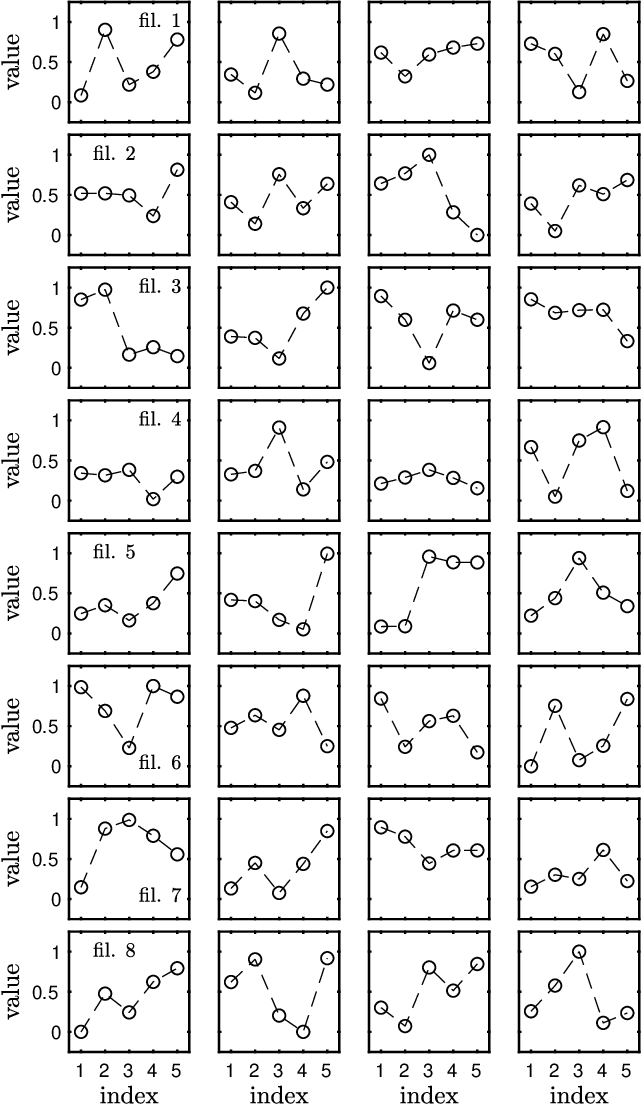}
        \caption{Weights for filters $1-8$ of the first layer of model
          $\mathbb{N}2$ are plotted, with each row comprising one filter and the
          individual kernels for inputs of induction time, density, soundspeed,
          and velocity spanning the columns. For plotting convenience, weights
          are normalized across filters for each input channel.}
        \label{fig:filtersfiltersfilters}
      \end{figure}

      In an effort to visualize to some extent the workings of the network, the
      $8$ filters of the first layer comprising the $\mathbb{N}2$ model are
      shown in \Cref{fig:filtersfiltersfilters}.
      In this figure each of the filters $1-8$ are `unwrapped' along the
      columns, showing the kernels of the filter corresponding to each input
      channel. In order, the columns show the learned weights of the kernels
      for induction time, density, soundspeed, and velocity.

    \subsection{Khokhlov strategy}

\begin{figure}[!ht]
        \center
        \includegraphics[scale=0.75]{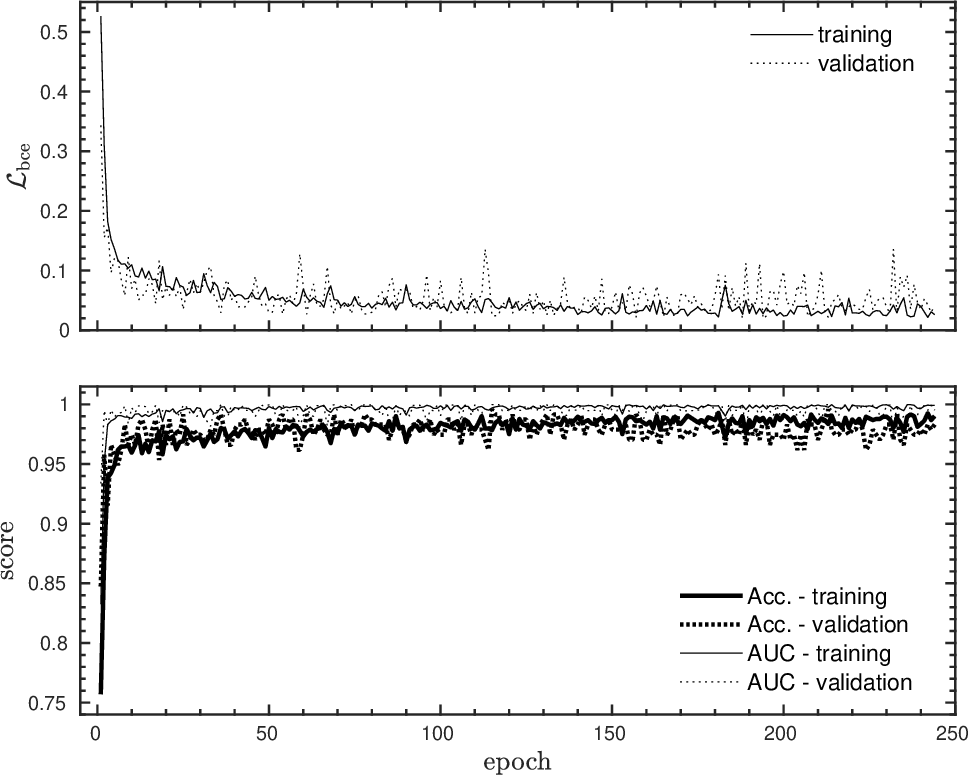}
        \caption{Training and validation performance of the $\mathbb{K}1$ model
        trained on the synthetic dataset, dataset
        \texttt{td.khokhlov.synthetic}. The network achieves a very high
        Accuracy score for the training (thick solid black line) and validation
        (thick dotted black line) datasets. Also the AUC score is near the
        maximum score possible of $1$ for both datasets (thin solid blue line
        and thin dotted blue line).}
        \label{fig:modelk1}
      \end{figure}

      Next, the performance of the model employing the Khokhlov
      strategy is examined. The first results presented are of the model $\mathbb{K}1$ trained on the synthetic
      dataset, \texttt{td.khokhlov.synthetic}. A $10$ percent data split is
      used for the validation set. In \Cref{fig:modelk1}
      the training and validation loss is plotted (top panel) as well as the
      accuracy scores (bottom panel). Next, the Accuracy (abbreviated Acc.  in
      the legend) and AUC scores are examined for the training and validation
      data.  An essentially perfect AUC score is observed on the validation
      data at epoch $194$. At the same epoch it is observed that
      $\mathrm{Accuracy} \approx 0.99$.

\begin{figure}[!ht]
        \center
        \includegraphics[scale=0.75]{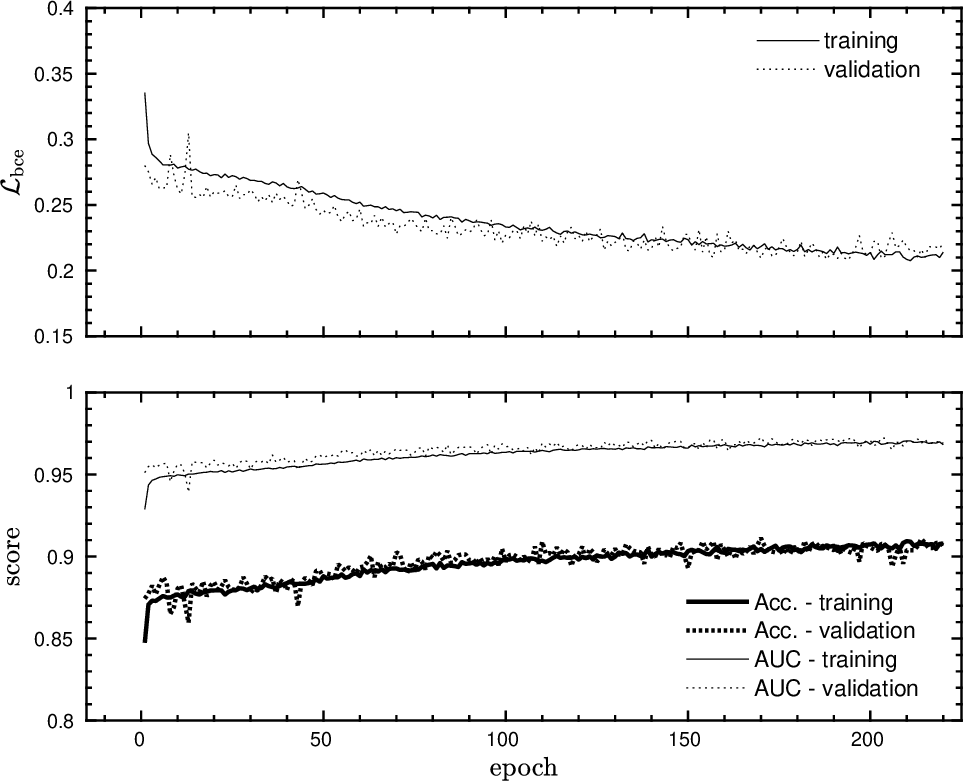}
        \caption{Training and validation performance of the $\mathbb{K}2$ model
          trained on the turbulence extracted dataset, dataset
          \texttt{td.khokhlov.turb}.}
        \label{fig:modelk2}
      \end{figure}

      Then the performance of model $\mathbb{K}2$ trained on dataset
      \texttt{td.khokhlov.turb} is presented. In \Cref{fig:modelk2}
      the training and validation loss is shown (top panel) along with the
      Accuracy and AUC scores (bottom panel). The training and validation loss
      display similar convergence, indicating that the model is generalizing
      adequately well and not `overfitting'. The same property is observed in
      the accuracy scores. Even for the more complicated dataset the current
      approach is able to predict about $91$ percent of validation samples
      correctly. The maximum AUC score attained is about $\mathrm{AUC} = 0.97$.

    \subsection{Performance summary}
    
      In \Cref{tbl:trainingsummary}
      \begin{table}[!htb]
        \centering
        \begin{tabular}{@{}lccccl@{}}\toprule
          Strategy & Dataset & Accuracy & Precision & Recall & AUC \\\midrule
          $\mathbb{N}1$ & \texttt{td.naive.synthetic} & 0.958 & 0.936 & 0.983 & 0.990 \\
          $\mathbb{N}2$ & \texttt{td.naive.turb} & 0.971 & 0.976 & 0.974 & 0.997 \\
          $\mathbb{N}3$ & \texttt{td.naive.turb} & 0.971 & 0.973 & 0.978 & 0.997 \\
          $\mathbb{N}4$ & \texttt{td.naive.turb} & 0.964 & 0.961 & 0.978 & 0.996 \\
          $\mathbb{N}5$ & \texttt{td.naive.turb} & 0.947 & 0.935 & 0.977 & 0.989 \\
          $\mathbb{K}1$ & \texttt{td.khokhlov.synthetic} & 0.987 & 0.983 & 0.991 & 1.0 \\
          $\mathbb{K}2$ & \texttt{td.khokhlov.turb} & 0.911 & 0.924 & 0.928 & 0.973 \\\bottomrule
        \end{tabular}
        \caption{Summary of results for the methods introduced in
          \Crefrange{eqn:n1}{eqn:n5} and \Crefrange{eqn:k1}{eqn:k2} on the
          validation samples from datasets \texttt{td.naive.synthetic},
          \texttt{td.naive.turb}, \texttt{td.khokhlov.synthetic}, and
          \texttt{td.khokhlov.turb}}
        \label{tbl:trainingsummary}
      \end{table}
      the training performance is summarized across the models and dataset
      combinations considered. Values are rounded to the nearest thousandth.
      Besides the slightly worse performance of the $\mathbb{N}5$ feature
      selection strategy, the naive strategies perform quite similarly. The
      Khokhlov feature selection strategy performs well on the realistic data,
      but not quite as well as the naive strategies do.


    \subsection{Verification of the subgrid-scale model in the \textit{a posteriori} setting}
    \label{sec:aposteriori}

      Verification testing of the trained neural network models is done in the
      \textit{a posteriori} setting. The networks are employed as SGS models in
      semi-realistic, one-dimensional reactive turbulence.  These turbulence
      simulations use the same setup as the \texttt{tburn} models except that
      they are of course one-dimensional. The turbulence is driven in $3$
      dimensions using the TurbGen code \citep{federrath2010}. The 3D driving
      routine computes velocities across all three components, with the idea
      being to introduce some of the axial physics of turbulence into the 1D
      evolution. The use of 1D turbulence simulations in the verification
      process is intended to save time over fully 3D simulations, allowing more
      verification studies to be run and better statistics of performance to be
      obtained.

\begin{figure}[!htb]
        \center
        \includegraphics[width=\textwidth]{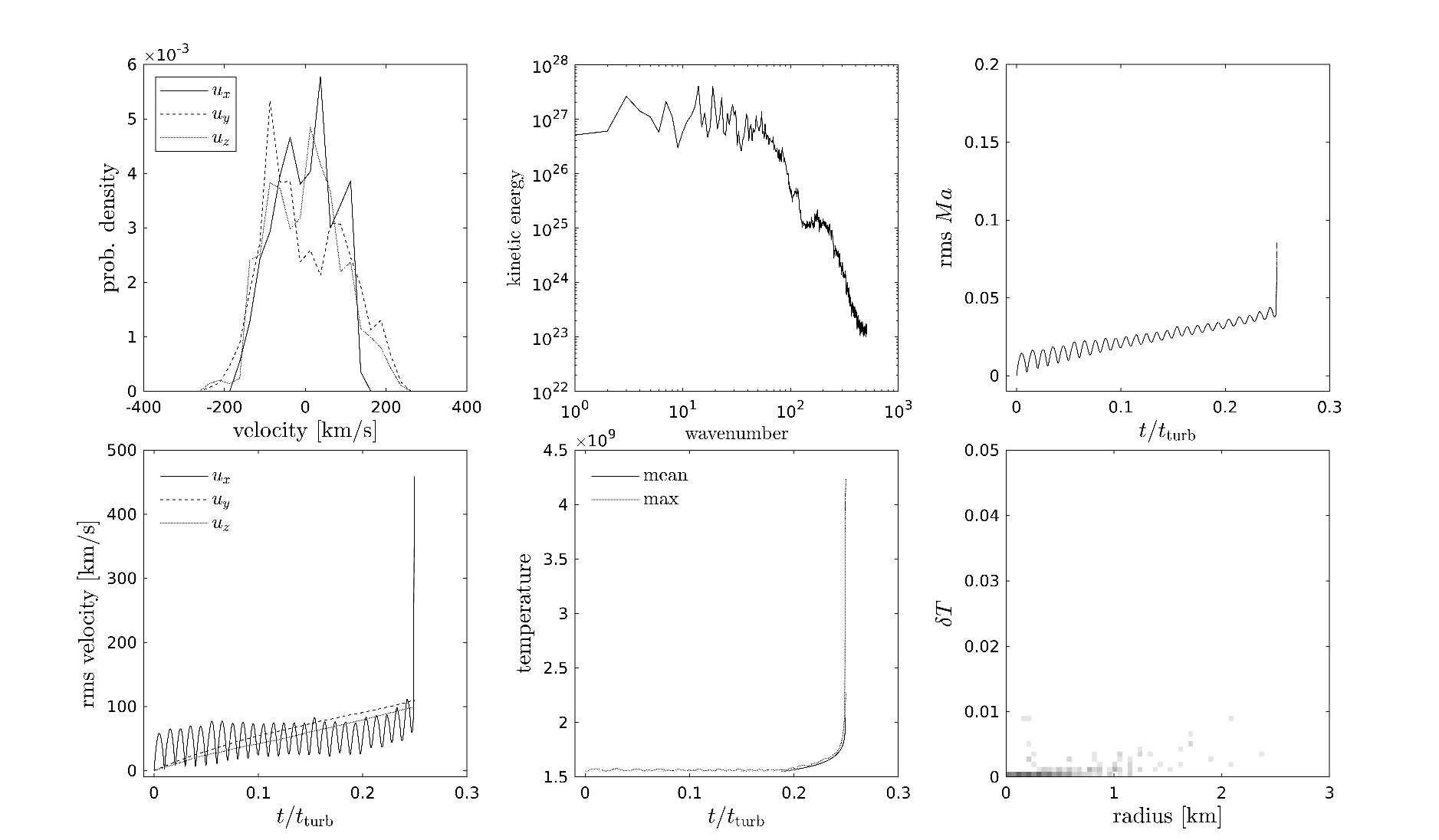}
        \caption{Overview of 1D turbulence used for verification studies. Shown
          is the probability density of velocities collected over $125$ time
          samples (top left panel); the kinetic energy spectrum (top right panel)
          collected over the same number of samples, compensated by multiplying
          with $k^{5/3}$, where $k$ is the wavenumber; the evolution of the RMS
          Mach number (middle left panel) the three velocity components (middle
          right panel), and the mean and maximum temperature (bottom left
          panel); and finally a bivariate histogram of amplitudes and widths of
          temperature fluctuations (bottom right panel) with higher frequency
          indicated by darker color.}
        \label{fig:turbmultipane}
      \end{figure}

      Through trial and error the turbulence driving parameters are adjusted to
      produce conditions that replicate some of the important statistics of 3D
      turbulence as closely as possible. Firstly, the turbulence
      is driven with a power law of $k^{-5/3}$, where $k$ is the wavenumber. The
      target velocity dispersion is $\unit[0.150 \times 10^{8}]{cm}$ and the
      target RMS Mach number is around $0.05$. While these velocities are below
      what is expected in 3D, they are about as high as they can be in 1D
      without producing strong shockwaves that initiate detonations before the
      Zel'dovich mechanism has a chance to occur. In \Cref{fig:turbmultipane}
      some of the salient quantities are plotted over about $0.25$
      autocorrelation times (here the autocorrelation time is
      $t_{\mathrm{turb}} \approx \unit[0.213]{s}$). The compensated kinetic
      energy spectrum (top right panel of \Cref{fig:turbmultipane}) scales well
      with the expected power law up to a wavenumber of about $30$, where
      numerical dissipation begins to take over, marking the end of the
      `intertial range'. The velocity components remain in a similar range due
      to the use of carefully selected amplitude adjustment factors. The
      temperature is initially set to $T = \unit[1.55\times10^{9}]{K}$.
      Self-heating raises the temperature closer to the ignition temperature
      during the evolution, and eventually a detonation occurs.

      The evaluation of the network models is performed during the reactive
      turbulence realizations.  The best performing models are selected for
      this verification process. In particular, the naive strategy,
      $\mathbb{N}2$, and the Khokhlov strategy, $\mathbb{K}2$. Both of these
      network models are trained using the dataset with realistic conditions.

      To establish meaningful statistics, the turbulent realizations are run
      $64$ times, each with a unique random seed. The network models evaluate
      potential detonation-forming hotspots during the evolution and compute a
      predicted probability of detonation for each one. The temporal accuracy
      of the predictions is assessed by measuring the time delay, $\delta_{t}$,
      between the time when the prediction is made and the time of the
      detonation birth. Naturally, some time delay is expected. The spatial
      accuracy is assesed by measuring the distance, $\delta_{x}$, from the
      point where the prediction is made to the \textit{predicted} initial
      location of the hotspot center. This predicted location is computed by
      tracing the characteristics of the detonation backwards.

\begin{figure}[!htb]
        \center
        \includegraphics[width=0.69\textwidth]{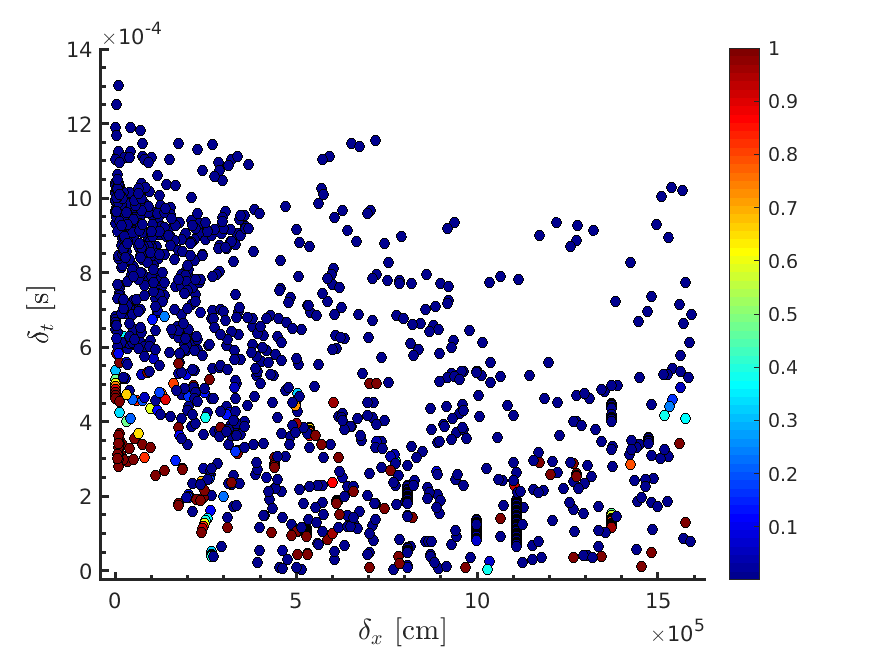}
        \caption{Results of the verification process for the naive model
          $\mathbb{N}2$. Shown are the network predicted detonation probabilities
          (shown in color) on the $\delta_{t}$-$\delta_{x}$ plane.}
        \label{fig:verificationN2}
      \end{figure}

      First, the model $\mathbb{N}2$ is considered. \Cref{fig:verificationN2}
      shows the network predicted detonation probability, with the value shown
      in color, on the $\delta_{t}$-$\delta_{x}$ plane. The predicted values at
      or near $1$ (indicating a strong confidence in a detonation occuring) are
      concentrated near the origin of the plot as they should be. The time
      delays are concentrated between $\unit[2.5\times10^{-4}]{s}$ and
      $\unit[4\times10^{-4}]{s}$. This is on the order of the amount of time
      for a detonation wave to develop (see \Cref{fig:spacetime} for
      reference). The points of high probability spread away from $\delta_{x}
      \approx 0$ may be false positives; those with $\delta_{t}$ being less
      than the detonation delay time may also be true positives, only that the
      simulation is terminated once the first detonation is born.

\begin{figure}[!htb]
        \center
        \includegraphics[width=0.69\textwidth]{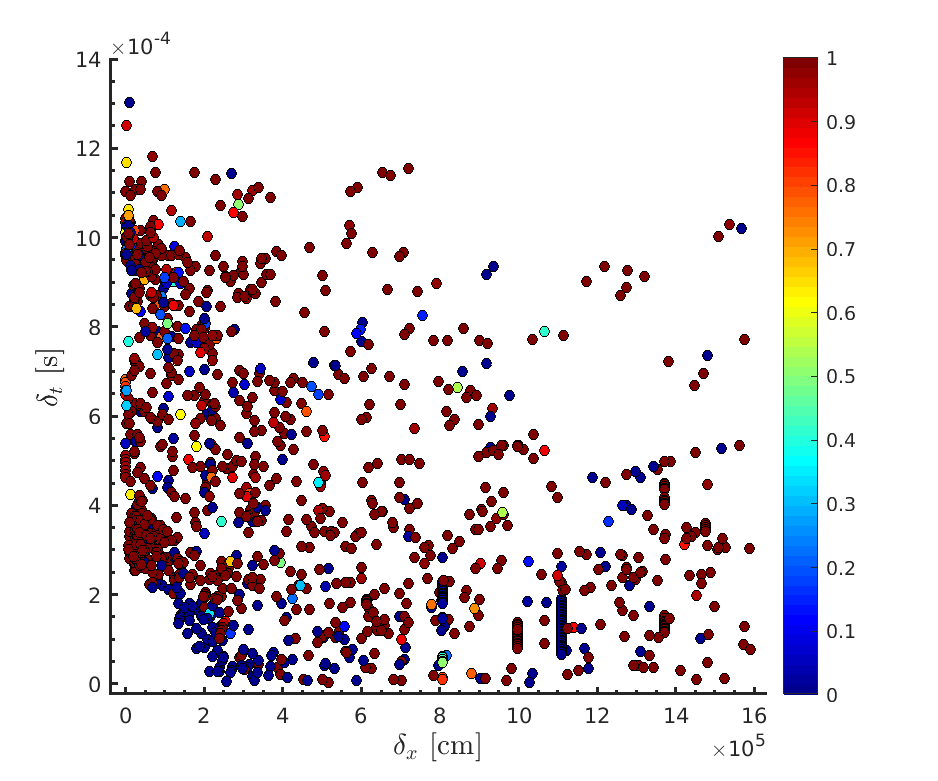}
        \caption{Results of the verification process for the Khokhlov model
          $\mathbb{K}2$. Shown are the network predicted detonation probabilities
          (shown in color) on the $\delta_{t}$-$\delta_{x}$ plane.}
        \label{fig:verificationK2}
      \end{figure}

      Next, the model $\mathbb{K}2$ is evaluated for the same series of
      reactive turbulence realizations. \Cref{fig:verificationK2}
      shows the predictions in the $\delta_{t}$-$\delta_{x}$ plane. Again, a
      concentration of predicted values of $1$ is seen near the origin as in
      \Cref{fig:verificationN2}. This indicates that the ability of the model
      to positively identify actual detonation-forming hotspots is
      satisfactory. However, there are undoubtedly many false positive
      predictions throughout the plane. It would seem that the Khokhlov
      strategy is less suited to distinguishing between detonation-forming and
      non-detonation-forming hotspots. Rather it seems to be identifying the
      necessary, but not sufficient condition for detonation to occur.

  \section{Discussion}


    In this chapter several novel strategies were introduced for predicting the
    onset of detonation formation among hotspots in reactive turbulence.  The
    results of several tens of thousands of direct numerical simulations of
    hotspot configurations were processed to produce training data for the
    NN-based strategies. The timescale analysis of Khokhlov
    \cite{khokhlov1991b} was used to parameterize the input space, allowing for
    a low-dimensional input space to be learned by the network.

    Some observations on the performance of the two feature selection
    strategies are provided. First, it seems as though the performance of the
    Khokhlov strategy on the realistic dataset is, comparable to, but
    ultimately inferior to the naive strategies. One possible explanation for
    this finding is that the Khokhlov strategy does not consider pointwise data
    but rather statistical quantities computed from data within the center of
    the hotspot region. Perhaps some information that is highly correlated to
    the potential for detonation formation is lost in the process of averaging,
    such as the local gradients. In the naive approach these local gradients
    are potentially learned during the network training.

    Another possibility is that the current realization of the Khokhlov
    strategy is simply lacking some additional statistical quantities. For
    example, it may be beneficial to supplement the input space with quantities
    such as the variance of densities in the region, or the mean Mach number to
    incorporate information about the mean flow. The potentially missing
    information can only be determined by additional exploratory studies such
    as the ones undertaken in the present work.

    \subsection{On the use of the classifier as a subgrid-scale model}

      Next the potential of the classifiers to be effectively used as SGS model
      in the \textit{a posteriori} setting is discussed. In the machine
      learning and `data science' communities there are some general standards
      for ruling that a classifier has sufficiently learned a task. For a
      dataset whose labels are not significantly skewed, an Accuracy score of
      above $80$ percent may be considered `good', and a value above $90$
      percent may be considered `excellent'. Note that the target application
      of reactive turbulence is an inherently stochastic process, and a given
      realization of reactive turbulence that is close to a ignition may
      produce multiple potential detonation-forming hotspots nearly
      simultaneously.  Thus there is no requirement to correctly classify every
      single one.  Instead the aim is for the neural network classifier to
      provide us with a general measure, or probability, of how likely that
      particular realization is to produce a DDT. In other words, the results
      of the present work, particularly the results of
      \Cref{fig:verificationN2} and \Cref{fig:verificationK2}, suggest that the
      output of the classifier be used as part of a statistical model rather
      than a deterministic one.


    \subsection{Notes on convergence of network accuracy}

      Finally some consideration is given to the issue of convergence of
      network accuracy with respect to the number of samples. This issue is
      pertinent for future work that considers the generation of a new,
      multidimensional dataset, as expanding the DNS of hotspot evolution to 2D
      or 3D will require roughly $4\times$ or $8\times$ computational cost.
      Thus it is important to provide an accurate assessment of the amount of
      computational work required to train a new network.

      The convergence of network accuracy is explored by running numerous
      training procedures for varying number of samples, $n_{s}$, while keeping
      the number of validation samples fixed at around $2000$ for each case.
      For each value of $n_{s}$, seven trials are run. The samples that are not
      in the validation set are eligible to be included in the training set
      during each trial, and that set is determined randomly. The same
      validation samples are used for each new trial.

\begin{figure}[!htb]
        \center
        \includegraphics[width=0.6\textwidth]{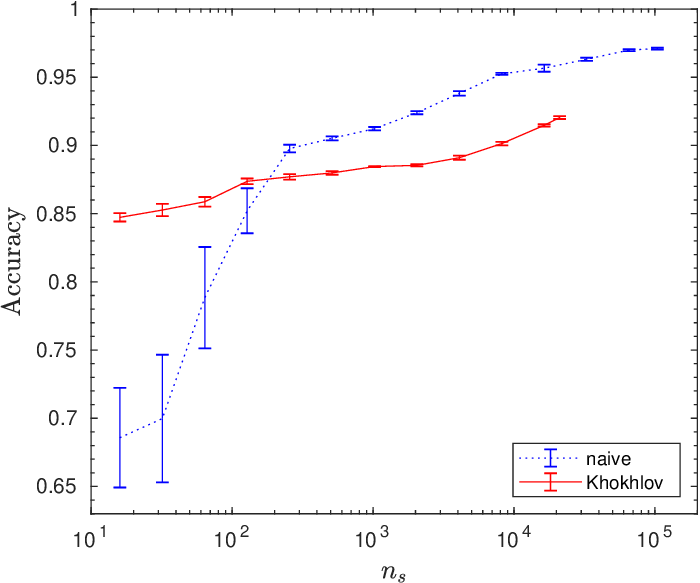}
        \caption{Convergence of network accuracy with the number of training samples.}
        \label{fig:convergence}
      \end{figure}

      In \Cref{fig:convergence}
      the convergence of network accuracy is shown with respect to the number
      of training samples. For the naive strategy, a sharp initial rate of
      increase is observed. The accuracy for the Khokhlov strategy is initially
      higher but the rate of increase is significantly less than that of the
      naive strategy initially. This suggests that Khokhlov strategy is able to
      generalize better with fewer number of training samples. However the
      naive strategy seems to differentiate better between
      non-detonation-forming and detonation-forming hotspots with increasing
      $n_{s}$. Note that the larger number of samples used for the naive
      approach is due to the data augmentation procedure introduced in
      \Cref{sec:realistic}.


\chapter{Summary}
\label{chp:summary}

    In this work the problem of detonation initiation within hotspots in
    degenerate matter is addressed. The leading theoretical and computational
    evidence suggests that hotspots are likely to cause detonation of carbon fuel
    during a Type Ia supernova under certain conditions and assumptions. Based on
    the existing literature and the present work, it is speculated that these
    hotspot-induced detonations occur due to the Zel'dovich reactivity gradient
    mechanism. The Zel'dovich reactivity mechanism postulates that critical
    gradients of induction time are required near the hotspot in order for a
    `spontaneous wave', generated near the center of the hotspot, to couple to the
    outgoing compressive wave, which is generated due to overpressure. If the
    critical gradient conditions are met then there is a chance for the coupling to
    succeed and create a positive-feedback loop where the reactive/spontaneous wave
    and the compressive wave reinforce each other through the compression and
    expansion of gas, forming a detonation.

    In this study, results available in the literature were confirmed, and 
    for the very first time simulations of hotspots with realistic background
    conditions were performed. This was done by extracting profiles from
    three-dimensional turbulence and using them as the initial conditions for
    direct numerical simulations of hotspots. The results indicate that despite
    best efforts, the outcome of the simulation cannot be decided by a simple
    parameterization.

    To `learn' the features of the hotspot that lead to the formation of a
    detonation wave, a novel neural-network based strategy was introduced. The
    networks were successfully trained on the large database of DNS studies
    used in the previous analysis.  The network learning procedures demonstrated
    high Accuracy and AUC scores for both the training and validation sets. Finally
    the use of the trained networks as a subgrid-scale model in reactive turbulence
    was investigated. It was found that the models were able to predict the onset
    of DDT.

    One of the major highlights of the present work is the large database of
    DNS studies of realistic hotspot configurations. Together with the newly
    proposed network approaches and future tweaking of the feature selection step
    as well as the network architectures, even better results could be achieved. An
    area where more work can be done if computational resources are available is in
    the generation of even more DNS studies of hotspot configurations. In
    particular, it would be most beneficial to expand the range of ambient
    densities in these models.

    Of course, one of the most obvious limitations of the studies is the
    restriction of the DNS of hotspots to one spatial dimension. While the
    Zel'dovich mechanism is inherently a one-dimensional mechanism, the presence of
    curves, or channels in the induction time field is expected to complicate the
    situation. These types of convoluted induction time channels are not uncommon
    in the distributed burning regime, and require special attention in any future
    work on this subject.

    When it comes to generating new training data for more advanced models, an
    active learning approach \cite{settles2012} may prove very beneficial,
    especially if multi-dimensional DNS of hotspots is to be performed. Such
    studies would be very costly, and rather than sample the input parameter space
    randomly, it would be more time-efficient and resource-efficient to samples
    in regions where the network has the greatest uncertainty. In this work, the
    convergence of the network with respect to the number of training samples was
    presented, providing a useful estimate of the cost of developing new databases
    of DNS hotspot simulations. 




\appendix
\chapter{Analysis of Turbulent Perturbations}
\label{chp:turbanalysis}

  In this appendix the statistics of temperature perturbations in
  three-dimensional reactive turbulence simulations is presented. This
  information is used to guide the design of computational hotspot experiments,
  as the hotspot configurations studied should reflect the physical system as
  closely as possible.

\begin{figure}[!htb]
    \center
    \includegraphics[width=0.95\textwidth]{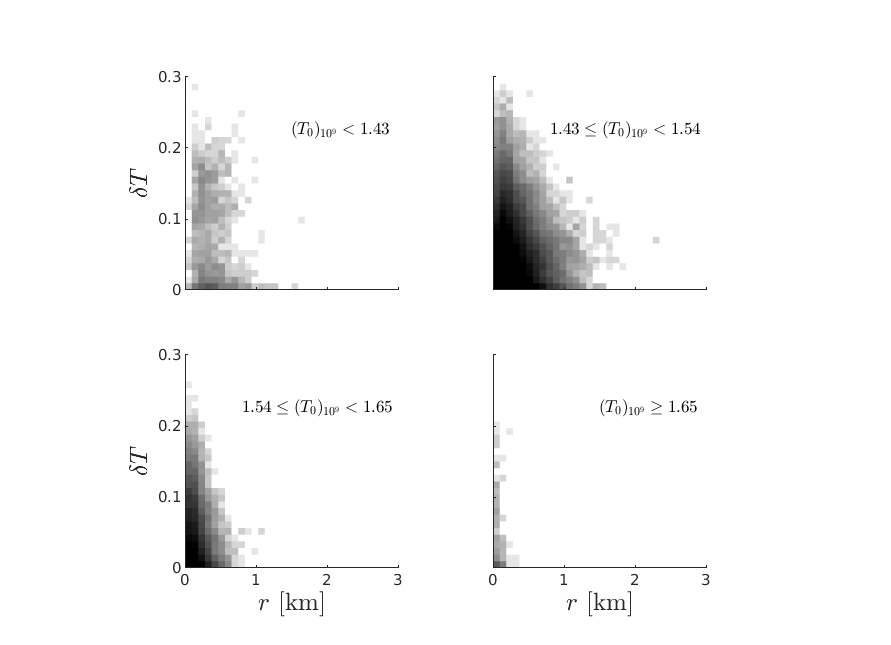}
    \caption{Statistics of temperature perturbations near the end of the
      self-heating phase in three-dimensional reactive turbulence
      simulations. The distributions are categorized into four temperature ranges.}
    \label{fig:wdstirperturbations}
  \end{figure}

  In \Cref{fig:wdstirperturbations}
  four bivariate histograms of the amplitudes and radii of temperature
  perturbations from a 3D reactive turbulence model are shown. The bivariate
  histograms each consider perturbations with ambient temperatures that are
  within a range of values. The first range is $T_{\mathrm{amb}} <
  \unit[1.43\times10^{9}]{K}$, the second is $\unit[1.43\times10^{9}]{K} \leq
  T_{\mathrm{amb}} < \unit[1.54\times10^{9}]{K}$, the third is
  $\unit[1.54\times10^{9}]{K} \leq T_{\mathrm{amb}} <
  \unit[1.65\times10^{9}]{K}$, and the fourth is $T_{\mathrm{amb}} \ge
  \unit[1.65\times10^{9}]{K}$.

  Given that the ambient density in the model is approximately $\unit[1 \times
  10^{7}]{g\ cm^{-3}}$ this means that temperatures below $\unit[1.43 \times
  10^{9}]{K}$ are very cold and not near ignition. For the perturbations
  with ambient temperatures in the range of $\unit[1.43 \times 10^{9}]{K}$
  through $\unit[1.54 \times 10^{9}]{K}$, the peak temperature can reach around
  $\unit[2\times10^{9}]{K}$. The perturbations with ambient temperatures in the
  range of $\unit[1.54 \times 10^{9}]{K}$ through $\unit[1.65 \times
  10^{9}]{K}$ are in the detonation range with peak temperatures around
  $\unit[2.1\times10^{9}]{K}$.

\begin{figure}[!htb]
    \center
    \includegraphics[width=0.875\textwidth]{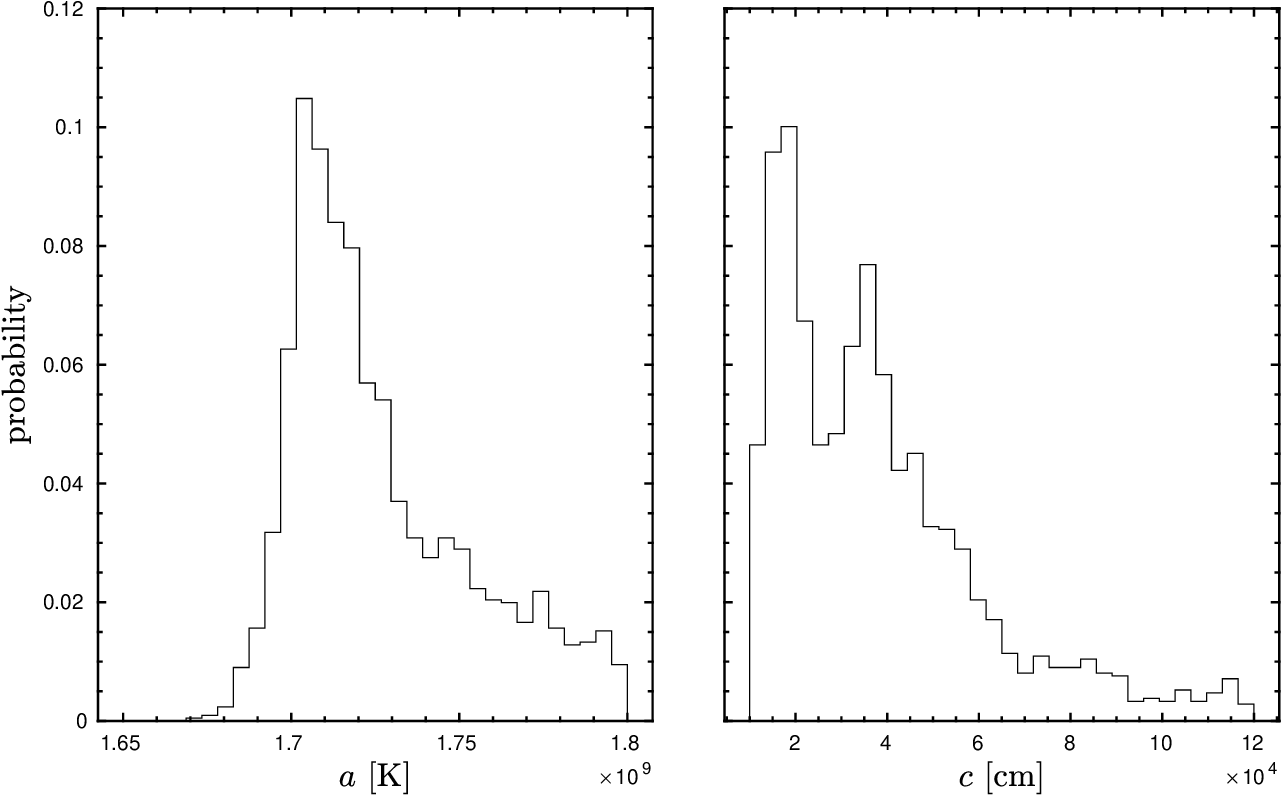}
    \caption{Histograms of the calculated amplitude (left panel) and standard deviation
      (right panel) of the fitted functions to the turbulence extracted
      profiles.}
    \label{fig:histoturbflucutations}
  \end{figure}

  Next the short induction time regions of the profiles extracted
  from three-dimensional turbulence used to form the \texttt{TE} database of
  simulations are analyzed. A regression analysis is performed using the Matlab fitting
  routine, \texttt{fit}, on each profile. The fit is performed with a Gaussian
  function with amplitude, $a$, mean value, $b$, and standard deviation, $c$.
  In \Cref{fig:histoturbflucutations}
  the probabilities of fitted amplitude and standard deviation values from the
  turbulent profiles are presented. The majority of fitted amplitudes are at
  around $\unit[1.7\times10^{9}]{K}$, with values reaching as high as
  $\unit[1.8\times10^{9}]{K}$ for the samples considered. The radii of the
  perturbations are of greater interest. It is observed that the greatest
  concentration of perturbations are small, with standard deviation values of
  roughly $\unit[2\times10^{4}]{cm}$. Then the distribution decays with the
  exception of another smaller concentration at around
  $\unit[3.5\times10^{4}]{cm}$. The tail of the distribution extends to about
  $\unit[1.2\times10^{5}]{cm}$; perturbations this large are in significantly
  smaller concentrations.

\begin{figure}[!htb]
    \center
    \includegraphics[width=0.6\textwidth]{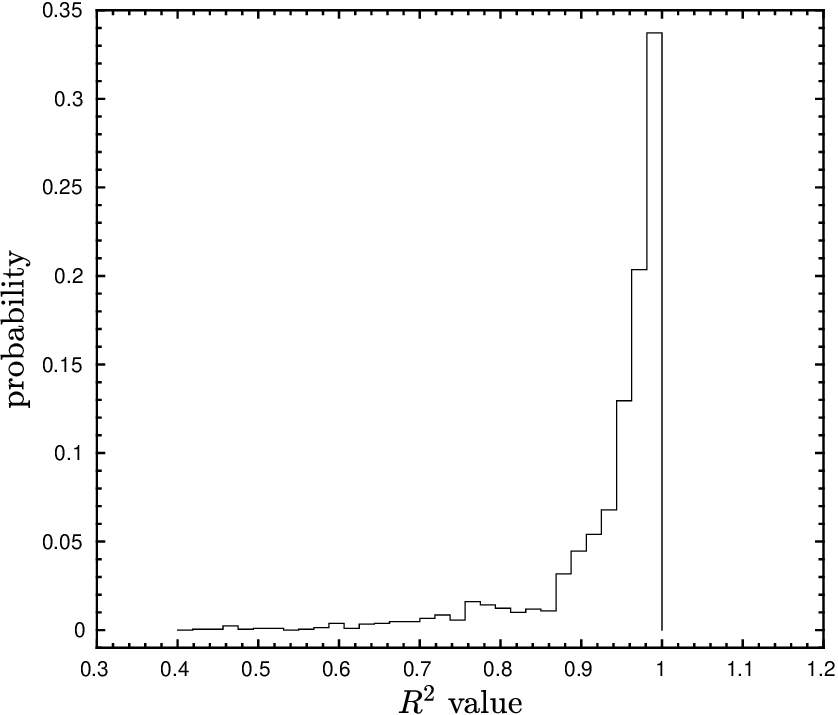}
    \caption{Histogram of the $R^{2}$ goodness-of-fit statistic for the fitted
    functions to the turbulence extracted profiles.}
    \label{fig:historsquared}
  \end{figure}

  The choice of a Gaussian for describing the natural perturbations in the
  reactive turbulence data is examined. The $R^{2}$ goodness-of-fit statistic
  is used for this purpose. In \Cref{fig:historsquared}
  we show the $R^{2}$ goodness-of-fit statistic, also computed in Matlab, for
  the fitting process just described. An $R^{2}$ value of $1$ is ideal, whereas
  lower values indicate greater deviation of the shape of the perturbation from
  a Gaussian. It is observed that the $R^{2}$ value is above $0.9$ for the
  majority of fits. This result provides good support for the use of a Gaussian
  as the initial condition for the synthetic hotspots. However the tail region
  of the distribution motivates the study of hotspots with realistic initial
  conditions.
\chapter{Data Preparation, Data Augmentation, and Hyperparameter Tuning}
\label{chp:nnappendix}

  \section{Data preparation}

    The range of values of the inputs to the neural networks is very large.
    Between the induction time and density for instance, values span roughly
    $12$ orders of magnitude. The induction time itself may range from around
    $\unit[10^{-5}]{s}$ to greater than $\unit[10^{-1}]{s}$ for a single input
    sample.  These input data must be appropriately scaled before being passed
    to a neural network. 

    The appropriate scales are determined by computing the maximum and minimum
    values for each variable over all of the input samples. The inputs are
    scaled between these values such that the minimum and maximum of the new
    values are between $0$ and $1$. To treat the wide range of values in the
    induction time and some other quantities, log-scaling is used.

  \section{Data augmentation}

    Data augmentation strategies are employed in a few instances in the present
    work. The first use is in the evaluation of hotspot configurations to be
    simulated. To construct the turbulence extracted (\texttt{TE}) database, a
    total of $933$ profiles are extracted from a 3D reactive turbulence
    dataset. These conditions are not simulated directly as their temperature
    in some cases is too low to achieve rapid burning. First a linear scaling
    of the temperature is applied to the original extracted temperature
    profiles, creating several offshoot models. The temperature is modulated
    such that the maximum lies between $\unit[2.0 \times 10^{9}]{K}$ and
    $\unit[2.3 \times 10^{9}]{K}$. These new profiles form the \texttt{TE}
    database.

    Before the process of training on the \texttt{td.naive.tburn} dataset,
    another form of data augmentation is utilized. This occurs during the
    downsampling process from the high-resolution DNS data to the
    $\Delta_{\mathrm{tb}}$ scale. The cell averaging window is shifted slightly
    during the downsampling of interpolated profiles in order to (1) create
    more offshoot training samples and (2) hopefully improve the
    generalizability of the trained networks.

  \section{Hyperparameter tuning}

    The tuning of CNN hyperparameters is performed using the Keras Tuner
    \citep{omalley2019} Python library. A subset of the training and validation
    samples was used for this tuning procedure, with the training set and
    validation set each having a ratio of detonating cases to non-detonating cases
    of about 65:100.

    The CNN architecture hyperparameter tuning routine considered as variables
    the number of convolutional layers, the number of convolutional filters
    (varying independently), the filter size (also varying independently), the
    number of fully connected layers, the number of nodes per layer (varying
    independently), and the percentage of dropout. The dropout is set to be the
    same for each layer. The summary of hyperparameters is included in
    \Cref{tbl:hyptuning}.
    \begin{table}
      \centering
      \begin{tabular}{@{}lcl@{}}\toprule
        Parameter & Description & Range \\\midrule
        \ \ \texttt{nclyrs} & number of conv. layers & $\left[1, 2\right]$ \\
        \ \ \texttt{nfltrs} & number of filters & $\left[3, 5, 7\right]$ \\
        \ \ \texttt{fltrsz} & filters size & $\left[8, 16, 32, 64, 128\right]$ \\
        \ \ \texttt{dopool} & use pooling or not & $\left[0, 1\right]$ \\
        \ \ \texttt{ndlyrs} & number of dense layers & $\left[8, 16, 32, 64, 128\right]$ \\
        \ \ \texttt{nnodes} & number of units/nodes& $\left[0, 128, 256, 512\right]$ \\
        \ \ \texttt{drpout} & percentage of dropout & $\left[0, 0.25, 0.5\right]$ \\\bottomrule
      \end{tabular}
      \caption{Summary of hyperparameters allowed to change freely, and the
        range of values that they are allowed to assume in the hyperparameter
        tuning process.}
      \label{tbl:hyptuning}
    \end{table}
    The \textit{BayesianOptimization} routine (see \citep{snoek2012}) is run
    for each of the four input models involved in the tuning process:
    $\mathbb{N}2$ $\mathbb{N}3$, $\mathbb{N}4$, and $\mathbb{N}5$. The AUC
    score averaged over several of the previous epochs is used as a tuning
    objective function.  The number of training epochs per trial is $40$.
\chapter{Neural Network Model Evaluation in Proteus}
\label{chp:sgsminproteus}

  In this appendix the custom-made Fortran and Python software routines used to
  track and evaluate prospective DDT hotspots (hereafter referred to as
  kernels) during runtime of the Proteus simulation are described. These
  routines are part of an add-on package for Proteus called
  \texttt{SubGridScaleDDT}.

  The \texttt{SubGridScaleDDT} software exists as a module of
  \texttt{physics/sourceTerms} in Proteus and can be included in the code by
  adding its path to the `Config' file before simulation setup. The purpose of
  the main routine, \texttt{SubGridScaleDDT.F90}, is to find all of the
  prospective kernels at the current timestep. To do this a call is made to
  \texttt{SubGridScaleDDT\_findAllKernels}. This routine searches through the
  simulation domain for computational cells satisfying a user-specified minimum
  induction time (cells with induction time values above this are unlikely to
  detonate and not worth tracking). The prospective DDT kernel is identified by
  evaluating first and second derivatives of induction time over the entire
  adaptive mesh refinement block to determine whether or not a point is a local
  minimum. Once a local minimum is found the kernel is added to the tracking
  list and given a unique tracking ID. The pseudo-code implementation is
  included in \Cref{alg:subgridscaleddt}.
  \begin{algorithm}
    \caption{Track and evaluate prospective DDT Kernels.}
    \label{alg:subgridscaleddt}
    \begin{algorithmic} 
      \STATE {\texttt{call SubGridScaleDDT\_findAllKernels()}}
      \FOR{all blocks}
        \STATE {\texttt{call} \texttt{SubGridScaleDDT\_findLocalMinima()}}
      \ENDFOR
      \FOR{all kernels}
        \STATE {\texttt{call} \texttt{SubGridScaleDDT\_evalKernel()}}
      \ENDFOR
    \end{algorithmic}
  \end{algorithm}
  Once it is determined that a kernel is a local minimum of induction time,
  then for the naive strategy an array consisting of nearby cells, centered
  about the minimum, is extracted by the routine
  \texttt{SubGridScaleDDT\_getLineout}. This extracted array is then passed to
  the routine \texttt{SubGridScaleDDT\_callNN}. This routine serves as a
  wrapper around the python caller.

  The neural network implementation uses the Python package, Keras. Keras serves
  as a front-end to the Tensorflow library. In order to embed the trained
  network in the runtime execution of the Proteus simulation code, the Python
  package CFFI (C-foreign function interface) was utilized.  The CFFI package
  generates C code that exports the user's application programming interface
  (API) to a shared object library file. This library file can then be linked
  to any Fortran or C program.

  The first step toward using the Python CFFI to build the C headers and
  dynamic shared object library file for Proteus is to edit the
  \texttt{nn\_builder.py} script. In this script the desired Keras model name and
  appropriate inputs and outputs must be defined. Running the
  \texttt{nn\_builder.py} script automatically generates the C header, C code,
  and shared object library file.

  Once the necessary files are built, there are several steps required before
  running Proteus with the SubGridScaleDDT module and neural network model
  active. First the shared object library file must be added to the shared
  library path shell variable, \texttt{LD\_LIBRARY\_PATH}. Second, the Keras
  model directory must be copied in the directory where the executable is
  expected to be run. Once these steps are completed the Proteus executable is
  ready to be run.






\bibliography{dissertation}
\bibliographystyle{abbrv}

\begin{biosketch}
  Brandon Gusto began his undergraduate studies at Florida State University in
  2013, majoring in mechanical engineering. He completed his Bachelor of
  Science degree in the Spring of 2017 with a double major in mechanical
  engineering and applied mathematics. He decided to stay at Florida State to
  continue his studies as a PhD student in the Department of Scientific
  Computing after meeting with professors and discussing research plans. While
  in the department Brandon studied finite volume schemes and wavelet-based
  adaptive mesh refinement in application to astrophysics problems under the
  supervision of his advisor, Dr. Plewa. In 2020 he was awarded the SMART
  scholarship from the Department of Defense. Upon successfully defending his
  dissertation, Brandon is expected to join the Naval Undersea Warfare Center
  Division Newport in Newport, Rhode Island as a civilian research scientist.
\end{biosketch}

\end{document}